\documentclass[onecolumn]{aa} 
\usepackage{graphicx}
\usepackage{float}
\usepackage{caption}
\usepackage{txfonts}
\usepackage{wasysym}
\usepackage{ulem}
\usepackage{bigdelim}
\usepackage{bigstrut}
\usepackage{multirow}
\usepackage[load-configurations=version-1]{siunitx}
\usepackage{booktabs,dcolumn}
\usepackage{lineno}
\begin{document}
   \title{Pre-conditioned Backward Monte Carlo solutions to radiative transport in planetary atmospheres
   }

   \subtitle{Fundamentals: Sampling of propagation directions in polarising media}

   \author{}
   \author{A. Garc\'ia Mu\~noz
          \inst{1}
          \and
          F. P. Mills\inst{2,3}
          }
   \institute{ESA Fellow, ESA/RSSD, ESTEC, 2201 AZ Noordwijk, The Netherlands\\   
    \email{tonhingm@gmail.com}
    \and
    Research School of Physics and
Engineering and Fenner School of Environment and Society, Australian
National University, Canberra, ACT 0200, Australia
\email{frank.mills@anu.edu.au}
\and
Space Science Institute, Boulder, CO 80301, USA
    }


 
  \abstract 
  {The interpretation of polarised radiation emerging from a planetary 
  atmosphere must rely on solutions to the vector Radiative Transport Equation (vRTE).
  Monte Carlo integration of the vRTE is a valuable approach 
  for its flexible treatment of complex
  viewing and/or illumination geometries 
  and because it can intuitively incorporate elaborate physics.
  }
  {  
  We present a novel Pre-Conditioned Backward Monte Carlo (PBMC) algorithm 
  for solving the vRTE
  and apply it to planetary atmospheres irradiated from above.  
  As classical BMC methods, our PBMC algorithm builds the solution 
  by simulating the photon trajectories from the detector towards the radiation source, 
  i.e. in the reverse order of the actual photon displacements.
  }
  {
  We show that the neglect of polarisation in the  
  sampling of photon propagation directions in classical BMC algorithms
  leads to unstable and biased solutions for conservative, optically-thick, 
  strongly-polarising media such as Rayleigh atmospheres.
  The numerical difficulty is avoided by pre-conditioning 
  the scattering matrix with information from 
  the scattering matrices of prior (in the BMC integration order) photon collisions. 
  Pre-conditioning introduces a sense of history in the photon 
  polarisation states through the simulated trajectories. 
  }
{The PBMC algorithm is robust and its accuracy is extensively 
demonstrated via comparisons with examples drawn from the literature for 
scattering in diverse media. Since the convergence rate 
for MC integration is independent of the integral's dimension, the scheme
is a valuable option for estimating the disk-integrated signal of stellar
radiation reflected from planets. 
Such a tool is relevant in the prospective investigation of exoplanetary phase curves. 
We lay out two frameworks for disk integration and, 
as an application, explore the impact of atmospheric stratification 
on planetary phase curves for large star-planet-observer phase angles.
By construction, backward integration provides a better control than forward integration 
over the planet region contributing to the solution, and this presents a clear advantage
when estimating the disk-integrated signal at moderate and large phase angles.
}


   \titlerunning{Backward Monte Carlo modelling of planet polarisation}
   \maketitle
%



\section{\label{intro_sec} Introduction}

The gases and aerosols that make up a planetary atmosphere 
leave characteristic signatures on the radiation emitted and/or 
reflected from the planet. 
The technique of polarimetry utilises the polarisation state of emergent 
radiation to investigate the planet's atmospheric optical properties. 
Polarimetry is relevant in the remote sensing of planetary atmospheres both 
as a stand-alone technique and in combination with photometry. 
In the Solar System, polarimetric observations made
from space-borne and ground-based telescopes 
have yielded insight into the gas and aerosol envelopes of
Earth (Dollfus, \cite{dollfus1957}; Hansen \& Travis, \cite{hansentravis1974}), 
Venus (Coffeen, \cite{coffeen1969}; Hansen \& Hovenier, \cite{hansenhovenier1974}), 
Mars (Santer et al., \cite{santeretal1985}), 
Jupiter and Saturn 
(Morozhenko \& Yanovitskii, \cite{morozhenkoyanovitskii1973};
Schmid et al., \cite{schmidetal2011}; 
West et al., \cite{westetal1983}),
Titan (Veverka, \cite{veverka1973}; West \& Smith, \cite{westsmith1991}), and
Neptune and Uranus (Joos \& Schmid, \cite{joosschmid2007}; 
Michalsky \& Stokes, \cite{michalskystokes1977};
Schmid et al., \cite{schmidetal2006}). 

Various spacecraft for Earth (ADEOS I and II, PARASOL)
and Solar System exploration (e.g. Voyager, Galileo, Cassini)
carried instrumentation  with (limited) polarimetric capabilities. 
Most modern ground-based observatories are equipped with polarimeters
for either spectroscopy or imaging. Ground-based observations of the outer planets, 
however, have only partial coverage of the Sun-target-Earth phase angle, which
limits the possible physical insight from polarimetric investigations. 
For the above reasons, it is generally agreed that polarimetry's potential 
for characterising the atmospheres of Earth and the rest of the 
Solar System planets remains underexploited. 
Interestingly, the discovery of planets 
orbiting stars other than our Sun has caused a renewed interest in 
polarimetry as both a detection and characterisation technique.
The key idea behind this new interest is that
stars are typically unpolarised or weakly polarised, whereas planets may be
partially polarised, 
which presents an advantage for the separation of the planet from the 
glare of its host star (e.g. Seager et al., \cite{seageretal2000};
Stam et al., \cite{stametal2004}).

The new-born field of exoplanet research is prompting significant effort
in the development of polarimetric facilities, 
as demonstrated by proposed space missions such as
ESA's SPICES (Boccaletti et al., \cite{boccalettietal2012}) or dedicated 
instrumentation for Gemini (Macintosh et al., \cite{macintoshetal2006})
or ESO's Very Large Telescope and European-Extremely Large Telescope
(Beuzit et al., \cite{beuzitetal2008}; Kasper et al., \cite{kasperetal2008}).  
Correspondingly, on the theoretical front, there has been work to 
investigate polarimetry's potential for identifying planets' orbital parameters 
as well as for characterising their main atmospheric and surface features
(e.g. 
Bailey, \cite{bailey2007}; 
Fluri \& Berdyugina, \cite{fluriberdyugina2010};
Seager et al., \cite{seageretal2000}; 
Stam and collaborators, \cite{stametal2004, stam2008}, 
but also
Karalidi \& Stam, \cite{karalidistam2012}, 
Karalidi et al.,
\cite{karalidietal2011,karalidietal2012,karalidietal2013};
Williams \& Gaidos, \cite{williamsgaidos2008}; 
Zugger et al. \cite{zuggeretal2010,zuggeretal2011}).  
As the number of exoplanets already surpasses the number of Solar System
planets, theoretical investigations that explore
 gas, cloud and surface properties,
possibly in the framework of a new generation of General Circulation Models, 
will continue to play a key role 
in the prediction and prospective characterisation of exoplanetary observables.

This paper is devoted to the numerical modelling of radiation scattered by
planetary atmospheres. Our approach relies on Backward Monte Carlo (BMC) 
integration of the vector Radiative Transport Equation (vRTE). 
Special attention is paid to the sampling of propagation 
directions in polarising media. 
We show that in classical BMC integration, failing to account for polarisation in the sampling of
propagation directions may destabilise and bias the numerical 
solution in conservative, optically-thick, strongly-polarising media. 
We propose a Pre-conditioned BMC (PBMC) algorithm and show that 
pre-conditioning the scattering matrix with information from 
prior collisions (in the order of backward integration) 
eliminates the numerical difficulties. 
Pre-conditioning is equivalent to providing information about the history and
polarisation state of photons through their simulated trajectories. 
We describe in detail the algorithm and its performance. 
Because it consistently delivers precisions of 10$^{-4}$ when compared 
to solutions that are accurate to at least that level, 
the algorithm may be considered 'exact' (in the de Haan et al., \cite{dehaanetal1987} sense)
or nearly so.
This paper is part of an ongoing effort to build a tool for the efficient 
simulation of the radiation emerging from both disk-resolved and disk-integrated
realistic planetary atmospheres. 
In its scalar form, the algorithm has already been used without description
(Garc\'ia Mu\~noz \& Pall\'e, \cite{garciamunozpalle2011};
Garc\'ia Mu\~noz \& Mills, \cite{garciamunozmills2012};
Garc\'ia Mu\~noz et al., \cite{garciamunozetal2011, garciamunozetal2012,garciamunozetal2014}). 
The cases investigated here focus on Rayleigh and Mie scattering, for which the scattering
matrix is easy to obtain. 
The theory is more general than that and should also 
apply to scattering particles with different scattering matrices.

The paper is structured as follows. In \S\ref{bmc_section}, 
we note some of the differences between forward and backward integration. 
BMC algorithms are very selective with the planet regions that they probe, and this is
a clear advantage, for instance, 
when producing the disk-integrated signal from a planet at a specified phase angle.
We overview the fundamentals of BMC algorithms and discuss the 
sampling of photon propagation directions in classical BMC algorithms and in our PBMC approach; 
we also present two different schemes for integration of the net
radiation reflected from a spherical-shell planet. 
In \S\ref{planeparallel_sec}, we assess the performance of the classical and pre-conditioned 
algorithms with test cases for plane parallel configurations. 
In \S\ref{disk_section}, we predict a number of planetary phase curves.
The extensive suite of test cases considered will hopefully help guide 
the decision of potential users of the PBMC algorithm.
Finally, in \S\ref{summary_section} we summarise the main conclusions and comment on follow-up work.

\section{\label{bmc_section}The BMC algorithm}

MC algorithms for radiative transport fall within the general class of Markov chain methods for 
the statistical simulation of photon collisions in scattering media
(Cashwell \& Everett, \cite{cashwelleverett1959}; 
Marchuk et al., \cite{marchuketal1980}).
By using appropriate statistical estimators, 
MC algorithms can estimate the radiation within and emerging from a medium.

MC algorithms are classified as forward or backward (FMC and BMC, respectively), 
depending on whether the solution is built by simulating the photon 
trajectories from the radiation source towards the observer or vice versa. 
FMC algorithms account easily for the photon's polarisation state in the 
sampling of the photon propagation direction following a collision
(e.g. 
Bartel \& Hielscher, \cite{bartelhielscher2000};
Bianchi et al., \cite{bianchietal1996}; 
Cornet et al., \cite{cornetetal2010}; 
Fischer et al., \cite{fischeretal1994}; 
Hopcraft et al., \cite{hopcraftetal2000}; 
Kastner, \cite{kastner1966}; 
Schmid, \cite{schmid1992}; 
Whitney, \cite{whitney2011}). 
That is not immediately possible in BMC algorithms because the 
scattering events are treated in the reverse order that they actually occur.
BMC algorithms generally treat the sampling of propagation directions 
by omitting the radiation's polarisation state 
and correcting subsequently for the bias introduced
(Collins et al., \cite{collinsetal1972} and works thereafter, e.g.:  
 Emde et al., \cite{emdeetal2010}; 
 Gay et al., \cite{gayetal2010}; 
 Oikarinen, \cite{oikarinen2001}).
As shown below, that approach may fail to render accurate solutions
in conditions for which the photon scattering directions are 
strongly influenced by their polarisation states. 
 
MC algorithms are 'exact' in the sense (de Haan et al., \cite{dehaanetal1987})  
that their accuracy is in principle limited only by the 
number of photon trajectory simulations.
Thus, MC algorithms are often used as standards in the validation of
other methods, particularly in cases that involve complex 
viewing and/or illumination geometries 
(Loughman et al., \cite{loughmanetal2004}; Postylyakov, \cite{postylyakov2004}). 

BMC algorithms are better suited for problems with small detectors and large
radiation sources, the opposite being true for FMC algorithms 
(Modest, \cite{modest2003}). 
This important distinction means that BMC integration 
turns out to be the appropriate choice for numerous
applications in the investigation of planetary atmospheres. 
By tracing the photon trajectories from the detector towards the planet (or towards 
a part of the planet that is known to be illuminated), 
BMC algorithms offer 
a more efficient approach to achieve a desired accuracy.
This is not directly possible in the FMC framework because there is no 
a priori knowledge about the directions the photons will exit the medium.
In FMC algorithms, further, 
estimating the emergent radiation typically requires averaging over a range of 
exiting directions. 
(Alternatively, variance reduction techniques such as the next-event point-estimator
can be utilised, e.g. Kaplan et al., \cite{kaplanetal2001} and 
Lux \& Koblinger, \cite{luxkoblinger1991}, their efficiency being strongly dependent on 
the detector's acceptance angle.)
These characteristics penalise the computational efficiency of FMC algorithms,
especially when only a specified number of viewing geometries with 
narrow acceptance angles are of interest.

Additional properties that make FMC/BMC algorithms appealing
in their application to planetary atmospheres include:
\begin{itemize}

\item They are easy to implement and debug. 
Their description can indeed be accomplished in less than one page 
(see Appendix \ref{appendixa}). 

\item The implementation of the scattering matrix does not require a series expansion of the
matrix elements. 

\item Curvature and twilight effects are naturally accounted for. Limb-viewing geometries
do not require a special treatment.

\item It is easy to separate the contributions 
from the atmosphere and surface, 
or from different atmospheric layers, 
or from various orders of scattering. 

\item 
Scattering by large particles, which lead to highly asymmetric
scattering phase functions,  
can be treated without significant computational penalty.

\item The computational cost for solving the vRTE and its scalar counterpart are comparable. 

\item The accuracy of the solution depends on the number
of photon trajectory simulations. 
Moderate-accuracy solutions can be obtained at small computational costs.

\item 
In BMC algorithms,  
each photon collision can be utilised to estimate the contribution to the detector
from various incident directions of the illuminating source.

\end{itemize}

Our implementation of the algorithm follows the basic layout by O'Brien (\cite{obrien1992, obrien1998}), 
that we extend to include polarisation. 
The implementation makes use of variance reduction techniques, 
which arise logically from the mathematical elaboration of the 
integrals that occur in the formal solution to the vRTE. 
O'Brien (\cite{obrien1992, obrien1998}) provides an excellent introduction to these ideas, and we follow to a 
large extent the nomenclature in those works.

\subsection{\label{fundamentals_section}Fundamentals}

\begin{figure}[h]
   \centering
   \includegraphics[width=6.cm]{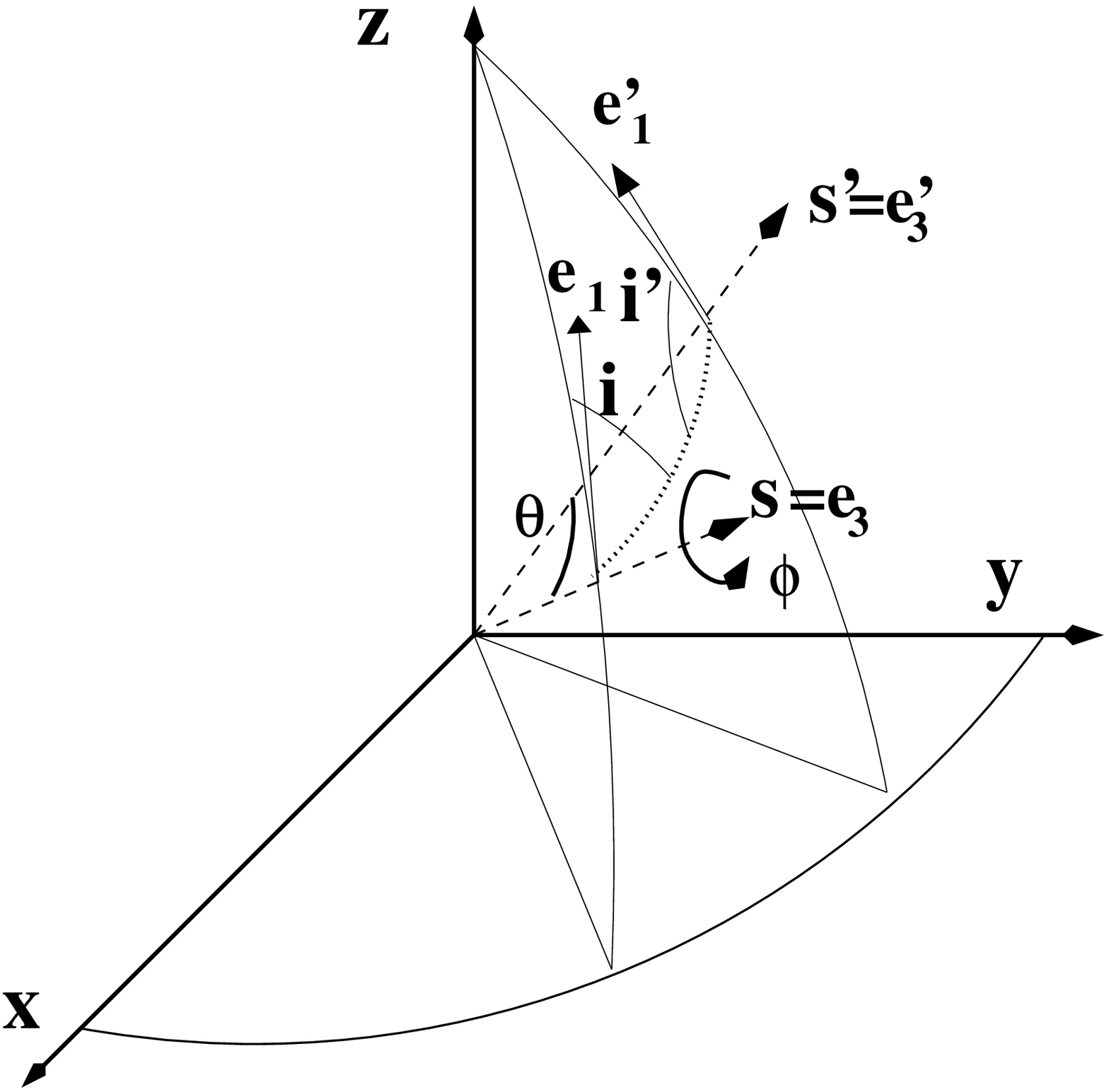} \\
   \includegraphics[width=6.cm]{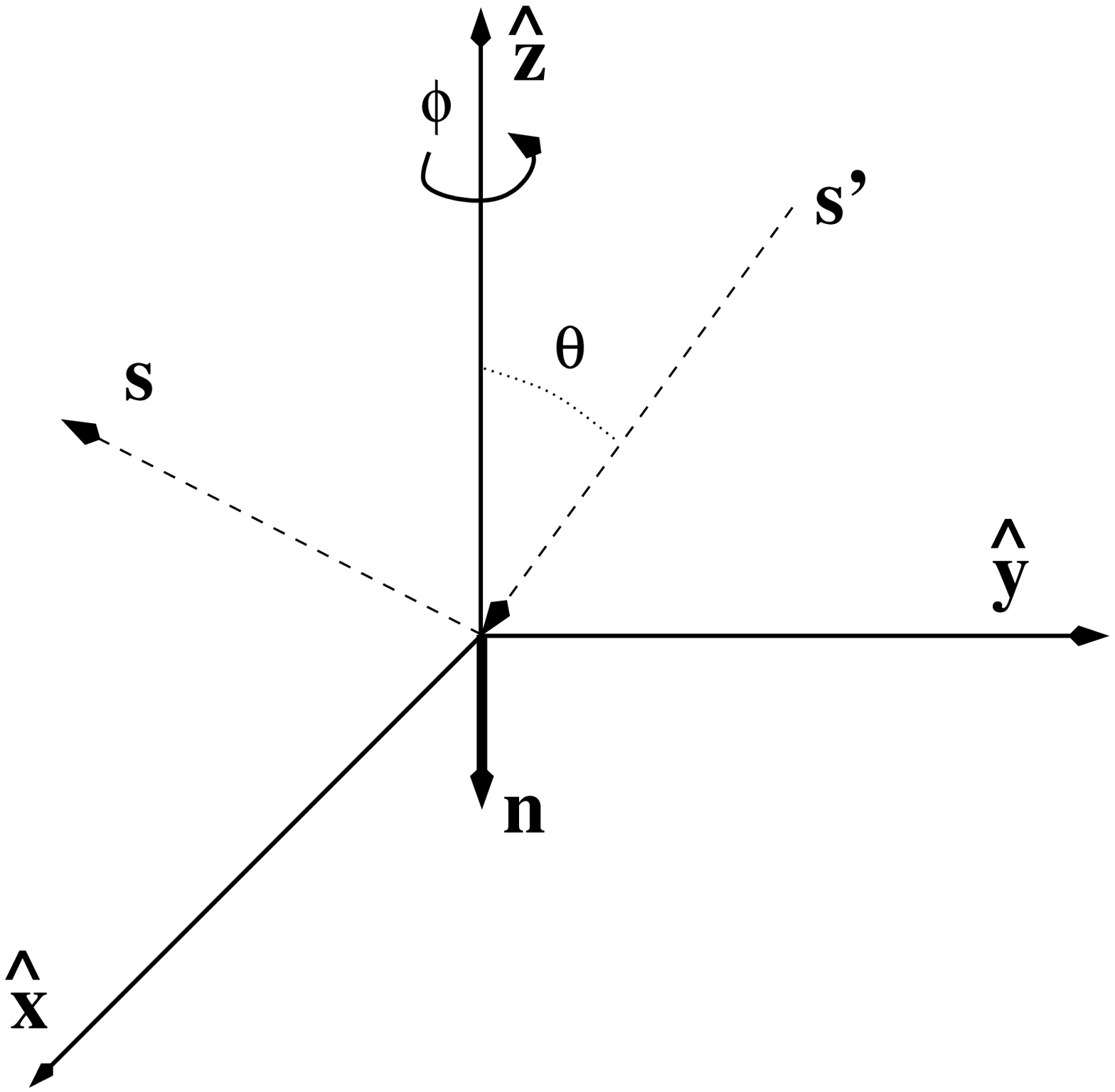} \\
    \caption{\label{sketch1_fig} Top. Definition of the \textit{incident}, $\mathbf{s'}$, and 
    \textit{emergent}, $\mathbf{s}$, photon directions at a scattering event within the atmosphere.
    The $xyz$ axes form a rest reference frame fixed to the planet. 
    The differential solid angle $d\Omega(\mathbf{s'})$=$\sin{\theta}d\theta d\phi$ 
    is defined with $\mathbf{s}$ serving as polar axis. 
    Angles $\theta$$\in$[0, $\pi$] and
    $\phi$$\in$[0, 2$\pi$].     
    In the backtracing of photons of BMC algorithms, 
    $\mathbf{s}$ is known at each collision and $\mathbf{s'}$ must
    be sampled from the relevant scattering phase function. 
    Angles $i'$ and $i$, both $\in$[0, $\pi$], are needed for consistent referencing 
    of the Stokes vector throughout the scattering process. 
    Vectors \{$\mathbf{e'_1}$, $\mathbf{e'_2}$ and $\mathbf{e'_3}$\} 
    and
    \{$\mathbf{e_1}$, $\mathbf{e_2}$ and $\mathbf{e_3}$\} 
    define right-handed coordinate systems at the meridional planes of the incident
    and emergent photon directions, respectively. 
    Bottom. Definition of the \textit{incident}, $\mathbf{s'}$, and 
    \textit{emergent}, $\mathbf{s}$, photon directions at a reflection event at the 
    local surface (plane $\hat{x}\hat{y}$). Here, $\mathbf{n}$ is the inward-pointing normal 
    vector at the surface, and $\hat{z}$ is oriented along $-\mathbf{n}$.
    The differential solid angle $d\Omega(\mathbf{s'})$=$\sin{\theta}d\theta d\phi$ 
    is defined with $\hat{z}$ serving as polar axis.  
    }
   \end{figure}

Our interest lies in the vRTE for a scattering and absorbing 
medium without volume or surface emission sources:
\begin{equation} 
\mathbf{s}\cdot \nabla \mathbf{I(x,s)}=-\gamma(\mathbf{x}) 
\mathbf{I(x,s)} + 
\beta(\mathbf{x}) 
 \int_{\Omega} d\Omega(\mathbf{s'}) \mathbb{P}\mathbf{(x,s,s')}\mathbf{I(x,s')},  
\label{integrodiff_eq}
\end{equation} 
where, $\mathbf{x}$ and $\mathbf{s}$ are vectors of position and direction, 
$\beta(\mathbf{x})$ and $\gamma(\mathbf{x})$ are the scattering and extinction
coefficients of the medium (independent of direction), 
and $d\Omega(\mathbf{s'})$ is the differential solid 
angle about direction $\mathbf{s'}$. 
The ratio 
$\varpi(\mathbf{x})$=$\beta(\mathbf{x})$/$\gamma(\mathbf{x})$ is the local 
single scattering albedo of the medium.
In terms of the $\theta$ and $\phi$ angles of Fig. (\ref{sketch1_fig}, Top), 
$d\Omega(\mathbf{s'})$=$\sin{\theta}d\theta d\phi$. 
$\mathbf{I(x,s)}$=$[I,Q,U,V]^{\rm{T}}$ is the Stokes vector that describes 
the polarisation state of radiation, and 
$\mathbb{P}\mathbf{(x,s,s')}$ is a 4$\times$4 matrix for deflection of radiation from 
the \textit{incident} direction $\mathbf{s'}$ to 
the \textit{emergent} direction $\mathbf{s}$. 
$\mathbb{P}\mathbf{(x,s,s')}$= $\mathbb{L}(\pi-i)$
$\mathbb{M}\mathbf{(x,s,s')}$ $\mathbb{L}(-i')$, and
$\mathbb{L}(\pi-{i})$ and $\mathbb{L}(-{i'})$ are rotation matrices for the conversion of the
Stokes vector from the meridional plane (the plane formed by the $z$ axis 
of a user-defined rest reference frame and the direction of 
photon propagation) to the scattering plane and vice versa. The rotation matrix is: 
\begin{equation}
\mathbb{L}(\kappa) = \left(
\begin{array}{rrrr}
1 & 0 & 0 & 0 \\
0 & \cos{2\kappa} & \sin{2\kappa} & 0 \\
0 & -\sin{2\kappa} & \cos{2\kappa} & 0 \\
0 & 0 & 0 & 1 \\
\end{array} \right)
\label{Limatrix_eq}
\end{equation}
with $\kappa$ being either $\pi-{i}$ or $-{i'}$, and angles 
$i'$ and $i$ defined as sketched in Fig. (\ref{sketch1_fig}, Top).
$\mathbb{M}\mathbf{(x,s,s')}$ is the scattering matrix, 
for which we assume  
$\mathbb{M}\mathbf{(x,s,s')}$=$\mathbb{M}\mathbf{(x,s\cdot s')}$=$\mathbb{M}\mathbf{(x,\cos{\theta})}$
and that it is normalised such that its (1, 1) entry verifies:
\begin{equation}
\int_{\Omega} d\Omega(\mathbf{s'}) \mathbb{M}_{1,1}(\mathbf{x,s\cdot s'}) = 1.
\label{normalisation_eq}
\end{equation}
In Mie scattering theory for spherical particles 
the matrix is fully prescribed by means of
four elements (Mishchenko et al., \cite{mishchenkoetal2002}): 
\begin{equation}
\mathbb{M}(\mathbf{x,\cos{\theta}}) = \frac{1}{4\pi}\left(
\begin{array}{rrrr}
a_1(\mathbf{x,{\theta}}) & b_1(\mathbf{x,{\theta}}) & 0 & 0 \\
b_1(\mathbf{x,{\theta}}) & a_1(\mathbf{x,{\theta}}) & 0 & 0 \\
0 & 0 & a_3(\mathbf{x,{\theta}}) & b_2(\mathbf{x,{\theta}}) \\
0 & 0 & -b_2(\mathbf{x,{\theta}}) & a_3(\mathbf{x,{\theta}}) \\
\end{array} \right).
\label{MMmatrix_eq}
\end{equation}
In the Rayleigh limit for particle sizes much smaller 
than the radiation wavelength, the four elements take on
analytical expressions that, neglecting anisotropy effects, are 
$a_1$=3($1+\cos^2{\theta}$)/4,
$b_1$=3($-1+\cos^2{\theta}$)/4,
$a_3$=3$\cos{\theta}$/2, and 
$b_2$=0. 

Equation (\ref{integrodiff_eq}) admits the formal solution 
for the Stokes vector at \{$\mathbf{x_{\rm{k}}, s_{\rm{k}}}$\}:
\begin{equation}
\mathbf{I(x_{\rm{k}},s_{\rm{k}})}=
t(\mathbf{x_{\rm{k}},x_{\rm{kb}}})
\mathbf{I(x_{\rm{kb}},s_{\rm{k}})}
\label{Ik_eq}
\end{equation}
$$
+\int_{\mathbf{x_{\rm{kb}}}}^{\mathbf{x_{\rm{k}}}} d\ell{_{\rm{ka}}}
t(\mathbf{x_{\rm{k}},x_{\rm{ka}}}) \beta(\mathbf{x_{\rm{ka}}})
 \int_{\Omega} d\Omega(\mathbf{s_{\rm{ka}}})
\mathbb{P(\mathbf{x_{\rm{ka}},s_{\rm{k}}, s_{\rm{ka}}})}\mathbf{I(x_{\rm{ka}},s_{\rm{ka}})}. 
$$
On the right hand side, the first term stands for radiation reflected 
from a point $\mathbf{x_{\rm{kb}}}$ at the boundary of the integration domain 
into direction $\mathbf{s_{\rm{k}}}$, whereas  
the second  term represents the radiation scattered within the medium
from \{$\mathbf{x_{\rm{ka}}, s_{\rm{ka}}}$\} to \{$\mathbf{x_{\rm{k}}, s_{\rm{k}}}$\}. 
Each term may include both diffuse and unscattered radiation components,
defined as the contributions from photons that have undergone at least one and zero
prior scattering collisions, respectively.
$d\ell{_{\rm{ka}}}$ stands for the arc-length along the path joining
$\mathbf{x_{\rm{kb}}}$ and $\mathbf{x_{\rm{k}}}$. 
The transmittance between $\mathbf{x_{\rm{ka}}}$ and $\mathbf{x_{\rm{k}}}$ is:
\begin{equation}
t(\mathbf{x_{\rm{k}},x_{\rm{ka}}})=
\exp{\lbrack -\int_{\mathbf{x_{\rm{ka}}}}^{\mathbf{x_{\rm{k}}}} 
d\ell'
\gamma(\mathbf{x}')\rbrack},
\end{equation}
and $t(\mathbf{x_{\rm{k}},x_{\rm{kb}}})$ is defined analogously.

It is useful to introduce the dimensionless variables:  
\begin{equation} 
\epsilon_{\rm{ka}}= \frac{t(\mathbf{x_{\rm{k}},x_{\rm{ka}}})-t(\mathbf{x_{\rm{k}},x_{\rm{kb}}})}{1-t(\mathbf{x_{\rm{k}},x_{\rm{kb}}})}, 
\label{chvar_eq} 
\end{equation} 
$a(\mathbf{x_{\rm{k}},x_{\rm{kb}}})$=1$-$$t(\mathbf{x_{\rm{k}},x_{\rm{kb}}})$, and 
$\varpi(\mathbf{x_{\rm{ka}}})$=$\beta(\mathbf{x_{\rm{ka}}})/\gamma(\mathbf{x_{\rm{ka}}})$,
that, by construction, range from 0 to 1, so that
the formal solution to Eq. (\ref{integrodiff_eq}) becomes:
\begin{equation}
\mathbf{I(x_{\rm{k}},s_{\rm{k}})}=
(1-a(\mathbf{x_{\rm{k}},x_{\rm{kb}}}))
\mathbf{I(x_{\rm{kb}},s_{\rm{k}})}
\label{Ik_eq2} 
\end{equation}
$$
+
a(\mathbf{x_{\rm{k}},x_{\rm{kb}}})
\int_{0}^{1} d\epsilon{_{\rm{ka}}}
\varpi(\mathbf{x_{\rm{ka}}})
\int_{\Omega} d\Omega(\mathbf{s_{\rm{ka}}})
\mathbb{P(\mathbf{x_{\rm{ka}},s_{\rm{k}}, s_{\rm{ka}}})}\mathbf{I(x_{\rm{ka}},s_{\rm{ka}})}.
$$
To evaluate Eq. (\ref{Ik_eq2}), 
boundary conditions at the top and bottom of the atmosphere are needed.  
We will here consider that the only source of illumination is stellar radiation from 
direction $\mathbf{s}_{\odot}$, for 
which the unimpeded, unpolarised irradiance is 
$\mathbf{F}_{\rm{\odot}}$=$\pi$$[1,0,0,0]^{\rm{T}}$$
\delta(\mathbf{s}'-\mathbf{s}_{\odot})$, 
with $\int d\Omega(\mathbf{s}')
\delta(\mathbf{s}'-\mathbf{s}_{\odot})$$\equiv$1.
Tacitly, the given $\mathbf{F}_{\rm{\odot}}$
assumes that the stellar size subtended from the planet is small so that 
the radiation incident on the planet is oriented in a single direction $\mathbf{s}_{\odot}$. 
We further assume Lambert reflection 
with albedo $r_g$ at the atmospheric bottom (the planet's surface) 
and a transparent atmospheric top for outgoing radiation. 

The surface reflectance properties
relate $\mathbf{I(x_{\rm{kb}},s_{\rm{k}})}$ to the incident Stokes vector at the boundary
$\mathbf{I(x_{\rm{kb}},s_{\rm{kb}})}$. For Lambert reflection at the atmospheric
bottom: 
\begin{equation} 
\mathbf{I}\mathbf{(x_{\rm{kb}},s_{\rm{k}})}=\frac{r_g(\mathbf{x_{\rm{kb}}})}{\pi} 
\int_{\Omega, \mathbf{s}_{\rm{kb}} \neq \mathbf{s}_{\rm{\odot}}}d\Omega(\mathbf{s_{\rm{kb}}})
\mathbf{n(x_{\rm{kb}})}\cdot\mathbf{s_{\rm{kb}}} 
\mathbb{D}\mathbf{I}\mathbf{(x_{\rm{kb}},s_{\rm{kb}} )} 
\label{Ikb_eq}
\end{equation}
$$
+ 
\frac{r_g(\mathbf{x_{\rm{kb}}})}{\pi} 
\mathbf{n(x_{\rm{kb}})}\cdot\mathbf{s_{\odot}} 
t\mathbf{(x_{\rm{kb}},x_{\rm{\odot}})} \mathbf{F}_{\odot}. 
$$
Here, $\mathbf{n(x_{\rm{kb}})}$ is the inward-pointing normal vector at the 
surface, $\mathbb{D}$ is the four-by-four depolarizing matrix with 
$\mathbb{D}_{1,1}$=1 as the only non-zero entry, 
and $\mathbf{I}\mathbf{(x_{\rm{kb}},s_{\rm{kb}} )}$ is the
Stokes vector for diffuse radiation reaching the surface. 
Figure (\ref{sketch1_fig}, Bottom) sketches the relevant geometrical parameters for 
photon collisions at the surface.
The two terms of Eq. (\ref{Ikb_eq}) are the separate contributions to 
$\mathbf{I}\mathbf{(x_{\rm{kb}},s_{\rm{k}})}$
from both diffuse radiation and from unscattered stellar radiation reaching the surface.
By definition of transparent atmospheric top, 
$\mathbf{I}\mathbf{(x_{\rm{kb}},s_{\rm{k}} )}$$\equiv$0 for 
$\mathbf{x_{\rm{kb}}}$ at the top of the atmosphere.

Similarly, 
it is convenient to separate
the diffuse and unscattered radiation within the atmospheric medium:
\begin{equation}
\int_{\Omega} d\Omega(\mathbf{s_{\rm{ka}}})
\mathbb{P(\mathbf{x_{\rm{ka}},s_{\rm{k}}, s_{\rm{ka}}})}\mathbf{I(x_{\rm{ka}},s_{\rm{ka}})} \rightarrow
\label{Ika_eq}
\end{equation}
$$
\int_{\Omega,\mathbf{s}_{\rm{ka}}\neq \mathbf{s}_{\rm{\odot}}} d\Omega(\mathbf{s_{\rm{ka}}})
\mathbb{P(\mathbf{x_{\rm{ka}},s_{\rm{k}}, s_{\rm{ka}}})}\mathbf{I(x_{\rm{ka}},s_{\rm{ka}})} + 
\mathbb{P(\mathbf{x_{\rm{ka}},s_{\rm{k}}, s_{\rm{\odot}}})}
t(\mathbf{x_{\rm{ka}}},\mathbf{x_{\rm{\odot}}}) \mathbf{F}_{\odot}.
$$
In both Eqs. (\ref{Ikb_eq}) and (\ref{Ika_eq}), $\mathbf{x_{\odot}}$ 
is the intersection at 
the top boundary of the rays traced in the $-\mathbf{s_{\odot}}$ direction
from  $\mathbf{x_{\rm{kb}}}$ and $\mathbf{x_{\rm{ka}}}$, respectively.  
Clearly, 
$t(\mathbf{x_{\rm{kb}}},\mathbf{x_{\rm{\odot}}})$ 
and $t(\mathbf{x_{\rm{ka}}},\mathbf{x_{\rm{\odot}}})$$\equiv$0 
if the stellar disk is not visible from either $\mathbf{x_{\rm{kb}}}$
and $\mathbf{x_{\rm{ka}}}$, respectively.

With the above considerations, Eq. (\ref{Ik_eq2}) is now expressed as:
\begin{eqnarray}
\mathbf{I(\mathbf{x_{\rm{k}},s_{\rm{k}}})}= 
(1-a(\mathbf{x_{\rm{k}},x_{\rm{kb}}}))
(\mathcal{L_B}(\mathbf{x_{\rm{k}},s_{\rm{k}}})+ 
\mathcal{B}\mathbf{I(\mathbf{x_{\rm{kb}},s_{\rm{kb}}})}) 
\label{Ikchvar_eq} 
\\
+a(\mathbf{x_{\rm{k}},x_{\rm{kb}}})
(\mathcal{L_A}(\mathbf{x_{\rm{k}},s_{\rm{k}}})+ 
\mathcal{A} \mathbf{I(\mathbf{x_{\rm{ka}},s_{\rm{ka}}})})
\nonumber
\end{eqnarray}
where: 
\begin{equation}
\mathcal{L_B}(\mathbf{x_{\rm{k}},s_{\rm{k}}}) =
\frac{r(\mathbf{x_{\rm{kb}}})}{\pi} 
\mathbf{n(x_{\rm{kb}})}\cdot\mathbf{s_{\odot}} 
t\mathbf{(x_{\rm{kb}},x_{\rm{\odot}})} \mathbf{F}_{\odot} 
\label{LB_eq}
\end{equation}	
\begin{equation}
\mathcal{L_A}(\mathbf{x_{\rm{k}},s_{\rm{k}}}) =
\int_{0}^{1} d\epsilon_{\rm{ka}}
\varpi(\mathbf{x_{\rm{ka}}}) t(\mathbf{x_{\rm{ka}},x_{\odot}})
\mathbf{{\mathbb{P}}(x_{\rm{ka}}, s_{\rm{k}}, s_{\odot} )} 
\mathbf{F_{\odot}} 
\end{equation}	
and
\begin{eqnarray}
\mathcal{B}\mathbf{I(\mathbf{x_{\rm{kb}},s_{\rm{kb}}})} = \\
\frac{r_g(\mathbf{x_{\rm{kb}}})}{\pi}
\int_{\Omega, \mathbf{s}_{\rm{kb}} \neq \mathbf{s}_{\rm{\odot}}}d\Omega(\mathbf{s_{\rm{kb}}})
\mathbf{n(x_{\rm{kb}})}\cdot\mathbf{s_{\rm{kb}}} 
\mathbb{D}\mathbf{I}\mathbf{(x_{\rm{kb}},s_{\rm{kb}} )}, 
\nonumber
\label{bikb_eq}
\end{eqnarray}

\begin{eqnarray}
\mathcal{A}\mathbf{I(\mathbf{x_{\rm{ka}},s_{\rm{ka}}})} =
\label{bika_eq}
\\
\int_{0}^{1} d\epsilon_{\rm{ka}} \varpi(\mathbf{x_{\rm{ka}}})
\int_{\Omega,\mathbf{s}_{\rm{ka}}\neq \mathbf{s}_{\rm{\odot}}} d\Omega(\mathbf{s_{\rm{ka}}})
\mathbb{P(\mathbf{x_{\rm{ka}},s_{\rm{k}}, s_{\rm{ka}}})}\mathbf{I(x_{\rm{ka}},s_{\rm{ka}})}.
\nonumber
\end{eqnarray}
In Eq. (\ref{Ikchvar_eq}), 
only the term preceded by $a(\mathbf{x_{\rm{k}},x_{\rm{kb}}})$ occurs for 
$\mathbf{x_{\rm{kb}}}$ at the atmospheric top but, for generality, 
we retain the two of them.
Both the $\mathcal{L_B}$ and $\mathcal{L_A}$ terms can be evaluated 
based on the optical properties of the medium, 
whereas the $\mathcal{B}$ and $\mathcal{A}$ terms need additional information 
in the form of the diffuse radiation vectors
$\mathbf{I(x_{\rm{kb}},s_{\rm{kb}})}$ and 
$\mathbf{I(x_{\rm{ka}},s_{\rm{ka}})}$.

\begin{figure}[h]
  \centering
  \includegraphics[width=9cm]{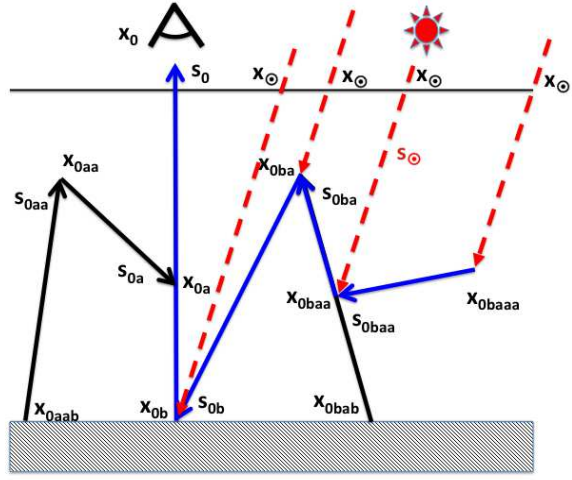}
  \caption{\label{sketch2_fig} 
  In black, sketch demonstrating the construction of the \{$\mathbf{x_{\rm{k}}, s_{\rm{k}}}$\} pairs 
  starting from \{$\mathbf{x_{\rm{0}}, s_{\rm{0}}}$\}
  as the photon is traced back from the detector through the medium. Vectors are pointed
  in the direction of photon propagation, which is the reverse of the direction of 
  integration in the BMC algorithm. 
  In principle, a \{$\mathbf{x_{\rm{k}}, s_{\rm{k}}}$\} pair can lead to two new 
  \{$\mathbf{x_{\rm{ka}}, s_{\rm{ka}}}$\} and \{$\mathbf{x_{\rm{kb}}, s_{\rm{kb}}}$\} pairs. 
  In the MC implementation of the algorithm, a scheme based on the coefficients of the
  $\mathcal{L_B}$+$\mathcal{B}$ and $\mathcal{L_A}$+$\mathcal{A}$ operators, 
  Eq. (\ref{Ikchvar_eq}), determines whether the photon's next move occurs 
  within the atmospheric medium or whether the photon moves onto the planet's surface. 
  In blue, one specific photon trajectory within the family of possible trajectories. For this
  specific trajectory, the red arrows denote the direction of the unscattered stellar
  photons. 
  }
   \end{figure}

Starting from \{$\mathbf{x_{\rm{0}}, s_{\rm{0}}}$\},
which determines the position of and entry direction into the detector, 
recurrent use of Eq. (\ref{Ikchvar_eq}), complemented by Eqs. (\ref{LB_eq})--(\ref{bika_eq}),  
leads to an expression for $\mathbf{I(x_{\rm{0}},s_{\rm{0}})}$
as an infinite summation series of integrals of 
increasingly higher dimensions (O'Brien, \cite{obrien1992,obrien1998}). 
Physically, higher dimension integrals account for additional orders of scattering of the 
simulated photons.
Figure (\ref{sketch2_fig}) shows the definition of the pairs
\{$\mathbf{x_{\rm{0a}}, s_{\rm{0a}}}$\}, \{$\mathbf{x_{\rm{0b}}, s_{\rm{0b}}}$\},
\{$\mathbf{x_{\rm{0aa}}, s_{\rm{0aa}}}$\}, \{$\mathbf{x_{\rm{0ab}}, s_{\rm{0ab}}}$\},
\{$\mathbf{x_{\rm{0ba}}, s_{\rm{0ba}}}$\}, and so forth, that appear in the recurrence law.
The series is convergent (and thus amenable to truncation)
provided that the optical thickness of the medium is finite
and/or the medium is not fully conservative (i.e. either $\varpi$ or $r_g$$\le$1). 
Effectively, the recurrence law builds the 
solution for $\mathbf{I(x_{\rm{0}},s_{\rm{0}})}$ by splitting each summation into a double
summation involving new $\mathcal{B}$ and $\mathcal{A}$ integrals at each step. 

Appendix \ref{appendixa} spells out the first few terms in the summation series
and summarises the practical implementation in the PBMC algorithm. 
Rewriting the integrals that appear in the summation in terms of appropriately normalised variables leads to improved 
convergence rates, an approach that is equivalent to 
so-called variance reduction techniques (O'Brien, \cite{obrien1992}).
In what follows, we address the integration in solid angle and the significance of 
polarisation in the sampling of photon propagation directions, which is the feature 
unique to our PBMC algorithm with respect to other BMC schemes.

\subsection{Monte Carlo integration} 

The essence of MC integration 
is to estimate multi-dimensional integrals through the evaluation of the 
integrand at properly
selected values of the integration variables: 
\begin{equation}
\int_{0}^{1}\int_{0}^{1}...\int_{0}^{1} 
f(u_1,u_2, ..., u_d)du_1du_2...du_d\approx 
\label{MCintegration_eq}
\end{equation}
$$
\frac{1}{N}\sum_{j=1}^{N} f(u_1^{[j]},u_2^{[j]},...,u_d^{[j]})
+ O\left(\frac{1}{\sqrt{N}}\right).
$$
Here, each $j$ represents a random draw from the uniform distribution functions
$u_k$$\in$[0, 1]. 
Importantly, MC integration
converges to the exact value at a rate that depends on the number of 
realisations, 
$N$, but not on the dimension of the integral, $d$. 

In a BMC framework, the evaluation of the summation series for 
$\mathbf{I(x_{\rm{0}},s_{\rm{0}})}$ is interpretable in terms of photons whose
trajectories are simulated in the backwards direction, i.e. from the detector through the 
medium, finally reaching the radiation source. 
Thus, we regularly refer to the determination of the 
\{$\mathbf{x_{\rm{k}}, s_{\rm{k}}}$\} pairs as simulated photon trajectories made up of
collision events at $\mathbf{x_{\rm{0a}}}$, $\mathbf{x_{\rm{0b}}}$, 
$\mathbf{x_{\rm{0aa}}}$, $\mathbf{x_{\rm{0ab}}}$, $\mathbf{x_{\rm{0ba}}}$, etc. 
Ultimately, the solution to the vRTE is built by simulating a number $n_{\rm{ph}}$ 
(=$N$ in Eq. (\ref{MCintegration_eq})) of photon trajectories.

\subsection{\label{solidangleint_sec}Integration in solid angle}

The summation series for $\mathbf{I(x_{\rm{0}},s_{\rm{0}})}$
obtained from recurrent use of Eq. (\ref{Ikchvar_eq})
contains multi-dimensional integrals in solid angle:
\begin{equation}
\int \int \int \int
d\Omega(\mathbf{s_{\rm{0a}}})
\mathbf{{\mathbb{P}}(s_{\rm{0}}, s_{\rm{0a}})}
\bigg\{
 d\Omega(\mathbf{s_{\rm{0aa}}})
\mathbf{{\mathbb{P}}(s_{\rm{0a}}, s_{\rm{0aa}})}
\bigg\{
\label{matrixprod_eq}
\end{equation}
$$
 d\Omega(\mathbf{s_{\rm{0aaa}}})
\mathbf{{\mathbb{P}}(s_{\rm{0aa}}, s_{\rm{0aaa}})}
\bigg\{
 d\Omega(\mathbf{s_{\rm{0aaaa}}})
\mathbf{{\mathbb{P}}(s_{\rm{0aaa}}, s_{\rm{0aaaa}})}
...
\bigg\} \bigg\} \bigg\},
$$
for collisions within the atmospheric medium. 
For simplicity in the notation, 
we removed all references to $\mathbf{x_{\rm{0k}}}$ within the $\mathbb{P}$
matrices.
The treatment of collisions at the bottom boundary is analogous. 
In a BMC framework $ d\Omega(\mathbf{s'})$ integration at a particular collision event
entails selecting an  \textit{incident} $\mathbf{s'}$ direction for a given 
\textit{emergent} $\mathbf{s}$ direction (see Fig. \ref{sketch1_fig}), 
according to an appropriate probability density function. 

\subsubsection{\label{classicalscheme_sec} The classical sampling scheme}
In classical BMC algorithms (Collins et al., \cite{collinsetal1972}, and
thereafter), 
evaluation of Eq. (\ref{matrixprod_eq}) proceeds by separating it into:
\begin{eqnarray}
\bigg\{
\int 
d\Omega(\mathbf{s_{\rm{0a}}})
\mathbf{{\mathbb{P}}(s_{\rm{0}}, s_{\rm{0a}})}
\bigg\}
\bigg\{
\int 
d\Omega(\mathbf{s_{\rm{0aa}}})
\mathbf{{\mathbb{P}}(s_{\rm{0a}}, s_{\rm{0aa}})}
\bigg\}
\bigg\{... 
\label{collins_eq}
\end{eqnarray}
and, subsequently, sampling the $\theta$ and $\phi$ angles
in each integral from the local $\mathbb{M}_{1,1}$ function and
from a uniform distribution between 0 and 2$\pi$, respectively.
Tacitly, the sampling scheme assumes that the relative orientations between 
$\mathbf{s}$ and $\mathbf{s'}$ must depend on the local properties of the medium 
but not on the propagation history of the photons, or that any bias introduced by
proceeding that way can be subsequently corrected for 
by dividing by the sampled $\mathbb{M}_{1,1}$. 
The assumption is exact in the treatment of the scalar RTE, 
but is fundamentally erroneous in polarising media. 
We refer to the simplified approach based on Eq. (\ref{collins_eq})
as the classical sampling scheme for photon propagation directions.

\subsubsection{\label{preconditionedscheme_sec} The pre-conditioned sampling scheme}
A more appropriate approach to the evaluation of Eq. (\ref{matrixprod_eq}) is to sample:
\begin{eqnarray}
\mathbf{s_{\rm{0a}}} & \mbox{from} & [\mathbf{{\mathbb{P}}(s_{\rm{0}}, s_{\rm{0a}})}]_{1,1}d\Omega(\mathbf{s_{\rm{0a}}}) \nonumber \\
\mathbf{s_{\rm{0aa}}} & \mbox{from} & 
[\mathbf{{\mathbb{P}}(s_{\rm{0}}, s_{\rm{0a}}){\mathbb{P}}(s_{\rm{0a}}, s_{\rm{0aa}})}]_{1,1}d\Omega(\mathbf{s_{\rm{0aa}}})
\label{PBMCsampling_eq}\\
\mathbf{s_{\rm{0aaa}}} & \mbox{from} & 
[\mathbf{{\mathbb{P}}(s_{\rm{0}}, s_{\rm{0a}}){\mathbb{P}}(s_{\rm{0a}}, s_{\rm{0aa}})
{\mathbb{P}}(s_{\rm{0aa}}, s_{\rm{0aaa}})}]_{1,1}d\Omega(\mathbf{s_{\rm{0aaa}}}) \nonumber\\
& \mbox{...and so on...} & \nonumber 
\end{eqnarray}
By proceeding sequentially, at each step
all the involved photon propagation directions but the one being sampled are known.
The scheme derives directly from Eq. (\ref{matrixprod_eq}), 
and preserves the history of the simulated photon trajectories through 
the ordered arrangement of the products of $\mathbf{{\mathbb{P}}}$ matrices.
At each collision event, the  matrices  
$\mathbf{{\mathbb{H}}(s_{\rm{0}}, s_{\rm{0a}})}$=$\mathbb{U}$ ($\equiv$unity matrix),
$\mathbf{{\mathbb{H}}(s_{\rm{0a}}, s_{\rm{0aa}})}$=$\mathbf{{\mathbb{P}}(s_{\rm{0}}, s_{\rm{0a}})}$,
$\mathbf{{\mathbb{H}}(s_{\rm{0aa}}, s_{\rm{0aaa}})}$=$\mathbf{{\mathbb{P}}(s_{\rm{0}}, s_{\rm{0a}}){\mathbb{P}}(s_{\rm{0a}}, s_{\rm{0aa}})}$, ...,
effectively pre-condition the local $\mathbb{P}$ matrix and, in turn, the probability for 
the scattering to occur in any of the possible $\mathbf{s_{\rm{0a}}}$, 
$\mathbf{s_{\rm{0aa}}}$, $\mathbf{s_{\rm{0aaa}}}$, ..., \textit{incident} directions. 
The pre-conditioning matrix evolves as the photon trajectory is being backtraced and, 
in this way, the photon history is preserved throughout the simulation.
Hereafter, we term this approach the pre-conditioned sampling scheme 
for photon propagation directions. This scheme is at the core of our PBMC algorithm.

Expanding Eq. (\ref{PBMCsampling_eq}) yields
insight into the pre-conditioned sampling scheme. 
For an arbitrary
$\mathbf{{\mathbb{H}}}\mathbf{{\mathbb{P}(\mathbf{s,s'})}}$$d\Omega(\mathbf{s'})$=$\mathbf{{\mathbb{H}}}\mathbb{L}(\pi-i)\mathbb{M}\mathbf{(x,\theta)} \mathbb{L}(-i')$
$d\Omega(\theta, \phi)$,
the (1, 1) entry leads to an expression proportional to $f(\theta, \phi)d\theta d\phi$=
\begin{equation}
=\big(a_1(\theta) + 
b_1(\theta) \big[ q \cos{(2\phi)} - u \sin{(2\phi)}
\big]\big) \frac{\sin{(\theta)} d\theta d\phi}{4\pi}, 
\label{fthetaphi_eq}
\end{equation}
where we defined $q$=${ \mathbb{H}_{1,2} }/{ \mathbb{H}_{1,1} }$ and 
$u$=${ \mathbb{H}_{1,3} }/{ \mathbb{H}_{1,1} }$. In the derivation of Eq. (\ref{fthetaphi_eq}), 
we used the geometrical relation between $i$ and $\phi$, for $\phi$ locally defined with respect to
the meridian plane (see Fig. \ref{sketch1_fig}, Top). 
Angle $i'$ is evaluated once both $\theta$ and $\phi$ are determined.

Two important properties apply to $f(\theta, \phi)$, namely:  
[1] it is $\ge$0 for $\theta$$\in$[0, $\pi$] and $\phi$$\in$[0, 2$\pi$], 
and [2] its integral over the $\theta$--$\phi$ domain is equal to one, 
which is straightforward to confirm from the normalisation of Eq. (\ref{normalisation_eq}).
Since $a_1(\theta)$$\ge$0 and $|b_1|$$\le$$|a_1|$ (Mishchenko et al., \cite{mishchenkoetal2002}), 
property [1] requires that $|q \cos{(2\phi)} - u \sin{(2\phi)}|$$\le$1. 
To prove that condition, it suffices to show that the first row of $\mathbb{H}$, 
$[\mathbb{H}_{1,1}, \mathbb{H}_{1,2}, \mathbb{H}_{1,3}, \mathbb{H}_{1,4}]$ 
(=$\mathbb{H}_{1,1}$$[1, q, u, v]$ in our own notation), 
forms from:
\begin{equation}
[1,0,0,0]\mathbf{{\mathbb{P}}(s_{\rm{0}}, s_{\rm{0a}}){\mathbb{P}}(s_{\rm{0a}}, s_{\rm{0aa}})
{\mathbb{P}}(s_{\rm{0aa}}, s_{\rm{0aaa}})}...
\label{adjoint_eq}
\end{equation}
The vector resulting from Eq. (\ref{adjoint_eq}) is indeed 
the transpose of:
\begin{equation}
...\mathbf{
{\mathbb{P}}^{\rm{T}}(-s_{\rm{0aaa}}, -s_{\rm{0aa}})
{\mathbb{P}}^{\rm{T}}(-s_{\rm{0aa}}, -s_{\rm{0a}})^{\rm{T}}
{\mathbb{P}}^{\rm{T}}(-s_{\rm{0a}}, -s_{\rm{0}})
} 
[1,0,0,0]^{\rm{T}}, 
\end{equation} 
which is the Stokes vector for an associated \textit{direct} problem 
of photons propagating onwards from the detector. 
In this \textit{direct} problem, the relevant scattering matrix
is $\mathbf{{\mathbb{M}}^{\rm{T}}}$. 
For Mie scattering, Eq. (\ref{MMmatrix_eq}), the matrix satisfies 
$\mathbf{{\mathbb{M}}^{\rm{T}}}$($b_2$)=$\mathbf{{\mathbb{M}}}$($-$$b_2$), which 
suggests a connection with the adjoint formulation 
based on vector Green's functions proposed by Carter et al. (\cite{carteretal1978}).
From the association of the backward problem with its \textit{direct} 
counterpart of scattering matrix $\mathbf{{\mathbb{M}}^{\rm{T}}}$, 
it becomes apparent that $q$ and $u$ are relative linear polarisations and
$v$ is the corresponding relative circular polarisation.  
As a result, $|q \cos{(2\phi)} - u \sin{(2\phi)}|$$\le$1.
 
Thus, $f(\theta, \phi)$ is a bivariate probability density function that 
can be used to sample the propagation directions in the 
backtracing of photons. 
Our pre-conditioned scheme of Eq. (\ref{PBMCsampling_eq}) is 
indeed similar in structure to the schemes utilised in some 
FMC algorithms 
(e.g. 
Bartel \& Hielscher, \cite{bartelhielscher2000};
Bianchi et al., \cite{bianchietal1996}; 
Cornet et al., \cite{cornetetal2010}; 
Fischer et al., \cite{fischeretal1994}; 
Hopcraft et al., \cite{hopcraftetal2000}; 
Kastner, \cite{kastner1966}; 
Schmid, \cite{schmid1992}; 
Whitney, \cite{whitney2011}). 

\begin{figure}
\centering
\includegraphics[width=7.3cm]{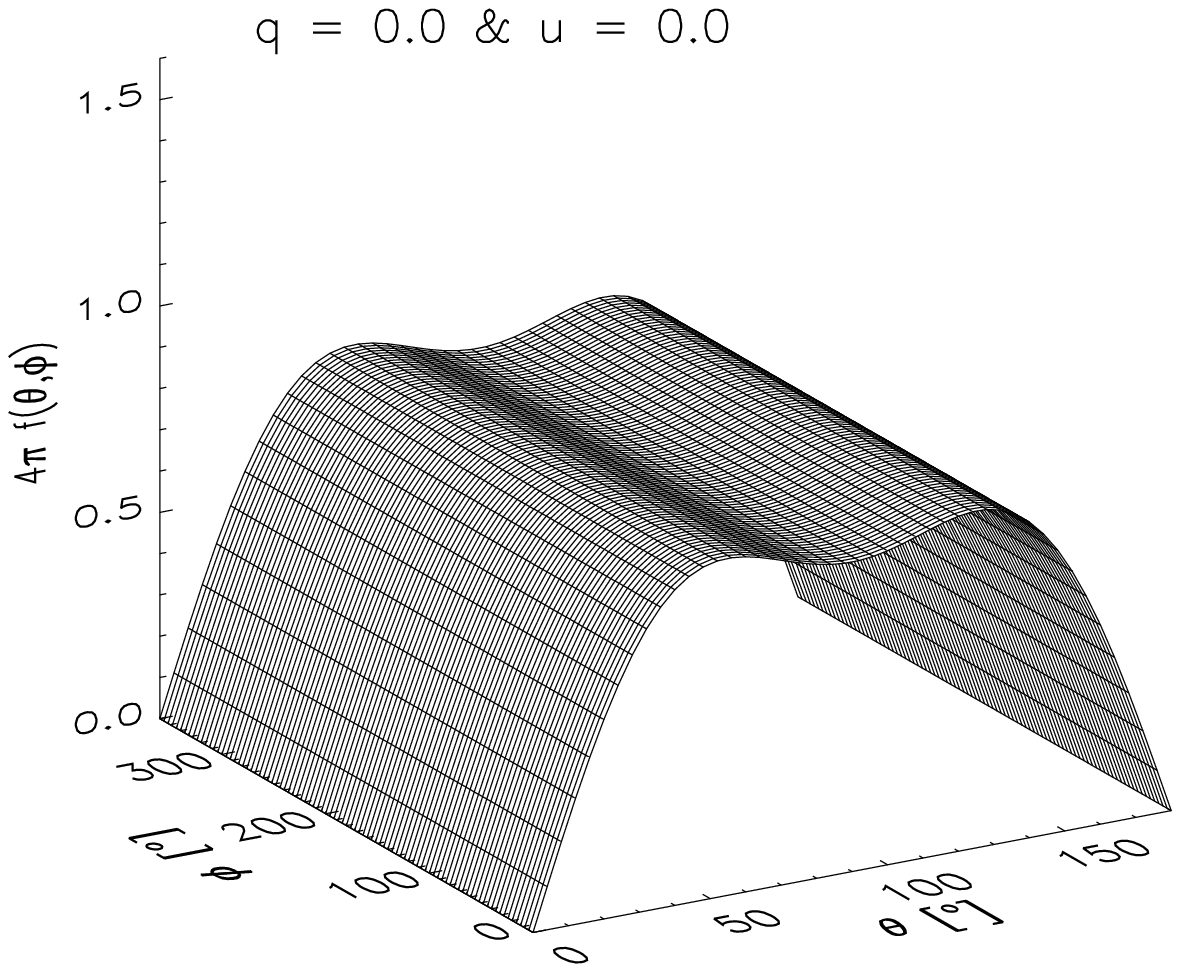} 
\includegraphics[width=7.3cm]{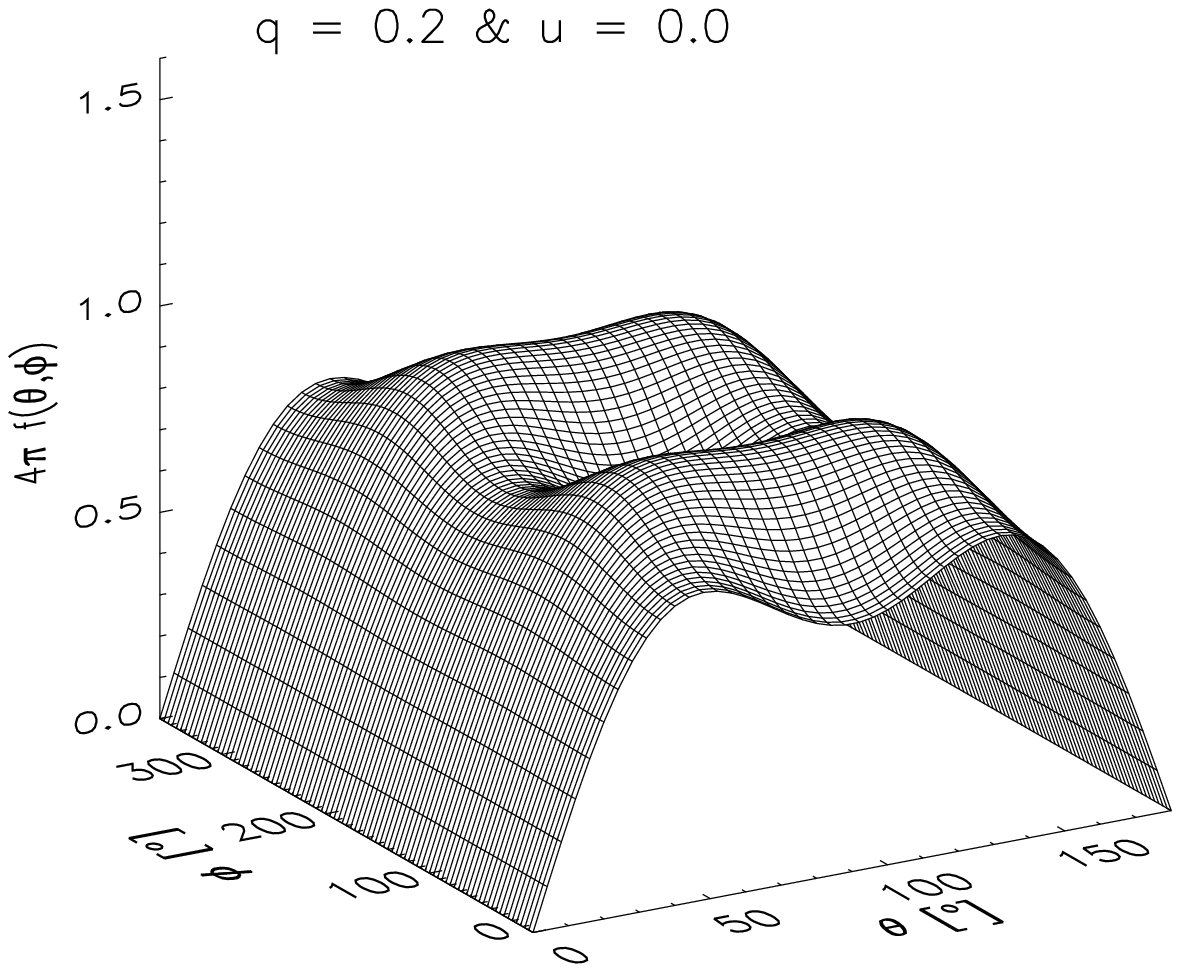} 
\includegraphics[width=7.3cm]{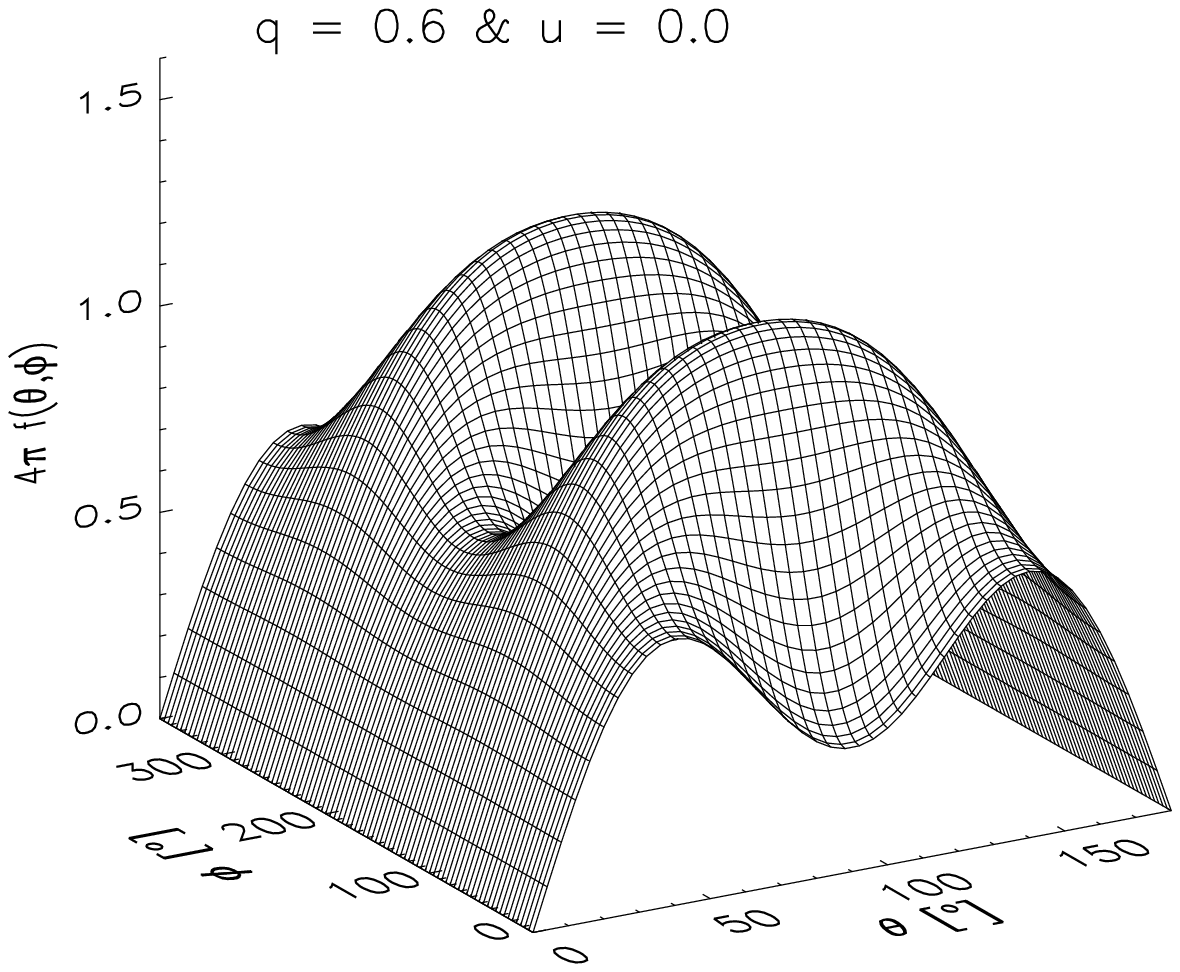} 
\includegraphics[width=7.3cm]{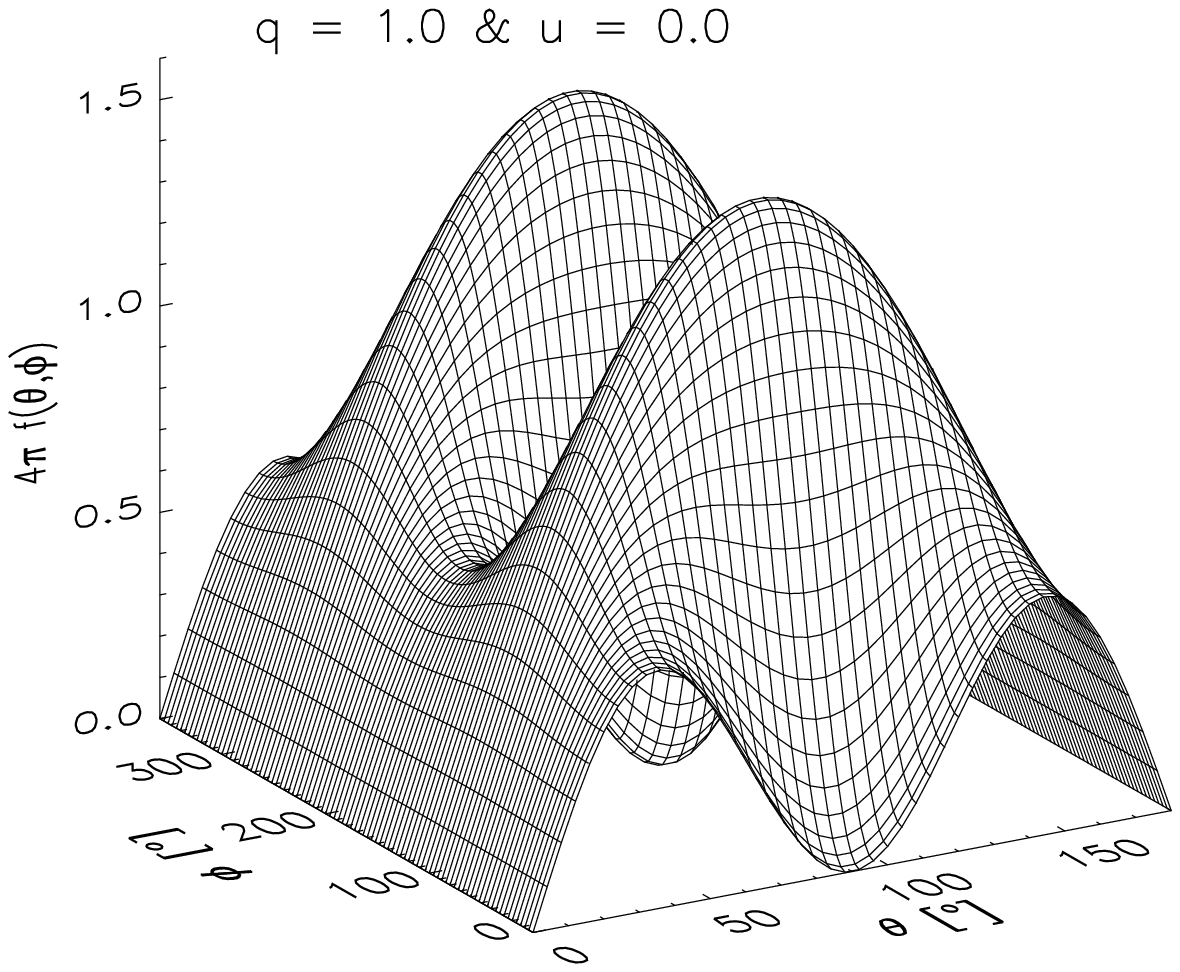} 
\caption{\label{pthetaphi_fig}
	Probability density function 
	$f(\theta,\phi)$={$f_{\theta}(\theta)$}{$f_{\phi|\theta}(\phi|\theta)$}
	in the pre-condioned sampling scheme of photon propagation directions,  
	Eqs. (\ref{ftheta_eq})--(\ref{fphitheta_eq}),  
	for a Rayleigh-scattering medium. Note the changes in $f(\theta,\phi)$
	with $q$, especially near the maximum of $|b_1(\theta)/a_1(\theta)|$ 
	for $\theta$=90$^{\circ}$. By ignoring polarisation, 
	the classical sampling scheme determines the $\theta$ and $\phi$ values
	of the \textit{incident} propagation direction from
	$f(\theta,\phi; q\equiv0)$. It is apparent that 
	the classical sampling scheme is more likely
	to fail in strongly-polarising media 
	that involve high $q$ values during the backtracing of photons. 
	}
\end{figure}

In practice, the sampling is facilitated by separating
$f(\theta, \phi)$= $f_{\theta}(\theta)$$f_{\phi|\theta}(\phi|\theta)$, with: 
\begin{equation}
f_{\theta}=a_1(\theta){\sin{(\theta)}}/{2}, \;\;\;\mbox{ and} 
\label{ftheta_eq}
\end{equation} 
\begin{equation}
f_{\phi|\theta}(\phi|\theta)=\big(1+b_1(\theta)/a_1(\theta)\big[ q \cos{(2\phi)} - u \sin{(2\phi)}\big]\big)/2\pi.
\label{fphitheta_eq}
\end{equation}
Here, $f_{\theta}$ is the conventional $\theta$-sampling function
implemented in most FMC and BMC algorithms, 
whether treating the scalar or vector RTE. 
Function $f_{\phi|\theta}(\phi|\theta)$ conveys that sampling in $\phi$ is
constrained by $\theta$ and, through $q$ and $u$, also
by the photon polarisation state and history.
Figure (\ref{pthetaphi_fig}) explores $f(\theta,\phi)$ for a few combinations of 
$q$ and $u$$\equiv$0 in the specific case of a Rayleigh medium. 
The classical sampling scheme is equivalent to drawing the
$\theta$ and $\phi$ from the probability density function $f(\theta,\phi; q\equiv 0)$. 
Doing so appears inappropriate in strongly-polarising media where 
$b_1(\theta)/a_1(\theta)$, $q$ and $u$ may take absolute values 
close to one through the photon simulations. 
The consequences of this are investigated below. 

\subsection{\label{diskschemes_ref} Disk-integration schemes}

We are interested in the radiation emerging from both disk-resolved and disk-integrated
planetary atmospheres. We here derive two disk-integration schemes and describe their 
incorporation into the PBMC algorithm.

\subsubsection{\label{visibledisk_sec}Integration over the 'visible' disk}

\begin{figure}[h]
\centering
\includegraphics[width=7cm]{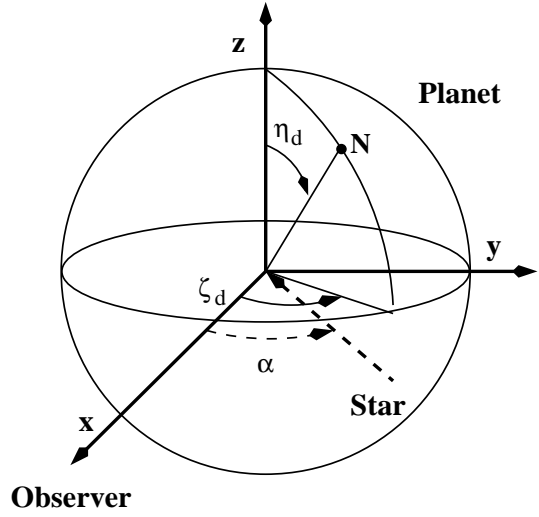}
\caption{\label{sketch5_fig} 
Geometrical parameters relevant to the integration over the planet's
\textit{visible} disk. For a pair of $u_i$ and $v_i$ values, 
N is the location on the disk where the observer's line of sight intercepts
the atmosphere. N is equivalent to $\mathbf{x_{\rm{0}}}$ in our implementation of the 
PBMC algorithm.
}
\end{figure}

Horak (\cite{horak1950}) laid out the expressions for evaluating the 
disk-integrated radiation scattered from a planet over its \textit{visible} disk.
In this context, \textit{visible} refers to the disk portion that appears 
illuminated by single-scattered photons as viewed from the observer's vantage point.
We refer to the sketch of Fig. (\ref{sketch5_fig}), that presents the relevant geometrical parameters. 
Provided that both the observer and the star are sufficiently far from the planet, 
Horak (\cite{horak1950}) arrives at the expression: 
\begin{equation} 
\mathbf{F}=\left(\frac{\rho}{\Delta} \right)^2 \int_{0}^{\pi} d\eta_d sin^2(\eta_d) 
\int_{\alpha-\pi/2}^{\pi/2} d\zeta_d cos(\zeta_d) \mathbf{I}(\zeta_d,\eta_d), 
\label{horak_eq} 
\end{equation} 
that we adapt to the vector case by using the Stokes vector $\mathbf{I}$,
in which case $\mathbf{F}$ (=[$F_I, F_Q, F_U, F_V$]$^{\rm{T}}$) 
is the irradiance Stokes vector. 
Here, $\rho$ (=$R_p$+$h_{\rm{TOA}}$ for planets with a solid core of radius
$R_p$ and an atmosphere extending up to altitudes of $h_{\rm{TOA}}$)
and $\Delta$ are the radius of the planet's
scattering disk and the observer-to-planet distance, respectively.
We normalize $\mathbf{F}$ by eliminating the $({\rho}/{\Delta})^2$ factor
from Eq. (\ref{horak_eq}).

Rather than working with longitudes, $\zeta_d$, and co-latitudes, $\eta_d$, it 
is convenient to introduce the two auxiliary variables:
\begin{equation}
u=\frac{1}{\pi}(\eta_d - \frac{1}{2}\sin(2\eta_d))
\label{u_etah_eq}
\end{equation}
\begin{equation}
v=\frac{1}{1+\cos(\alpha)} ( \sin(\zeta_d) + \cos(\alpha)   ),
\label{v_zetah_eq}
\end{equation}
such that, after some manipulations, Eq. (\ref{horak_eq}) transforms into:
\begin{equation}
\mathbf{F}= \frac{\pi}{2}(1+\cos{(\alpha)}) 
\int_{0}^{1} 
\int_{0}^{1} du dv \;\mathbf{I}(u,v).
\label{horak2_eq}
\end{equation}
The pre-multiplying factor before the double integral is the projected size 
of the planet's \textit{visible} disk. 
The double integral may be seen as an average radiance Stokes vector 
over that domain.

In the form of Eq. (\ref{horak2_eq}), it is straightforward to insert the evaluation of
$\mathbf{F}$ into the PBMC algorithm as the sum:
\begin{equation}
\mathbf{F} = \frac{\pi}{2}(1+\cos{(\alpha)}) 
\frac{1}{n_{\rm{ph}}}
\sum_{i=1}^{n_{\rm{ph}}} <\mathbf{I}(u_i,v_i)>,
\label{horak3_eq}
\end{equation}
where $u_i$, $v_i$ are picked from the random uniform distributions $u$, $v$$\in$[0, 1].
Each $u_i$, $v_i$ yields the location 
on the planet's disk where the observer's line of sight intercepts the planet's
atmosphere
or, equivalently, $\mathbf{x_{\rm{0}}}$ in the implementation of the algorithm 
of Appendix \ref{appendixa}. 
For a sufficiently remote observer, $\mathbf{s_{\rm{0}}}$ is, 
according to Fig. (\ref{sketch5_fig}), 
permanently oriented along the $x$ axis. 
The application of Eq. (\ref{horak3_eq}) requires the inversion of Eqs. (\ref{u_etah_eq})--(\ref{v_zetah_eq}).
For 
$u$$\rightarrow$$\eta_d$, we interpolate from pre-calculated tabulations of
$u$=$u$($\eta_d;\alpha$).
For $v$$\rightarrow$$\zeta_d$, the inversion is done analytically. 
Since in our formulation $\mathbf{I}$ is by default referenced to the meridian plane 
containing the $z$ axis and $\mathbf{s_{\rm{0}}}$,
and $\mathbf{s_{\rm{0}}}$ is fixed in space, 
there is no need to rotate the emergent $<$$\mathbf{I}(u_i,v_i)$$>$ 
Stokes vectors, 
which can be directly added into Eq. (\ref{horak3_eq}). 
In our normalisation, 
the first of the $\mathbf{F}$ elements is $A_g$$\Phi$($\alpha$), 
with $A_g$ being the planet's geometric albedo and $\Phi$(0)$\equiv$1.

\subsubsection{\label{entiredisk_sec}Integration over the entire disk}

Alternatively to the integration over the \textit{visible} disk, 
one can proceed to integrate over the entire disk. 
Introducing $r$ and $\Theta$ as the polar coordinates that determine the projection of N 
in Fig. (\ref{sketch5_fig}) on the $yz$ plane, 
and the normalised variables $u'$=$\Theta$/2$\pi$ and 
$v'$=($r/\rho$)$^2$, integration over the projected surface element
$r$$d$$r$$d$$\Theta$ leads to:
\begin{equation} 
\mathbf{F}=
\frac{1}{\Delta^2} \int_{0}^{\rho} r dr \int_{0}^{2\pi}  d\Theta \mathbf{I}(r, \Theta)
=
\left(\frac{\rho}{\Delta} \right)^2 
\pi
\int_{0}^{1} 
\int_{0}^{1} du' dv' \mathbf{I}(u', v'), 
\label{entiredisk_eq} 
\end{equation} 
which, after eliminating the $({\rho}/{\Delta})^2$ factor, translates into:
\begin{equation} 
\mathbf{F}=
\frac{\pi}{n_{\rm{ph}}}
\sum_{i=1}^{n_{\rm{ph}}} <\mathbf{I}(u'_i,v'_i)>, 
\label{entirediskMC_eq} 
\end{equation} 
in the PBMC algorithm. 
Again, $u_i'$, $v_i'$ are picked from uniform distributions 
$u'$, $v'$$\in$[0, 1]. 

Some of the advantages of the latter implementation with respect to that in 
\S\ref{visibledisk_sec} include: [1] it makes no
assumption on the extent of the effectively-scattering disk and, therefore, 
 properly handles the full range of phase angles from superior to 
inferior conjunctions; 
[2] each photon trajectory simulation can simultaneously 
contribute to various specified phase angles. 
A drawback of the latter implementation (shared with FMC algorithms) is that 
for a given number of photon realisations
$n_{\rm{ph}}$ the solution statistics becomes poorer for the larger phase angles 
because fewer of the simulated photon trajectories actually connect the observer
and the direction of illumination.
We explore in \S\ref{disk_section} some of
these issues in the application of the two disk-integration schemes 
to both Rayleigh and Venus-like atmospheres.

\section{\label{planeparallel_sec} Comparison of the PBMC algorithm
against solutions from other methods} 

We assessed the performance of our PBMC algorithm against a suite of test cases 
for which reliable solutions are either available in the literature or can be produced with
existing models. 
The suite includes solutions to both the scalar and vector RTE, 
different viewing/illumination geometries and a variety of optical
properties for the scattering particles. 
Here in \S\ref{planeparallel_sec}, 
we focus on scattering in plane-parallel atmospheres.

\begin{figure}[h]
\centering
\includegraphics[width=8cm]{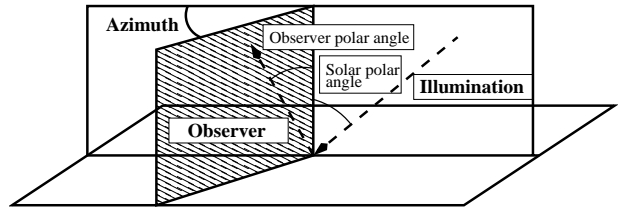}
 \caption{\label{sketch3_fig} Illumination and viewing angles for the
 plane-parallel atmosphere test cases discussed in \S\ref{planeparallel_sec}.
 }
\end{figure}

Figure (\ref{sketch3_fig}) sketches the relevant angles. In particular, 
an azimuth of 0 corresponds to the observer facing the Sun,  
and 180$^{\circ}$ to the observer looking away from the Sun. 
In the scalar RTE calculations, 
polarisation is omitted by zeroing all entries but $\mathbb{M}_{1,1}$ in the 
$\mathbb{M}$ scattering matrix.

\subsection{\label{rayleigh_disort}Non-polarised Rayleigh scattering}

In a first assessment, we compared our PBMC algorithm in its scalar mode
against DISORT (Stamnes et al., \cite{stamnesetal1988}) solutions in Rayleigh scattering media. 
The exercise includes 15,876 test cases that explore both optically thin and thick atmospheres
with viewing/illumination angles from zenith inclination to nearly horizontal pointing
(see Table \ref{disort_table}).
The comparison, the details of which are given in the Online Material, shows an 
excellent match between the two approaches.

\begin{table}[h]
\caption{\label{disort_table} Parameters in the investigation of conservative Rayleigh
scattering in plane-parallel atmospheres. 
The total number of test cases amounts to 12$\times$3$\times$7$\times$9$\times$7=15,876. 
Throughout the exercises of 
\S\ref{rayleigh_disort} and \ref{pol_ray_rad_sec}, 
we assumed an atmospheric single scattering albedo $\varpi$$\equiv$1.
}
\begin{flushleft}
\begin{tabular}{l}
\hline
\hline
\multicolumn{1}{l}{Optical thickness, $\tau$:} \\
$\;\;\;\;\;$0.02, 0.05, 0.1, 0.15, 0.25, 0.5, 1, 2, 4, 8, 16, 32\\
\multicolumn{1}{l}{Lambert surface albedo, $r_g$:} \\
$\;\;\;\;\;$0, 0.25, 0.8 \\
\multicolumn{1}{l}{Cosine of solar polar angle (SPA): } \\
$\;\;\;\;\;$0.1, 0.2, 0.4, 0.6, 0.8, 0.92, 1 \\
\multicolumn{1}{l}{Cosine of observer polar angle (OPA):} \\
$\;\;\;\;\;$0.02, 0.06, 0.1, 0.2, 0.4, 0.64, 0.84, 0.92, 1 \\
\multicolumn{1}{l}{Azimuth between Solar and observer planes, $\Delta$$\phi$:} \\
$\;\;\;\;\;$0, 30, 60, 90, 120, 150, 180$^{\circ}$ \\
\hline
\hline
\end{tabular}
\end{flushleft}
\end{table}

\subsection{\label{pol_ray_rad_sec}Polarised Rayleigh scattering} 
  
Coulson et al. (\cite{coulsonetal1960}) tabulated solutions for the elements of the Stokes vector 
in conservative, polarising, Rayleigh-scattering 
atmospheres above Lambert reflecting surfaces.
More recently, Natraj and collaborators (\cite{natrajetal2009,natrajhovenier2012}) 
extended the calculations to arbitrarily large optical thicknesses. 
The newly tabulated Stokes vectors (that we adopt as reference) are 
claimed to be accurate to within one unit in the eighth decimal place.
We computed the 15,876 cases summarised in Table (\ref{disort_table})
for $n_{\rm{ph}}$ up to 10$^7$ with our PBMC algorithm in its vRTE mode. 
For comparison, we utilised both the classical and pre-conditioned sampling 
schemes introduced in \S\ref{solidangleint_sec}.

\begin{figure}
   \centering
   \includegraphics[width=10cm]{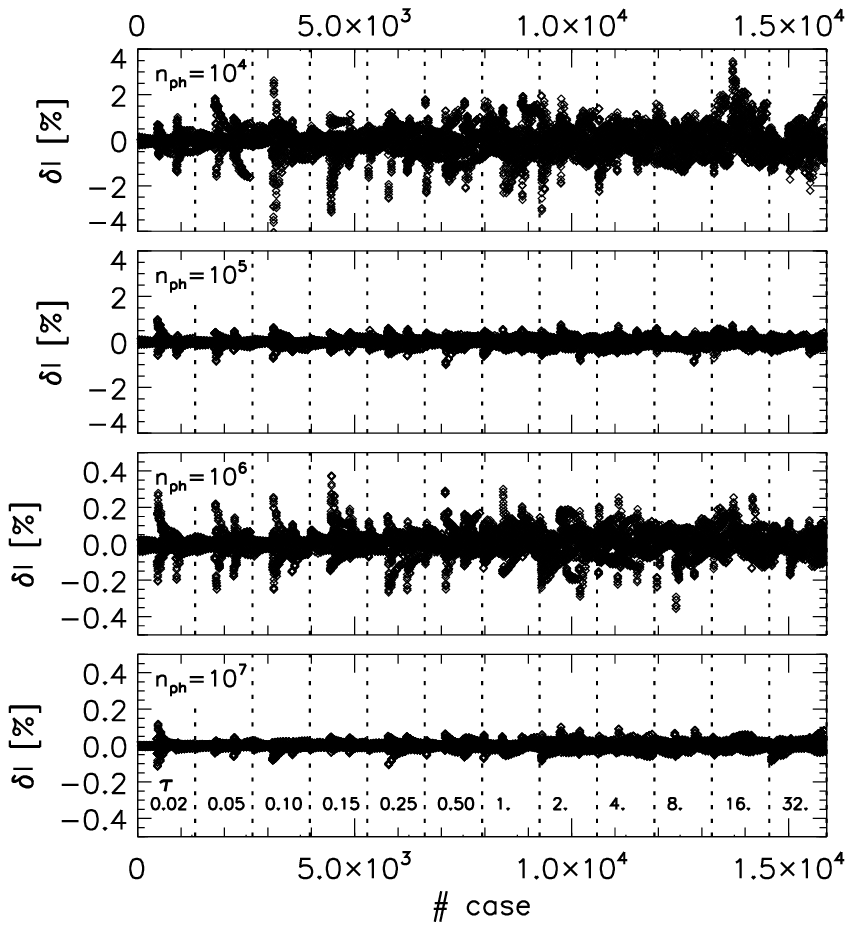}\\
   \includegraphics[width=10cm]{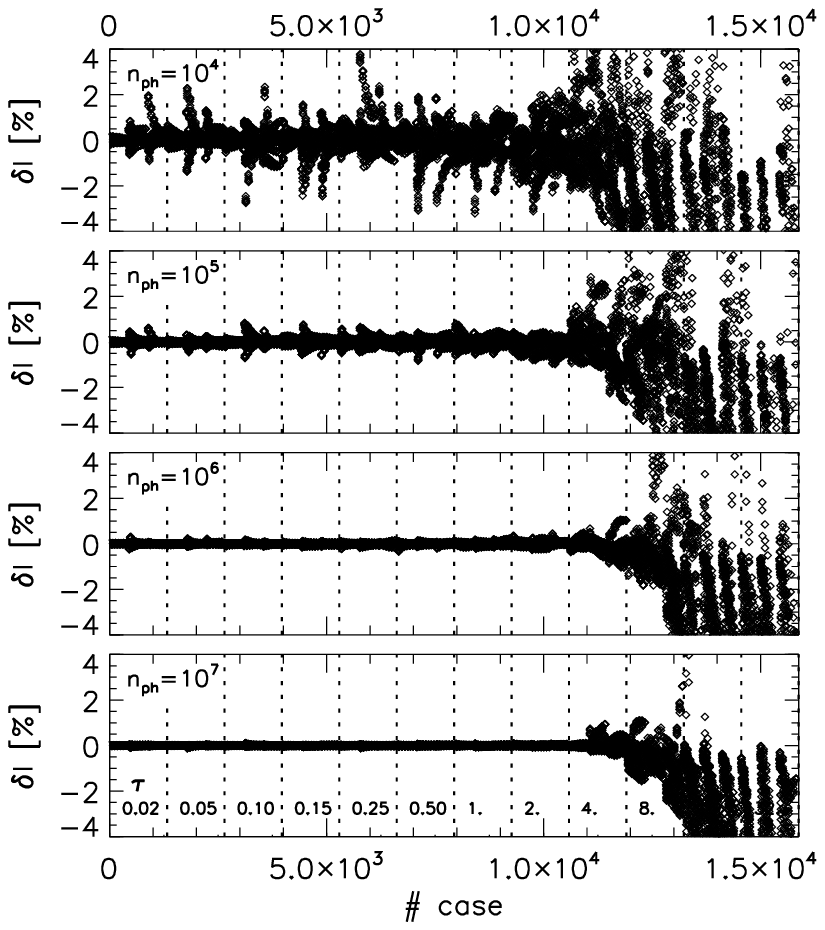}
      \caption{\label{plot_dI_vRTE_fig} Differences in intensity, $\delta I$, 
      for the solution of the vRTE in conservative, Rayleigh-scattering atmospheres.
      The algorithm uses the pre-conditioned (top) and classical (bottom) 
      sampling schemes for photon propagation directions.
      }
   \end{figure}

\begin{figure}
   \centering
   \includegraphics[width=10cm]{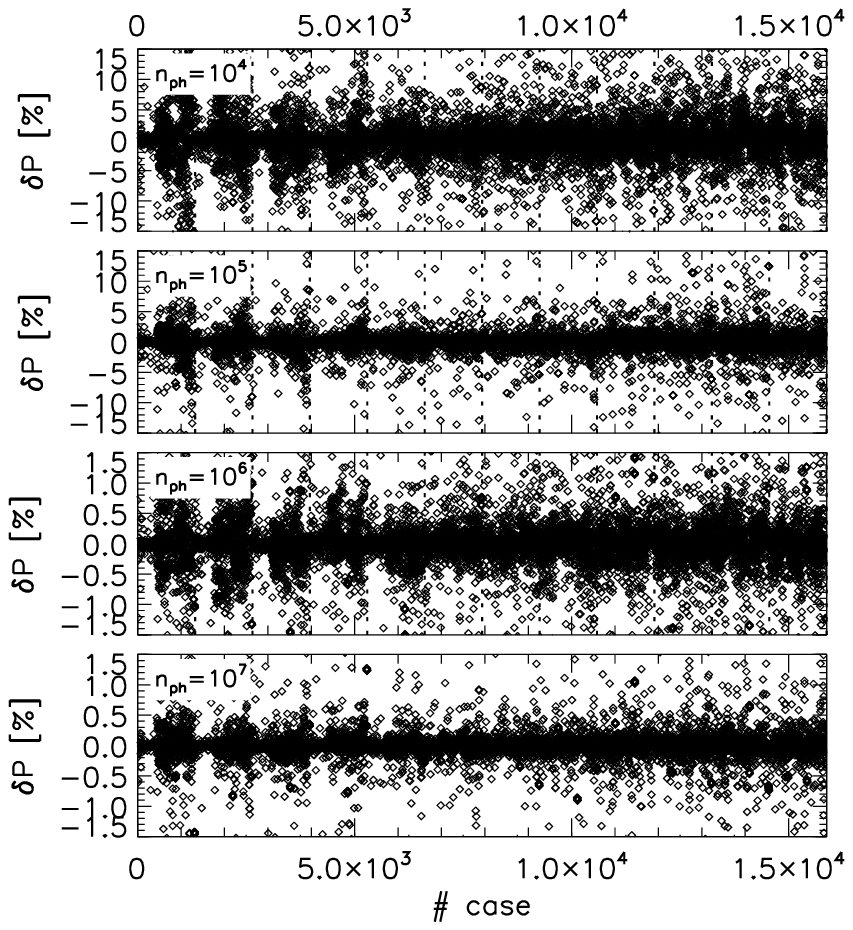}
      \caption{\label{plot_dP_vRTE_fig} Differences in polarisation, $\delta P$, 
      for the solution of the vRTE in conservative, Rayleigh-scattering atmospheres.
      The algorithm uses the pre-conditioned sampling scheme for photon propagation
      directions. 
}
   \end{figure}

Figure (\ref{plot_dI_vRTE_fig}) shows
$\delta$$I$ (=$(I_{\rm{BMC}}$$-$$I_{\rm{ref}})/I_{\rm{ref}}$$\times$100) for the pre-conditioned (top) and classical sampling
schemes (bottom). For the latter, 
Fig. (\ref{plot_dP_vRTE_fig}) shows $\delta P$ (=$(P_{\rm{BMC}}$$-$$P_{\rm{ref}})/P_{\rm{ref}}$$\times$100), 
where $P$=$\sqrt{Q^2+U^2}$.

The $\delta$$I$ graphs reveal that the two sampling schemes
perform generally well for optical thicknesses $\le$4, 
but that the classical scheme destabilises and/or biases the solutions for
larger thicknesses. 
A similar behaviour occurs also for $\delta P$.
Median values for $|\delta I|$ as calculated with the pre-conditioned scheme 
are listed in Table (\ref{convergence_table}). 
For the PBMC solutions, 
the convergence rate is comparable to that for the solution of the scalar RTE.

\begin{figure}[h]
\centering
\includegraphics[width=9cm]{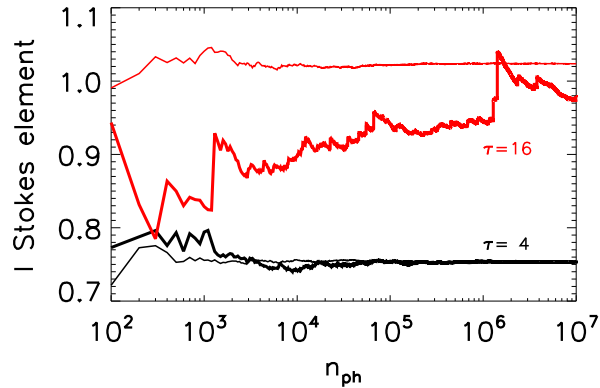}
\caption{\label{CLASSICvsPREC_fig} 
Convergence history of the $I$ Stokes element 
for a conservative Rayleigh atmosphere over a black surface 
with cos(SPA)=cos(OPA)=1 and two different optical thicknesses. 
Thin and thick curves represent the solutions obtained with the pre-conditioned
and classical sampling schemes, respectively. 
Abrupt changes in the solution with the classical sampling scheme for optical thickness of 4 
occur, but they are not discernible at the scale of the graph.
}
\end{figure}

Figure (\ref{CLASSICvsPREC_fig}) offers some insight into the stability issue with 
the classical sampling scheme.
It shows the convergence history for the $I$ Stokes
element for a cos(OPA)=cos(SPA)=1 viewing/illumination geometry 
and varying optical thicknesses above a black surface. 
(OPA/SPA stand for observer/solar polar angles, Table \ref{disort_table}.)
The most striking feature of Fig. (\ref{CLASSICvsPREC_fig}) is that the classical sampling scheme
 produces  abrupt changes in the solution with effects 
that may not go away even after many photon simulations. 
The instabilities become more frequent and noticeable 
for the larger optical thicknesses.
Referring to Eq. (\ref{fphitheta_eq}) and Fig. (\ref{pthetaphi_fig}), the neglect of polarisation 
in the classical sampling scheme is likely to favor some 
propagation directions rather than others and, in turn, erroneously bias the solution.
Inspection of some of the abrupt changes indicates that they are associated with 
a sequence of photon collision events each with scattering angle $\theta$ 
near 90$^{\circ}$ and therefore likely to be misrepresented by the classical sampling scheme. 
The disturbance  becomes more apparent in optically thick,
conservative media because they allow for many more collisions before the photon is lost. 
Further evidence for the latter comes from the fact that Rayleigh
calculations with $\varpi$$\sim$0.95 or less (not shown)
show no stability issues for any optical thickness in the range tested.
The bottom line is that the primary assumption of the classical sampling scheme, i.e.
that the multi-dimensional integral of Eq. (\ref{matrixprod_eq}) can be approximated by
separate integrals as given  by Eq. (\ref{collins_eq}) plus a subsequent correction, becomes
inappropriate for specific configurations.

\begin{figure}[h]
\centering
\includegraphics[width=9cm]{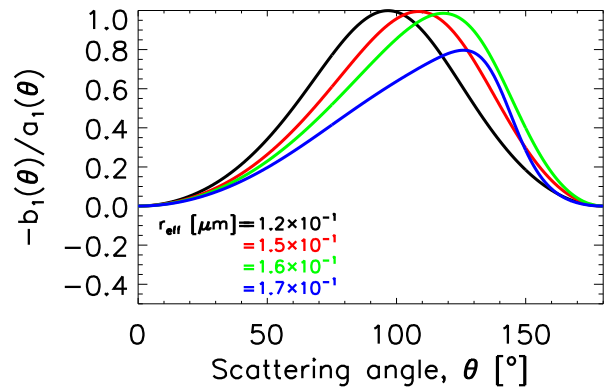}
\caption{\label{B1A1_fig} 
Polarisation in single scattering for monodisperse droplets of various radii and
real refractive index equal to 1.53 at $\lambda$=0.63 $\mu$m. 
}
\end{figure}

\begin{figure}[h]
\centering
\includegraphics[width=9cm]{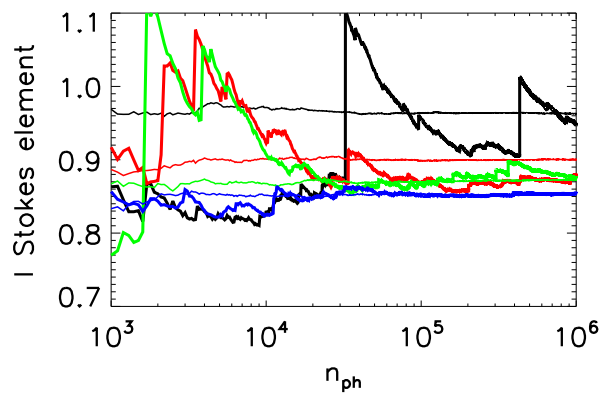}
\caption{\label{CLASSICvsPREC_mie_fig} 
Convergence history for scattering by the monodisperse droplets of Fig. (\ref{B1A1_fig}).
Thin and thick curves represent the solutions obtained with the pre-conditioned
and classical sampling schemes, respectively. 
The classical scheme produces inconsistent solutions for strongly polarising conditions.
The calculations assumed optical thickness equal to 16 and cos(OPA)=cos(SPA)=1.
}
\end{figure}

The idea is confirmed by investigating the solution to the vRTE in other polarising media.
For this purpose, we produced scattering matrices at $\lambda$=0.63 $\mu$m 
for monodisperse droplets of 
real refractive index equal to 1.53 and a few radii from
1.2$\times$10$^{-1}$ to 1.7$\times$10$^{-1}$ $\mu$m. 
Figure (\ref{B1A1_fig}) shows the corresponding 
$-$$b_1$($\theta$)/$a_1$($\theta$) ratios,  
which are properties of the media but also the corresponding degrees of polarisation for
photons scattered one single time. 
Referring to the structure of Eqs. (\ref{ftheta_eq})--(\ref{fphitheta_eq}), it is apparent
that smaller $|$$b_1$($\theta$)/$a_1$($\theta$)$|$ ratios distort  the 
probability density function $f(\theta,\phi; q \neq 0)$ less with respect to the case for $q$=0.
The convergence history for the solutions to the multiple scattering problem 
in a medium of optical thickness equal to 16 and cos(OPA)=cos(SPA)=1 are 
shown in Fig. (\ref{CLASSICvsPREC_mie_fig}).
They reveal that the classical scheme performs poorly in the more strongly
polarising media, 
but performs similarly to the pre-conditioned sampling algorithm
in less polarising conditions.

To the best of our knowledge, there have been no previous reports of 
difficulties using BMC algorithms 
with classical sampling, probably because benchmarking solutions
for optically thick Rayleigh atmospheres had not been readily available.
This example serves to highlight the importance of 
benchmarking solutions in the literature.
 
\subsection{\label{pol_mie_rad_sec}Polarised Mie scattering}

We tested our PBMC algorithm against a number of vRTE solutions in 
Mie-scattering media.
The Stokes vectors for radiation emerging from a conservative atmosphere with 
so-called haze-L scattering particles 
have been tabulated by de Haan et al. (\cite{dehaanetal1987}) 
from calculations based on the doubling-adding method.
Table (\ref{dehaanetal1987_table5}) (and Table \ref{dehaanetal1987_table6} of the Online Material) 
show some of their solutions and the corresponding PBMC calculations. 
The $I$ Stokes element from both calculation methods generally agrees 
to the fourth decimal place for $n_{\rm{ph}}$=10$^9$. 
Typically, solutions accurate to within one per cent in $I$ are obtained for 
$n_{\rm{ph}}$=10$^5$. 
Polarisation is small in all cases 
investigated in Table (\ref{dehaanetal1987_table5}). As a consequence, the 
convergence of the $Q$, $U$ and $V$ elements is slower than for $I$. 
In the Online Material, we extend the comparison by considering the
results published by Garcia \& Siewert (\cite{garciasiewert1986}) for scattering within a 
Venus-like atmosphere. 
The good match of our PBMC results 
attests to the capacity of our algorithm to 
produce accurate solutions to elaborate scattering problems.

\begin{table*}[th] 
\centering
\caption{\label{dehaanetal1987_table5} Solutions to the Stokes vector in a
conservative, 
haze-L atmosphere of optical thickness equal to 1 and 
cos(SPA)=0.5. 
The de Haan et al. (\cite{dehaanetal1987}) solutions 
are extracted from their Table 5.
}
\begin{tabular}{*{7}{l}}
\hline
\{cos(OPA);  & de Haan et al. & \multicolumn{4}{c}{PBMC, $n_{\rm{ph}}$=}  \\
Azimuth\} & \multicolumn{1}{c}{(\cite{dehaanetal1987})} & 
\multicolumn{1}{c}{10$^{5}$} &
\multicolumn{1}{c}{10$^{6}$} &
\multicolumn{1}{c}{10$^{7}$} &
\multicolumn{1}{c}{10$^{8}$} &
\multicolumn{1}{c}{10$^{9}$}
 \\
\hline
$\left. \begin{tabular}{@{\ }l@{}} 
   \{0.1; 0.\} \\ 
  \end{tabular}\right.$ 
&
$\left( \begin{tabular}{@{\ }l@{}} 
   {$+$}{1.10269} \\    
   {$+$}{0.004604} \\ 
   {$+$}{0.} \\ 
   {$+$}{0.}
  \end{tabular}\right)$ 
  & 
$\left( \begin{tabular}{@{\ }l@{}} 
   {$+$}{1.101852} \\    
   {$+$}{0.004629} \\ 
   {$+$}{0.000002} \\ 
   {$+$}{0.000002}
  \end{tabular}\right)$ 
  &
$\left( \begin{tabular}{@{\ }l@{}} 
   {$+$}{1.103338} \\    
   {$+$}{0.004588} \\ 
   {$+$}{0.000000} \\ 
   {$+$}{0.000000}
  \end{tabular}\right)$ 
&
$\left( \begin{tabular}{@{\ }l@{}} 
   {$+$}{1.102915} \\    
   {$+$}{0.004601} \\ 
   {$+$}{0.000004} \\ 
   {$-$}{0.000002}
  \end{tabular}\right)$ 
&
$\left( \begin{tabular}{@{\ }l@{}} 
   {$+$}{1.102821} \\    
   {$+$}{0.004603} \\ 
   {$-$}{0.000001} \\ 
   {$-$}{0.000001}
  \end{tabular}\right)$ 
&
$\left( \begin{tabular}{@{\ }l@{}} 
   {$+$}{1.102866} \\    
   {$+$}{0.004605} \\ 
   {$+$}{0.000000} \\ 
   {$+$}{0.000000}
  \end{tabular}\right)$ 
 \\
$\left. \begin{tabular}{@{\ }l@{}} 
   \{0.5; 0.\} \\ 
  \end{tabular}\right.$ 
&
$\left( \begin{tabular}{@{\ }l@{}} 
   {$+$}{0.31943} \\    
   {$-$}{0.002881} \\ 
   {$+$}{0.} \\ 
   {$+$}{0.}
  \end{tabular}\right)$ 
  & 
$\left( \begin{tabular}{@{\ }l@{}} 
   {$+$}{0.319251} \\  
   {$-$}{0.002927} \\ 
   {$+$}{0.000001} \\ 
   {$+$}{0.000006}
  \end{tabular}\right)$ 
  &
$\left( \begin{tabular}{@{\ }l@{}} 
   {$+$}{0.320067} \\  
   {$-$}{0.002894} \\ 
   {$-$}{0.000005} \\ 
   {$+$}{0.000002}
  \end{tabular}\right)$ 
&
$\left( \begin{tabular}{@{\ }l@{}} 
   {$+$}{0.319427} \\    
   {$-$}{0.002871} \\ 
   {$-$}{0.000001} \\ 
   {$+$}{0.000001}
  \end{tabular}\right)$ 
  &
$\left( \begin{tabular}{@{\ }l@{}} 
   {$+$}{0.319394} \\     
   {$-$}{0.002877} \\ 
   {$-$}{0.000001} \\ 
   {$+$}{0.000000}
  \end{tabular}\right)$ 
&
$\left( \begin{tabular}{@{\ }l@{}} 
   {$+$}{0.319410} \\     
   {$-$}{0.002880} \\ 
   {$-$}{0.000001} \\ 
   {$+$}{0.000000}
  \end{tabular}\right)$ 
 \\    
$\left. \begin{tabular}{@{\ }l@{}} 
   \{1.0; 0.\} \\ 
  \end{tabular}\right.$ 
&
$\left( \begin{tabular}{@{\ }l@{}} 
   {$+$}{0.033033} \\    
   {$-$}{0.002979} \\ 
   {$+$}{0.} \\ 
   {$+$}{0.}
  \end{tabular}\right)$ 
  & 
$\left( \begin{tabular}{@{\ }l@{}} 
   {$+$}{0.032659} \\    
   {$-$}{0.002955} \\ 
   {$+$}{0.000040} \\ 
   {$+$}{0.000006}
  \end{tabular}\right)$ 
  &
$\left( \begin{tabular}{@{\ }l@{}} 
   {$+$}{0.032963} \\      
   {$-$}{0.002976} \\ 
   {$-$}{0.000005} \\ 
   {$+$}{0.000002}
  \end{tabular}\right)$ 
&
$\left( \begin{tabular}{@{\ }l@{}} 
   {$+$}{0.033024} \\      
   {$-$}{0.002977} \\ 
   {$+$}{0.000001} \\ 
   {$+$}{0.000000}
  \end{tabular}\right)$ 
&
$\left( \begin{tabular}{@{\ }l@{}} 
   {$+$}{0.033019} \\    
   {$-$}{0.002977} \\ 
   {$-$}{0.000001} \\ 
   {$+$}{0.000000}
  \end{tabular}\right)$ 
&
$\left( \begin{tabular}{@{\ }l@{}} 
   {$+$}{0.033034} \\  
   {$-$}{0.002979} \\ 
   {$-$}{0.000001} \\ 
   {$+$}{0.000000}
  \end{tabular}\right)$ 
 \\   
 \hline
$\left. \begin{tabular}{@{\ }l@{}} 
   \{0.1; 30.\} \\ 
  \end{tabular}\right.$ 
&
$\left( \begin{tabular}{@{\ }l@{}} 
   {$+$}{0.66414} \\    
   {$+$}{0.000303} \\ 
   {$-$}{0.002770} \\ 
   {$+$}{0.000038}
  \end{tabular}\right)$ 
  & 
$\left( \begin{tabular}{@{\ }l@{}} 
   {$+$}{0.662924} \\    
   {$+$}{0.000390} \\ 
   {$-$}{0.002766} \\ 
   {$+$}{0.000054}
  \end{tabular}\right)$ 
  &
$\left( \begin{tabular}{@{\ }l@{}} 
   {$+$}{0.663431} \\    
   {$+$}{0.000301} \\ 
   {$-$}{0.002736} \\ 
   {$+$}{0.000039}
  \end{tabular}\right)$ 
&
$\left( \begin{tabular}{@{\ }l@{}} 
   {$+$}{0.664643} \\   
   {$+$}{0.000310} \\ 
   {$-$}{0.002769} \\ 
   {$+$}{0.000038}
  \end{tabular}\right)$ 
&
$\left( \begin{tabular}{@{\ }l@{}} 
   {$+$}{0.664342} \\     
   {$+$}{0.000302} \\ 
   {$-$}{0.002770} \\ 
   {$+$}{0.000038}
  \end{tabular}\right)$ 
&
$\left( \begin{tabular}{@{\ }l@{}} 
   {$+$}{0.664298} \\     
   {$+$}{0.000302} \\ 
   {$-$}{0.002770} \\ 
   {$+$}{0.000038}
  \end{tabular}\right)$ 
 \\    
$\left. \begin{tabular}{@{\ }l@{}} 
   \{0.5; 30.\} \\ 
  \end{tabular}\right.$ 
&
$\left( \begin{tabular}{@{\ }l@{}} 
   {$+$}{0.25209} \\   
   {$-$}{0.001444} \\ 
   {$-$}{0.004141} \\ 
   {$+$}{0.000017}
  \end{tabular}\right)$ 
  & 
$\left( \begin{tabular}{@{\ }l@{}} 
   {$+$}{0.253527} \\  
   {$-$}{0.001471} \\ 
   {$-$}{0.004180} \\ 
   {$-$}{0.000003}
  \end{tabular}\right)$ 
  &
$\left( \begin{tabular}{@{\ }l@{}} 
   {$+$}{0.252656} \\  
   {$-$}{0.001428} \\ 
   {$-$}{0.004139} \\ 
   {$+$}{0.000012}
  \end{tabular}\right)$ 
&
$\left( \begin{tabular}{@{\ }l@{}} 
   {$+$}{0.252107} \\   
   {$-$}{0.001445} \\ 
   {$-$}{0.004135} \\ 
   {$+$}{0.000017}
  \end{tabular}\right)$ 
&
$\left( \begin{tabular}{@{\ }l@{}} 
   {$+$}{0.252055} \\   
   {$-$}{0.001444} \\ 
   {$-$}{0.004137} \\ 
   {$+$}{0.000018}
  \end{tabular}\right)$ 
&
$\left( \begin{tabular}{@{\ }l@{}} 
   {$+$}{0.252060} \\      
   {$-$}{0.001444} \\ 
   {$-$}{0.004140} \\ 
   {$+$}{0.000018}
  \end{tabular}\right)$ 
 \\
$\left. \begin{tabular}{@{\ }l@{}} 
   \{1.0; 30.\} \\ 
  \end{tabular}\right.$ 
&
$\left( \begin{tabular}{@{\ }l@{}} 
   {$+$}{0.033033} \\    
   {$-$}{0.001489} \\ 
   {$-$}{0.002580} \\ 
   {$+$}{0.}
  \end{tabular}\right)$ 
  & 
$\left( \begin{tabular}{@{\ }l@{}} 
   {$+$}{0.032689} \\     
   {$-$}{0.001472} \\ 
   {$-$}{0.002629} \\ 
   {$-$}{0.000002}
  \end{tabular}\right)$ 
  &
$\left( \begin{tabular}{@{\ }l@{}} 
   {$+$}{0.032989} \\    
   {$-$}{0.001506} \\ 
   {$-$}{0.002578} \\ 
   {$+$}{0.000001}
  \end{tabular}\right)$ 
&
$\left( \begin{tabular}{@{\ }l@{}} 
   {$+$}{0.033071} \\   
   {$-$}{0.001492} \\ 
   {$-$}{0.002580} \\ 
   {$+$}{0.000000}
  \end{tabular}\right)$ 
&
$\left( \begin{tabular}{@{\ }l@{}} 
   {$+$}{0.033041} \\   
   {$-$}{0.001488} \\ 
   {$-$}{0.002580} \\ 
   {$+$}{0.000000}
  \end{tabular}\right)$ 
&
$\left( \begin{tabular}{@{\ }l@{}} 
   {$+$}{0.033050} \\   
   {$-$}{0.001490} \\ 
   {$-$}{0.002581} \\ 
   {$+$}{0.000000}
  \end{tabular}\right)$ 
 \\         
\hline
\end{tabular}
\end{table*}

\section{\label{disk_section}Planetary phase curves} 

The fact that the convergence rate of MC integration is 
independent of the dimension of the integral, Eq. (\ref{MCintegration_eq}), 
can be used to efficiently estimate 
the net radiation scattered from the planet.
In the Solar System, Venus represents a unique demonstration 
of how disk-integrated
polarisation can be used to infer a planet's cloud composition
(Coffeen, \cite{coffeen1969}; Hansen \& Hovenier, \cite{hansenhovenier1974}).
At remote distances from Earth, 
exoplanets will not be spatially resolvable in the near future and, thus, 
their investigation will necessarily rely on disk-integrated measurements. 
Initial attempts to investigate the optical properties of 
exoplanet atmospheres in reflected light by means of 
polarisation have been made (Berdyugina et al., 
\cite{berdyuginaetal2008,berdyuginaetal2011}, 
Wiktorowicz, \cite{wiktorowicz2009}). 
Foreseeably, a new generation of telescopes and instruments will provide the 
technical capacity to detect and characterise a variety of exoplanets.

To explore the disk-integration schemes of \S\ref{diskschemes_ref}, we utilised
a few configurations relevant to both Rayleigh and Venus-like atmospheres. 
Essentially, the disk-integration scheme selects the entry point of the photon into the atmosphere. 
The three-dimensional photon trajectory is then traced through the medium. 
The PBMC algorithm is implemented over a spherical shell description of the planet's 
atmosphere, which allows us to investigate phenomena related to atmospheric curvature and 
stratification at large star-planet-observer phase angles.

\subsection{\label{disk_ray_section}Rayleigh phase curves}

Buenzli \& Schmid (\cite{buenzlischmid2009}) have investigated with a 
FMC algorithm the phase curves of Rayleigh-scattering planets.
Their study expands on earlier work (e.g. Kattawar \& Adams, \cite{kattawaradams1971}) 
by systematically exploring the parameter space 
(optical thickness, atmospheric single scattering albedo 
and Lambert surface albedo).
Madhusudhan \& Burrows (\cite{madhusudhanburrows2012}) have also produced
Rayleigh phase curves on the basis
of analytical solutions to the plane-parallel problem. 
Rayleigh scattering may provide a first approximation to the
interpretation of a planet's phase curve.  
It is, however, of limited usefulness in the general understanding of 
possibly occurring atmospheres. 
In such cases, more flexible treatments including Mie and other non-Rayleigh
forms of scattering are needed.
Thus,  
Stam et al. (\cite{stametal2006}) have 
devised an efficient technique for disk-integration based on the 
plane-parallel approximation that can deal with arbitrary scattering particles
for planets with horizontally-uniform atmospheres.

We produced Rayleigh phase curves for the
configurations listed in Table (\ref{buenzlischmid_table})
with the \textit{visible}-disk integration scheme of \S\ref{visibledisk_sec}
and compared them to 
those published by Buenzli \& Schmid (\cite{buenzlischmid2009}). 
Specific properties of the curves such as the geometric or spherical 
albedo, or the value and position of the polarisation peak
have been discussed in that work, 
and are not discussed further here. 
With the PBMC algorithm all properties are evaluated at the
specified phase angles without having to bin (and possibly extrapolate) 
in phase angle.

\begin{table} 
\caption{\label{buenzlischmid_table} 
Parameters in the investigation of disk integration 
for both conservative and non-conservative Rayleigh atmospheres.
The total number of cases amounts to 13$\times$9$\times$3=351. 
}
\begin{flushleft}
\begin{tabular}{l}
\hline
\hline
\multicolumn{1}{l}{Optical thickness:} \\
0.01, 0.05, 0.1, 0.2, 0.3, 0.4, 0.6, 0.8, 1, 2, 5, 10, 30\\
\multicolumn{1}{l}{Single scattering albedo:} \\
0.1, 0.2, 0.4, 0.6, 0.8, 0.9, 0.95, 0.99, 1\\
\multicolumn{1}{l}{Surface albedo:} \\
0, 0.3, 1 \\
\hline
\hline
\end{tabular}
\end{flushleft}
\end{table}

The agreement between Buenzli \& Schmid (\cite{buenzlischmid2009}) 
and our PBMC calculations is very good. 
For $n_{\rm{ph}}$=10$^6$, the median of the absolute differences in $F_I$ between the
two approaches over the $\alpha$=7.5--132.5$^{\circ}$ range is about 0.1\%, 
which is consistent with the accuracy targeted by Buenzli \& Schmid (\cite{buenzlischmid2009}). 
Since the computational time is dictated by the number of photon realisations, it
turns out that the computational cost 
is comparable in both the spatially-resolved problems of \S\ref{pol_ray_rad_sec} and in
the spatially-unresolved problems discussed here.
In other words, integrating over the disk involves a computational time comparable to 
obtaining the solution over a localised region of the planet. 
With the \textit{visible}-disk integration scheme, further, 
the convergence properties of the algorithm become independent of phase angle. 

\begin{figure}
\centering
\includegraphics[width=7cm]{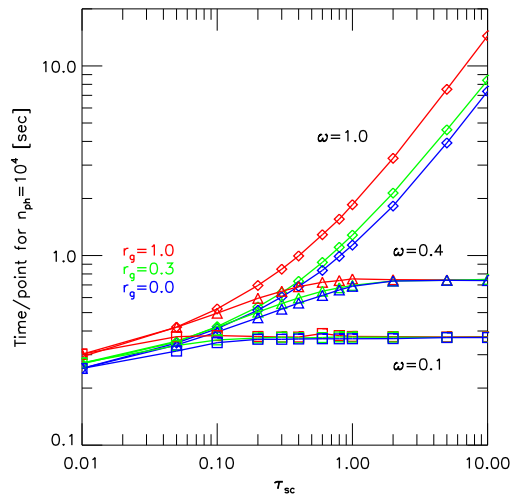}
\caption{\label{time_fig} Computational time per phase curve
point for $n_{\rm{ph}}$=10$^4$ and Rayleigh-scattering calculations.
For a full phase curve with, for instance, 34 evenly-separated points 
from 7.5 to 172.5$^{\circ}$ (as for Fig. \ref{rayleighLC_fig}), 
the computational time is 34 times what is indicated in
the plot. The curves correspond to different values of the single scattering albedo of the
medium, $\varpi$, and the Lambert surface albedo, $r_g$.
}
\end{figure}

Figure (\ref{rayleighLC_fig}) shows phase curves for $F_I$ and $F_Q$/$F_I$ with 
$\varpi$=1, $r_g$=1 and $\tau$=0.1, 0.4, 1, 2, 5 and 10, and $n_{\rm{ph}}$=10$^4$. 
Because we use the $xz$ meridional plane to refererence the Stokes vector 
$\mathbf{I}$, the ratio $F_Q$/$F_I$ is consistent with positive polarisation in
the $xz$ plane perpendicular to the scattering plane. 
Figure (\ref{time_fig}) shows the computational times on a 
2.6 GHz desktop computer for an average point of a phase curve 
and $n_{\rm{ph}}$=10$^4$. 
Computational times are platform-dependent and not often published in the literature, 
which prevents a comparison with the performance of other algorithms.

\subsection{\label{disk_venus_section} Venus phase curves}

Venus is a well-known example that demonstrates the potential of disk-integrated polarimetry 
in the remote investigation of clouds
(Coffeen, \cite{coffeen1969}; Hansen \& Hovenier, \cite{hansenhovenier1974}). 
The disk-integrated polarisation of Venus is small 
but sensitive to wavelength and phase angle, 
facts that were exploited by Hansen \& Hovenier (\cite{hansenhovenier1974}) to
characterise the droplets that make up the Venus upper clouds. 
Venus sets a valuable precedent for the eventual investigation of exoplanetary 
clouds with polarimetry. 

As a further assessment of our PBMC algorithm, we looked into the Venus polarisation 
phase curves. This analysis has the added value of comparing with real planetary measurements. 
From the visible through the near-infrared, the Venus clouds are optically thick
and close to fully conservative.  
The analysis, thus, provides insight into the performance of the PBMC 
algorithm in conditions that require many photon collisions per simulation.

\begin{figure}
   \centering
   \includegraphics[width=10.5cm]{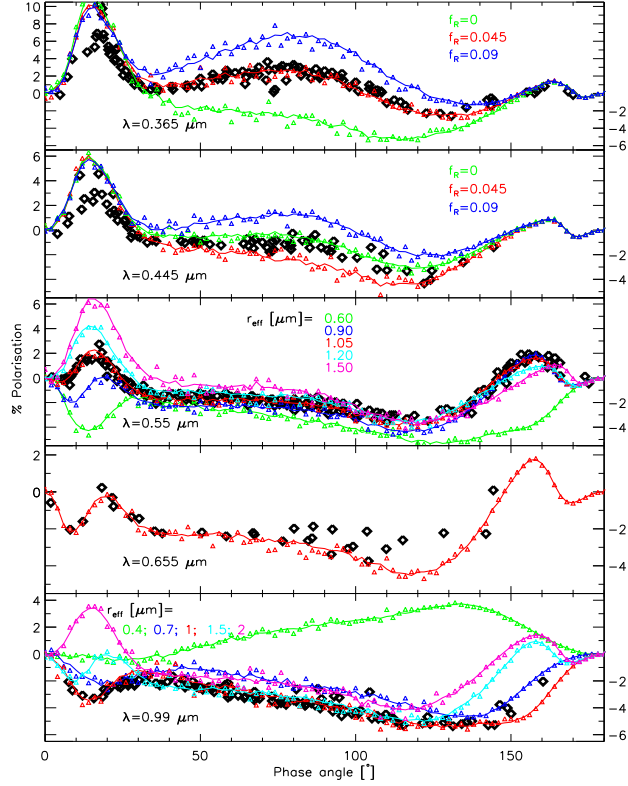}
      \caption{\label{venusfig1_fig} Polarisation phase curves for Venus. Black diamonds
      are the measurements used in Hansen \& Hovenier (\cite{hansenhovenier1974}). 
      Color symbols and curves are our PBMC calculations for $n_{\rm{ph}}$=10$^4$
      and 10$^5$, respectively. This figure can be compared to
      Figs. 4, 8--9, and  11--12 in Hansen \& Hovenier 
      (\cite{hansenhovenier1974}).
      }
   \end{figure}

Figure (\ref{venusfig1_fig}) shows (black diamonds) the digitised
data points for the degree of linear polarisation utilised by 
Hansen \& Hovenier (\cite{hansenhovenier1974}) 
at 0.365, 0.445, 0.55, 0.655 and 0.99 $\mu$m. 
The color symbols are our PBMC 
calculations for $n_{\rm{ph}}$=10$^4$, and the underlying solid curves are the 
calculations for $n_{\rm{ph}}$=10$^5$. 
For the modelling, we use the prescriptions for particle size distributions 
(gamma-distribution, effective radius $r_{\rm{eff}}$=1.05, 
effective variance $v_{\rm{eff}}$=0.07), 
refractive indices and
atmospheric single scattering albedo inferred by 
Hansen \& Hovenier (\cite{hansenhovenier1974}). 
We assume that the atmosphere is made up of 
a single slab of optical thickness equal to 30 overlaying a fully-reflective
Lambert surface. 
The legends in the panel give additional information about the 
Rayleigh-scattering component, $f_{\rm{R}}$ (see 
Hansen \& Hovenier, \cite{hansenhovenier1974}), which becomes important 
at UV wavelengths, and various $r_{\rm{eff}}$ values. 
For $n_{\rm{ph}}$=10$^4$ and the \textit{visible}-disk scheme
of \S\ref{visibledisk_sec}, 
the computational time per point in the phase curve 
is about 8 secs. For the curves of Fig. (\ref{venusfig1_fig}), 
we took 2$^{\rm{o}}$-increments in $\alpha$, 
which entails that the full phase curve for $n_{\rm{ph}}$=10$^4$
is produced in about 12 min. The statistical dispersion of the PBMC calculations is  
smaller than the dispersion associated with the measurements and, from a practical 
viewpoint, it seems practical to truncate to $n_{\rm{ph}}$=10$^4$.
The phase curves of Fig. (\ref{venusfig1_fig}) can be directly compared to the 
model calculations of Figs. 4, 8--9, and  11--12 in Hansen \& Hovenier 
(\cite{hansenhovenier1974}), which confirms the good agreement between both approaches. 

\begin{figure}
   \centering
   \includegraphics[width=7cm]{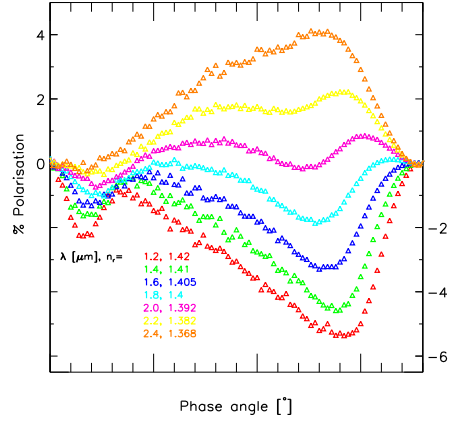}
      \caption{\label{venusfig2_fig} Polarisation phase curves for Venus calculated with our
      PBMC algorithm with $n_{\rm}$=10$^5$ at near-infrared wavelengths for a 
       H$_2$SO$_4$/H$_2$O dilution at 75\%. We adopted a single scattering albedo of 1 in
       all cases. $n_r$ is the refractive index in each case.}
   \end{figure}

In addition, we produced polarisation phase curves at 
wavelengths from 1.2 to 2.4 $\mu$m for the droplets' size distribution given
above 
and real-only refractive indices based on a 75\% H$_2$SO$_4$/H$_2$O solution by mass
(Hansen \& Hovenier, \cite{hansenhovenier1974}).
They are shown in Fig. (\ref{venusfig2_fig}), that illustrates further 
the sensitivity of polarisation to wavelength.

\begin{figure}
   \centering
   \includegraphics[width=7cm]{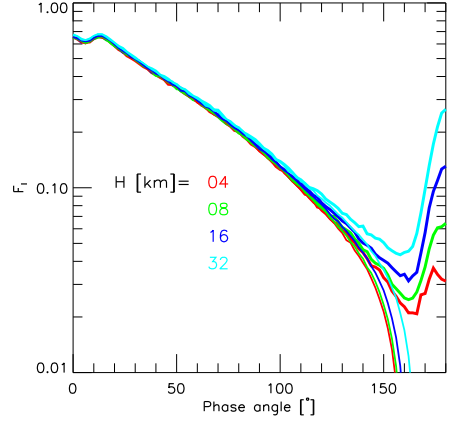}
      \caption{\label{venusfig3_fig} Phase curves  $F_I$ in the optical for Venus-like planets 
      with atmospheres stratified according to the given scale heights.
      }
   \end{figure}

As a final exercise, we explored the appropriateness of integrating over the
\textit{visible} disk, \S\ref{visibledisk_sec}, against the more comprehensive
approach of \S\ref{entiredisk_sec}. 
For this, we took as a basis the Venus atmosphere at 0.55 $\mu$m described 
above. Differences are expected to arise when the atmosphere is vertically
extended. Thus, we stratified the total optical thickness of the atmosphere (=30) 
with scale heights $H$ (the e-folding length for changes in the
$\gamma$ extinction coefficient in the vertical) of 4, 8, 16 and 32 km. 
Both disk-integration schemes produce nearly identical results (not shown)
for the $F_Q$/$F_I$ ratio. 
In contrast, the differences in $F_I$, Fig. (\ref{venusfig3_fig}), can be significant
as the planet approaches inferior conjunction, especially 
for the larger scale heights. 
Figure (\ref{venusfig3_fig}) provides valuable clues to choose the
appropriate disk-integration scheme for specific applications. 
The figure does also show that 
stratification and curvature effects become important for sufficiently 
large phase angles and $H$/$R_p$ ratios.

\section{\label{summary_section}Summary and future work}

We have presented a novel Pre-conditioned Backward Monte Carlo (PBMC) algorithm
based on Backward Monte Carlo integration 
to solve the vector Radiative Transport Equation (vRTE) in planetary atmospheres. 
A unique feature of our PBMC algorithm is that it pre-conditions the 
scattering matrix before sampling the \textit{incident} propagation direction at a photon collision. 
Pre-conditioning retains some of the information associated with the
polarisation state of photons, a feature shown to be critical 
for the correct treatment of 
conservative, optically-thick, strongly-polarising media. 
This is, to the best of our knowledge, the first investigation to report
the numerical difficulties that may occur in BMC algorithms (and possibly FMC as well) when
polarisation is ignored in sampling propagation directions. 
We give extensive evidence that 
our proposed pre-conditioned sampling scheme ensures the stability of the PBMC algorithm.

We explored the performance of the PBMC algorithm, 
showing that it consistently produces solutions accurate to better 
than 0.01\% in the Stokes element $I$ when compared to published benchmarks 
provided that enough photon trajectories are simulated. 
Our extensive assessment exercise, that includes accuracies and computational times,  
should help potential users assess the pros/cons of the method. 
We believe that similar exercises should become common place in the investigation of 
vRTE solvers.

In its spherical shell version, 
our PBMC algorithm is well suited to evaluate the net radiation scattered by a spatially-unresolved planet. 
This feature is particularly interesting in the investigation of the phase curves of 
Solar System planets and exoplanets. We proposed two disk-integration schemes 
and showed that integration over the \textit{visible} disk incurs a computational cost comparable 
to solving the vRTE over a localised region of the planet, provided that the 
spatial details of the emerging radiation can be overlooked. 
Thus far, we have focused on planetary atmospheres that may be vertically stratified but are
otherwise homogeneous in the horizontal direction. 
Future work will extend the disk-integration scheme to horizontally
inhomogeneous planets, thus accounting for a full three-dimensional description of the planet. 
With more and more three-dimensional simulations of the 
chemical, dynamical and energetic configuration of exoplanets becoming available from
General Circulation Models, 
there is a niche of opportunity for our PBMC algorithm in the framework of exoplanet studies. 
Similar ideas will also be explored for disk-integrated thermal emission
and for the simultaneous spectral-and-disk integration of both scattered and 
thermally-emitted radiation. 

A one-slab, plane-parallel version of our PBMC algorithm is available upon request. 
Making the algorithm publicly available will hopefully encourage comparative investigations 
of vRTE solvers. 

\begin{acknowledgements}
We gratefully acknowledge N.I. Ignatiev for confirmation of the definition of angles 
in DISORT,  Richard Turner for assistance with the computer cluster 'The Grid' at ESA/ESTEC, 
and the Grupo de Ciencias Planetarias at UPV/EHU, Spain, for access to
their computational resources.

\end{acknowledgements}

\begin{appendix}

\section{\label{appendixa} PBMC algorithm implementation}

The implementation of our PBMC algorithm follows the formulation 
by O'Brien (\cite{obrien1992,obrien1998}), that we extend to include polarisation. 
Other BMC formulations exist, 
which differ mainly in their 
definition of the statistical estimator's kernel 
(e.g. Collins et al., \cite{collinsetal1972}; 
Marchuk et al., \cite{marchuketal1980}; 
Postylyakov, \cite{postylyakov2004}).  
For completeness, we here sketch the practical details of the algorithm. 

Starting from \{$\mathbf{x_{\rm{0}}, s_{\rm{0}}}$\},
recurrent use of Eq. (\ref{Ikchvar_eq}) for the first few pairs
\{$\mathbf{x_{\rm{k}}, s_{\rm{k}}}$\} leads to:
\begin{eqnarray*}
\mathbf{I(\mathbf{x_{\rm{0}},s_{\rm{0}}})}= 
(1-a(\mathbf{x_{\rm{0}},x_{\rm{0b}}}))
(\mathcal{L_B}(\mathbf{x_{\rm{0}},s_{\rm{0}}})+ 
\mathcal{B}\mathbf{I(\mathbf{x_{\rm{0b}},s_{\rm{0b}}})}) 
\nonumber
\\
+a(\mathbf{x_{\rm{0}},x_{\rm{0b}}})
(\mathcal{L_A}(\mathbf{x_{\rm{0}},s_{\rm{0}}})+ 
\mathcal{A} \mathbf{I(\mathbf{x_{\rm{0a}},s_{\rm{0a}}})})
\nonumber
\\
\mathbf{I(\mathbf{x_{\rm{0b}},s_{\rm{0b}}})}= 
(1-a(\mathbf{x_{\rm{0b}},x_{\rm{0bb}}}))
(\mathcal{L_B}(\mathbf{x_{\rm{0b}},s_{\rm{0b}}})+ 
\mathcal{B}\mathbf{I(\mathbf{x_{\rm{0bb}},s_{\rm{0bb}}})}) 
\nonumber
\\
+a(\mathbf{x_{\rm{0b}},x_{\rm{0bb}}})
(\mathcal{L_A}(\mathbf{x_{\rm{0b}},s_{\rm{0b}}})+ 
\mathcal{A} \mathbf{I(\mathbf{x_{\rm{0ba}},s_{\rm{0ba}}})})
\nonumber
\\
\mathbf{I(\mathbf{x_{\rm{0a}},s_{\rm{0a}}})}= 
(1-a(\mathbf{x_{\rm{0a}},x_{\rm{0ab}}}))
(\mathcal{L_B}(\mathbf{x_{\rm{0a}},s_{\rm{0a}}})+ 
\mathcal{B}\mathbf{I(\mathbf{x_{\rm{0ab}},s_{\rm{ab}}})}) 
\nonumber
\\
+a(\mathbf{x_{\rm{0a}},x_{\rm{0ab}}})
(\mathcal{L_A}(\mathbf{x_{\rm{0a}},s_{\rm{0a}}})+ 
\mathcal{A} \mathbf{I(\mathbf{x_{\rm{0aa}},s_{\rm{0aa}}})})
\nonumber
\\
\mathbf{I(\mathbf{x_{\rm{0bb}},s_{\rm{0bb}}})}= 
(1-a(\mathbf{x_{\rm{0bb}},x_{\rm{0bbb}}}))
(\mathcal{L_B}(\mathbf{x_{\rm{0bb}},s_{\rm{0bb}}})+ 
\mathcal{B}\mathbf{I(\mathbf{x_{\rm{0bbb}},s_{\rm{0bbb}}})}) 
\nonumber
\\
+a(\mathbf{x_{\rm{0bb}},x_{\rm{0bbb}}})
(\mathcal{L_A}(\mathbf{x_{\rm{0bb}},s_{\rm{0bb}}})+ 
\mathcal{A} \mathbf{I(\mathbf{x_{\rm{0bba}},s_{\rm{0bba}}})})
\nonumber
\\
\mathbf{I(\mathbf{x_{\rm{0ba}},s_{\rm{0ba}}})}= 
(1-a(\mathbf{x_{\rm{0ba}},x_{\rm{0bab}}}))
(\mathcal{L_B}(\mathbf{x_{\rm{0ba}},s_{\rm{0ba}}})+ 
\mathcal{B}\mathbf{I(\mathbf{x_{\rm{0bab}},s_{\rm{0bab}}})}) 
\nonumber
\\
+a(\mathbf{x_{\rm{0ba}},x_{\rm{0bab}}})
(\mathcal{L_A}(\mathbf{x_{\rm{0ba}},s_{\rm{0ba}}})+ 
\mathcal{A} \mathbf{I(\mathbf{x_{\rm{0baa}},s_{\rm{0baa}}})})
\nonumber
\\
\mathbf{I(\mathbf{x_{\rm{0ab}},s_{\rm{0ab}}})}= 
(1-a(\mathbf{x_{\rm{0ab}},x_{\rm{0abb}}}))
(\mathcal{L_B}(\mathbf{x_{\rm{0ab}},s_{\rm{0ab}}})+ 
\mathcal{B}\mathbf{I(\mathbf{x_{\rm{0abb}},s_{\rm{0abb}}})}) 
\nonumber
\\
+a(\mathbf{x_{\rm{0ab}},x_{\rm{0abb}}})
(\mathcal{L_A}(\mathbf{x_{\rm{0ab}},s_{\rm{0ab}}})+ 
\mathcal{A} \mathbf{I(\mathbf{x_{\rm{0aba}},s_{\rm{0aba}}})})
\nonumber
\\
\mathbf{I(\mathbf{x_{\rm{0aa}},s_{\rm{0aa}}})}= 
(1-a(\mathbf{x_{\rm{0aa}},x_{\rm{0aab}}}))
(\mathcal{L_B}(\mathbf{x_{\rm{0aa}},s_{\rm{0aa}}})+ 
\mathcal{B}\mathbf{I(\mathbf{x_{\rm{0aab}},s_{\rm{0aab}}})}) 
\nonumber
\\
+a(\mathbf{x_{\rm{0aa}},x_{\rm{0aab}}})
(\mathcal{L_A}(\mathbf{x_{\rm{0aa}},s_{\rm{0aa}}})+ 
\mathcal{A} \mathbf{I(\mathbf{x_{\rm{0aaa}},s_{\rm{0aaa}}})})
\nonumber
\end{eqnarray*}
A summation series for $\mathbf{I(\mathbf{x_{\rm{0}},s_{\rm{0}}})}$ 
is obtained by sequentially inserting the 
$\mathbf{I(\mathbf{x_{\rm{kb}},s_{\rm{kb}}})}$ and
$\mathbf{I(\mathbf{x_{\rm{ka}},s_{\rm{ka}}})}$ into the corresponding 
$\mathbf{I(\mathbf{x_{\rm{k}},s_{\rm{k}}})}$. 
In doing so, each $\mathbf{I(\mathbf{x_{\rm{k}},s_{\rm{k}}})}$ turns into a double summation
of increasingly higher dimension integrals. 
In the PBMC framework, each of those integrals is estimated by their integrands
at properly selected values of the integration variables. 

The overall process, however, is greatly simplified if at each step only
one of the two summations is pursued. 
The structure of Eq. (\ref{Ikchvar_eq}), with 
coefficients $1$$-$$a(\mathbf{x_{\rm{k}},x_{\rm{kb}}})$ and $a(\mathbf{x_{\rm{k}},x_{\rm{kb}}})$,
suggests the way to proceed. 
In the more general case, it is convenient to draw a random number $\varrho$$\in$[0, 1] and follow 
the $\mathcal{A}$ summation if $a(\mathbf{x_{\rm{k}},x_{\rm{kb}}})$$\ge$$\varrho$$>$0 
and the $\mathcal{B}$ summation if $a(\mathbf{x_{\rm{k}},x_{\rm{kb}}})$$<$$\varrho$$<$1.

The ultimate goal of the PBMC algorithm is to estimate the Stokes vector at the detector 
from a number $n_{\rm{ph}}$ of single photon experiments. 
For a fixed \{$\mathbf{x_{\rm{0}}, s_{\rm{0}}}$\}, this is done by evaluating:
\begin{equation}
\mathbf{I}\mathbf{(x_{\rm{0}},s_{\rm{0}})}= 
\frac{1}{n_{\rm{ph}}} \sum_{i_{\rm{ph}}=1}^{n_{\rm{ph}}}
<\mathbf{I}^{i_{\rm{ph}}}\mathbf{(x_{\rm{0}},s_{\rm{0}})}>,
\label{ioave_eq}
\end{equation} 
where each $<$$\mathbf{I}^{i_{\rm{ph}}}\mathbf{(x_{\rm{0}},s_{\rm{0}})}$$>$ 
is an estimate based on a single photon simulation.
The estimate becomes statistically meaningful by repeating the process 
$n_{\rm{ph}}$ of times. 
When the integration is over the planetary disk, Eq. (\ref{ioave_eq}) is replaced by
Eqs. (\ref{horak3_eq}) or (\ref{entirediskMC_eq}), and the position $\mathbf{x}_{\rm{0}}$
of entry of the simulated photon into the atmosphere is determined with the corresponding scheme.

The process that yields $<$$\mathbf{I}\mathbf{(x_{\rm{0}},s_{\rm{0}})}$$>$ 
(index $i_{\rm{ph}}$ omitted)  starts 
by tracing the ray from $\mathbf{x}_{\rm{0}}$  in the $-\mathbf{s}_{\rm{0}}$ direction, 
following the instructions:
\begin{enumerate}
\item Initialise 
$<$$\mathbf{I}$($\mathbf{x}_{\rm{0}},\mathbf{s}_{\rm{0}}$)$>$=$\mathbf{0}$, 
$\mathbf{x}_{\rm{k}}$=$\mathbf{x}_{\rm{0}}$, 
$\mathbf{s}_{\rm{k}}$=$\mathbf{s}_{\rm{0}}$, 
$\mathbf{{\mathbb{H}}}_{\rm{k}}$=$\mathbf{{\mathbb{H}}}_{\rm{0}}$ ($\equiv$unity matrix)
and $\textit{w}_{\rm{k}}$=1.

\item Determine $\mathbf{x_{\rm{kb}}}$ and $a(\mathbf{x_{\rm{k}},x_{\rm{kb}}})$. Then:

\begin{enumerate}
\item If $r_g(\mathbf{x_{\rm{kb}}})$=0 or if vector $-\mathbf{s}_{\rm{k}}$ does not intersect the
planet's surface:
\begin{itemize}
\item $g$=$a(\mathbf{x_{\rm{k}},x_{\rm{kb}}})$. 
\item Go for A at step 3.
\end{itemize}

\item Otherwise:
\begin{itemize}
\item $g$=1. 
\item Draw a random number $\varrho$$\in$[0, 1]. Then: 
\begin{itemize}
\item If $a(\mathbf{x_{\rm{k}},x_{\rm{kb}}})$ $\ge$ $\varrho$ $>$ 0, go for A at step 3.
\item If $a(\mathbf{x_{\rm{k}},x_{\rm{kb}}})$ $<$ $\varrho$ $<$ 1,  go for B at step 4.
\end{itemize}
\end{itemize}

\end{enumerate}

\item Going for A: Collision in between boundaries. 

\begin{itemize}
\item Draw a random number $\epsilon_{\rm{ka}}$$\in$[0, 1] and 
displace the photon from $\mathbf{x}_{\rm{k}}$ to $\mathbf{x}_{\rm{ka}}$
along $-\mathbf{s}_{\rm{k}}$ according to Eq. (\ref{chvar_eq}). 
\item At $\mathbf{x}_{\rm{ka}}$, draw a random number $\zeta_{\rm{ka}}$$\in$[0, 1] 
and find $\theta_{\rm{ka}}$ from the probability distribution function of Eq. (\ref{ftheta_eq}). 
This is done by tabulation and subsequent inversion of 
$\int$$f_{\theta}(\theta)$$d\theta$$\in$[0, 1].

\item At $\mathbf{x}_{\rm{ka}}$, find $\phi_{\rm{ka}}$ from the probability distribution function 
of Eq. (\ref{fphitheta_eq}). This is done by means of the rejection method 
and the fact that by construction 2$\pi$$f_{\phi|\theta}$($\phi|\theta$) $\le$2.  

\item Update:

\begin{itemize}

\item
$<$$\mathbf{I}$($\mathbf{x}_{\rm{0}},\mathbf{s}_{\rm{0}}$)$>$$\leftarrow$$<$$\mathbf{I}$($\mathbf{x}_{\rm{0}},\mathbf{s}_{\rm{0}}$)$>$+
$w_{\rm{k}}$$g$$\varpi(\mathbf{x_{\rm{ka}}}) t(\mathbf{x_{\rm{ka}},x_{\odot}})
\mathbb{H}_{\rm{k}} \mathbf{{\mathbb{P}}(x_{\rm{ka}}, s_{\rm{k}}, s_{\odot} )} 
 \mathbf{F_{\odot}} $

\item $w_{\rm{k}}$$\leftarrow$$w_{\rm{k}}$$g$$\varpi(\mathbf{x_{\rm{ka}}})$ 

\item $\mathbb{H}_{\rm{k}}$$\leftarrow$
$\mathbb{H}_{\rm{k}}$${\mathbf{{\mathbb{P}}(x_{\rm{ka}},s_{\rm{k}}, s_{\rm{ka}})}}$/($\mathbb{H}_{\rm{k}}$${\mathbf{{\mathbb{P}}(x_{\rm{ka}},s_{\rm{k}}, s_{\rm{ka}})}}$)$_{1,1}$. 

\item $\mathbf{x}_{\rm{k}}$$\leftarrow$$\mathbf{x}_{\rm{ka}}$ and $\mathbf{s}_{\rm{k}}$$\leftarrow$$\mathbf{s}_{\rm{ka}}$. 

\end{itemize}

\item If $w_{\rm{k}}$$\ge$$\varepsilon_{\rm{ph}}$, go to step 2. Otherwise, go to step 5.

\end{itemize}

\item Going for B: Collision at the bottom boundary. 
Draw random numbers $\zeta_{\rm{kb}}, \eta_{\rm{kb}}$$\in$[0, 1].
\begin{itemize}
\item Displace the photon from $\mathbf{x}_{\rm{k}}$ to $\mathbf{x}_{\rm{kb}}$ 
along $-\mathbf{s}_{\rm{k}}$.
\item At $\mathbf{x}_{\rm{kb}}$, evaluate $\phi_{\rm{kb}}$=$\zeta_{\rm{kb}}$*2$\pi$
and cos($\theta_{\rm{kb}}$)=$\sqrt{\eta_{\rm{kb}}}$. 

\item Update:

\begin{itemize}

\item
$<$$\mathbf{I}$($\mathbf{x}_{\rm{0}},\mathbf{s}_{\rm{0}}$)$>$$\leftarrow$$<$$\mathbf{I}$($\mathbf{x}_{\rm{0}},\mathbf{s}_{\rm{0}}$)$>$+\\
 ${w}_{\rm{k}}$$g$${r_g(\mathbf{x_{\rm{kb}}})}$ 
 $(\mathbf{n(x_{\rm{kb}})}\cdot\mathbf{s_{\odot}})/\pi$ 
$t\mathbf{(x_{\rm{kb}},x_{\rm{\odot}})}
\mathbb{H}_{\rm{k}} \mathbf{F}_{\odot}$

\item $w_{\rm{k}}$$\leftarrow$$w_{\rm{k}}$$g$${r_g(\mathbf{x_{\rm{kb}}})}$ 

\item $\mathbb{H}_{\rm{k}}$$\leftarrow$
$\mathbb{H}_{\rm{k}}$${\mathbf{{\mathbb{P}}(x_{\rm{kb}},s_{\rm{k}}, 
s_{\rm{kb}})}}$/($\mathbb{H}_{\rm{k}}$${\mathbf{{\mathbb{P}}(x_{\rm{kb}},s_{\rm{k}}, s_{\rm{kb}})}}$)$_{1,1}$. 

\item $\mathbf{x}_{\rm{k}}$$\leftarrow$$\mathbf{x}_{\rm{kb}}$ and $\mathbf{s}_{\rm{k}}$$\leftarrow$$\mathbf{s}_{\rm{kb}}$. 

\end{itemize}

\item If $w_{\rm{k}}$$\ge$$\varepsilon_{\rm{ph}}$, go to step 2. Otherwise, go to step 5.

\end{itemize}
 
\item End of $<$$\mathbf{I}$($\mathbf{x}_{\rm{0}},\mathbf{s}_{\rm{0}}$)$>$ loop. 
 
\end{enumerate}

The loop ends when the \textit{weight} $w_{\rm{k}}$ reaches a user-defined threshold $\varepsilon_{\rm{ph}}$
that truncates the summation series. 
$\varepsilon_{\rm{ph}}$ has an impact on both the solution's accuracy 
and execution time. 
Values in the range 10$^{-4}$--10$^{-5}$ are 
adequate for required accuracies of about 0.1\%
in the $I$ Stokes element.

BMC algorithms with classical sampling schemes for photon propagation 
directions have a structure similar to the above. 
In the classical sampling scheme, the \textit{incident} photon directions 
$\mathbf{s'}$ are sampled from the local $f(\theta, \phi)$ for $q$$\equiv$0
(Fig. \ref{pthetaphi_fig}, top panel) rather 
than from the full $f(\theta, \phi)$ of Eqs. (\ref{ftheta_eq})--(\ref{fphitheta_eq})
(Fig. \ref{pthetaphi_fig}, panels for $|q|$$>$0). 
The classical BMC algorithm can be seen as a variation to the above algorithm, 
with the main difference being the definition of $\mathbb{H}_{\rm{k}}$. 

\end{appendix}

\Online


\appendix

\section{Comparison with de Haan et al. (\cite{dehaanetal1987})}

\begin{table}[h] 
\centering
\caption{\label{dehaanetal1987_table6} Solutions to the Stokes vector in a
conservative, 
haze-L atmosphere of optical thickness equal to 1 and 
cos(SPA)=0.1. 
The de Haan et al. (\cite{dehaanetal1987}) solutions 
are extracted from their Table 6.
}
\begin{tabular}{*{7}{l}}
\hline
\{cos(OPA);  & de Haan et al. & \multicolumn{4}{c}{PBMC, $n_{\rm{ph}}$=}  \\
Azimuth\} & \multicolumn{1}{c}{(\cite{dehaanetal1987})} & 
\multicolumn{1}{c}{10$^{5}$} &
\multicolumn{1}{c}{10$^{6}$} &
\multicolumn{1}{c}{10$^{7}$} &
\multicolumn{1}{c}{10$^{8}$} &
\multicolumn{1}{c}{10$^{9}$}  \\
\hline
$\left. \begin{tabular}{@{\ }l@{}} 
   \{0.1; 0.\} \\ 
  \end{tabular}\right.$ 
&
$\left( \begin{tabular}{@{\ }l@{}} 
   {$+$}{2.93214} \\    
   {$+$}{0.009900} \\ 
   {$+$}{0.} \\ 
   {$+$}{0.}
  \end{tabular}\right)$ 
  & 
$\left( \begin{tabular}{@{\ }l@{}} 
   {$+$}{2.927667} \\      
   {$+$}{0.009888} \\ 
   {$+$}{0.000015} \\ 
   {$+$}{0.000007}
  \end{tabular}\right)$ 
  &
$\left( \begin{tabular}{@{\ }l@{}} 
   {$+$}{2.929702} \\   
   {$+$}{0.009908} \\ 
   {$+$}{0.000008} \\ 
   {$+$}{0.000000}
  \end{tabular}\right)$ 
&
$\left( \begin{tabular}{@{\ }l@{}} 
   {$+$}{2.932289} \\  
   {$+$}{0.009902} \\ 
   {$+$}{0.000001} \\ 
   {$+$}{0.000000}
  \end{tabular}\right)$ 
&
$\left( \begin{tabular}{@{\ }l@{}} 
   {$+$}{2.932468} \\   
   {$+$}{0.009899} \\ 
   {$-$}{0.000001} \\ 
   {$+$}{0.000000}
  \end{tabular}\right)$ 
&
$\left( \begin{tabular}{@{\ }l@{}} 
   {$+$}{2.932400} \\    
   {$+$}{0.009899} \\ 
   {$+$}{0.000000} \\ 
   {$+$}{0.000000}
  \end{tabular}\right)$ 
 \\
$\left. \begin{tabular}{@{\ }l@{}} 
   \{0.5; 0.\} \\ 
  \end{tabular}\right.$ 
&
$\left( \begin{tabular}{@{\ }l@{}} 
   {$+$}{0.22054} \\    
   {$+$}{0.000976} \\ 
   {$+$}{0.} \\ 
   {$+$}{0.}
  \end{tabular}\right)$ 
  & 
$\left( \begin{tabular}{@{\ }l@{}} 
   {$+$}{0.222888} \\   
   {$+$}{0.000974} \\ 
   {$+$}{0.000015} \\ 
   {$+$}{0.000001}
  \end{tabular}\right)$ 
  &
$\left( \begin{tabular}{@{\ }l@{}} 
   {$+$}{0.220427} \\       
   {$+$}{0.000975} \\ 
   {$+$}{0.000006} \\ 
   {$+$}{0.000000}
  \end{tabular}\right)$ 
&
$\left( \begin{tabular}{@{\ }l@{}} 
   {$+$}{0.220107} \\     
   {$+$}{0.000975} \\ 
   {$-$}{0.000002} \\ 
   {$+$}{0.000000}
  \end{tabular}\right)$ 
&
$\left( \begin{tabular}{@{\ }l@{}} 
   {$+$}{0.220366} \\    
   {$+$}{0.000976} \\ 
   {$+$}{0.000001} \\ 
   {$+$}{0.000000}
  \end{tabular}\right)$ 
&
$\left( \begin{tabular}{@{\ }l@{}} 
   {$+$}{0.220391} \\   
   {$+$}{0.000976} \\ 
   {$+$}{0.000000} \\ 
   {$+$}{0.000000}
  \end{tabular}\right)$ 
 \\    
$\left. \begin{tabular}{@{\ }l@{}} 
   \{1.0; 0.\} \\ 
  \end{tabular}\right.$ 
&
$\left( \begin{tabular}{@{\ }l@{}} 
   {$+$}{0.009287} \\    
   {$-$}{0.000815} \\ 
   {$+$}{0.} \\ 
   {$+$}{0.}
  \end{tabular}\right)$ 
  & 
$\left( \begin{tabular}{@{\ }l@{}} 
   {$+$}{0.009324} \\      
   {$-$}{0.000809} \\ 
   {$+$}{0.000010} \\ 
   {$+$}{0.000004}
  \end{tabular}\right)$ 
  &
$\left( \begin{tabular}{@{\ }l@{}} 
   {$+$}{0.009345} \\       
   {$-$}{0.000816} \\ 
   {$-$}{0.000012} \\ 
   {$+$}{0.000001}
  \end{tabular}\right)$ 
&
$\left( \begin{tabular}{@{\ }l@{}} 
   {$+$}{0.009328} \\     
   {$-$}{0.000817} \\ 
   {$-$}{0.000001} \\ 
   {$+$}{0.000000}
  \end{tabular}\right)$ 
&
$\left( \begin{tabular}{@{\ }l@{}} 
   {$+$}{0.009292} \\     
   {$-$}{0.000815} \\ 
   {$+$}{0.000000} \\ 
   {$+$}{0.000000}
  \end{tabular}\right)$ 
&
$\left( \begin{tabular}{@{\ }l@{}} 
   {$+$}{0.009287} \\     
   {$-$}{0.000815} \\ 
   {$+$}{0.000000} \\ 
   {$+$}{0.000000}
  \end{tabular}\right)$ 
 \\   
 \hline
$\left. \begin{tabular}{@{\ }l@{}} 
   \{0.1; 30.\} \\ 
  \end{tabular}\right.$ 
&
$\left( \begin{tabular}{@{\ }l@{}} 
   {$+$}{0.76910} \\    
   {$-$}{0.003758} \\ 
   {$+$}{0.003124} \\ 
   {$+$}{0.000012}
  \end{tabular}\right)$ 
  & 
$\left( \begin{tabular}{@{\ }l@{}} 
   {$+$}{0.766669} \\       
   {$-$}{0.003721} \\ 
   {$+$}{0.003114} \\ 
   {$+$}{0.000011}
  \end{tabular}\right)$ 
  &
$\left( \begin{tabular}{@{\ }l@{}} 
   {$+$}{0.767652} \\      
   {$-$}{0.003747} \\ 
   {$+$}{0.003125} \\ 
   {$+$}{0.000012}
  \end{tabular}\right)$ 
&
$\left( \begin{tabular}{@{\ }l@{}} 
   {$+$}{0.768724} \\     
   {$-$}{0.003750} \\ 
   {$+$}{0.003124} \\ 
   {$+$}{0.000012}
  \end{tabular}\right)$ 
&
$\left( \begin{tabular}{@{\ }l@{}} 
   {$+$}{0.769202} \\    
   {$-$}{0.003757} \\ 
   {$+$}{0.003124} \\ 
   {$+$}{0.000012}
  \end{tabular}\right)$ 
&
$\left( \begin{tabular}{@{\ }l@{}} 
   {$+$}{0.769190} \\     
   {$-$}{0.003759} \\ 
   {$+$}{0.003124} \\ 
   {$+$}{0.000012}
  \end{tabular}\right)$ 
 \\    
$\left. \begin{tabular}{@{\ }l@{}} 
   \{0.5; 30.\} \\ 
  \end{tabular}\right.$ 
&
$\left( \begin{tabular}{@{\ }l@{}} 
   {$+$}{0.132828} \\    
   {$+$}{0.000220} \\ 
   {$-$}{0.000525} \\ 
   {$+$}{0.000007}
  \end{tabular}\right)$ 
  & 
$\left( \begin{tabular}{@{\ }l@{}} 
   {$+$}{0.131144} \\     
   {$+$}{0.000180} \\ 
   {$-$}{0.000512} \\ 
   {$+$}{0.000002}
  \end{tabular}\right)$ 
  &
$\left( \begin{tabular}{@{\ }l@{}} 
   {$+$}{0.132665} \\     
   {$+$}{0.000218} \\ 
   {$-$}{0.000523} \\ 
   {$+$}{0.000007}
  \end{tabular}\right)$ 
&
$\left( \begin{tabular}{@{\ }l@{}} 
   {$+$}{0.132663} \\    
   {$+$}{0.000221} \\ 
   {$-$}{0.000528} \\ 
   {$+$}{0.000007}
  \end{tabular}\right)$ 
&
$\left( \begin{tabular}{@{\ }l@{}} 
   {$+$}{0.132708} \\   
   {$+$}{0.000220} \\ 
   {$-$}{0.000526} \\ 
   {$+$}{0.000007}
  \end{tabular}\right)$ 
&
$\left( \begin{tabular}{@{\ }l@{}} 
   {$+$}{0.132740} \\     
   {$+$}{0.000220} \\ 
   {$-$}{0.000525} \\ 
   {$+$}{0.000007}
  \end{tabular}\right)$ 
 \\
$\left. \begin{tabular}{@{\ }l@{}} 
   \{1.0; 30.\} \\ 
  \end{tabular}\right.$ 
&
$\left( \begin{tabular}{@{\ }l@{}} 
   {$+$}{0.009287} \\    
   {$-$}{0.000408} \\ 
   {$-$}{0.000706} \\ 
   {$+$}{0.}
  \end{tabular}\right)$ 
  & 
$\left( \begin{tabular}{@{\ }l@{}} 
   {$+$}{0.009749} \\      
   {$-$}{0.000429} \\ 
   {$-$}{0.000718} \\ 
   {$+$}{0.000002}
  \end{tabular}\right)$ 
  &
$\left( \begin{tabular}{@{\ }l@{}} 
   {$+$}{0.009306} \\  
   {$-$}{0.000412} \\ 
   {$-$}{0.000702} \\ 
   {$+$}{0.000000}
  \end{tabular}\right)$ 
&
$\left( \begin{tabular}{@{\ }l@{}} 
   {$+$}{0.009283} \\    
   {$-$}{0.000407} \\ 
   {$-$}{0.000706} \\ 
   {$+$}{0.000000}
  \end{tabular}\right)$ 
&
$\left( \begin{tabular}{@{\ }l@{}} 
   {$+$}{0.009279} \\    
   {$-$}{0.000407} \\ 
   {$-$}{0.000706} \\ 
   {$+$}{0.000000}
  \end{tabular}\right)$ 
&
$\left( \begin{tabular}{@{\ }l@{}} 
   {$+$}{0.009286} \\   
   {$-$}{0.000408} \\ 
   {$-$}{0.000706} \\ 
   {$+$}{0.000000}
  \end{tabular}\right)$ 
 \\         
\hline
\end{tabular}
\end{table}

\clearpage

\section{Comparison with DISORT}

For an assessment of the PBMC algorithm against the problem of
radiation emerging from a conservative, non-polarising Rayleigh atmosphere
above a Lambert reflecting surface, 
we built a battery of solutions with DISORT 
(Stamnes et al., \cite{stamnesetal1988}). 
DISORT is a well-documented and thoroughly-tested solver of the scalar RTE for monochromatic 
radiation in multiple-scattering media based on the discrete-ordinate method. 

\begin{figure}[b]
   \centering
   \includegraphics[width=10cm]{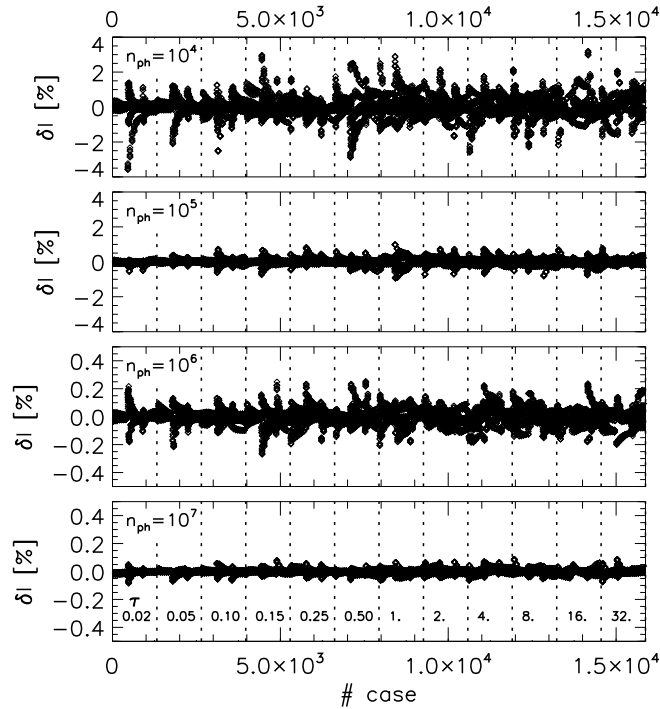}
      \caption{\label{DISORTerror_fig} Relative differences in intensity, $\delta I$,  
      between DISORT and our PBMC algorithm for conservative, 
      non-polarising, Rayleigh atmospheres. 
      The full set of cases is summarised in Table (\ref{disort_table}) of the main text. 
      Dashed vertical lines separate cases run with different atmospheric optical thickness
      ($\tau$, in the graph). 
      The PBMC algorithm runs simultaneously 7$\times$7=49 configurations for the 
      illumination geometry. 
      }
   \end{figure}

Table (\ref{disort_table}) summarises the model parameters and their ranges 
for the comparison exercise. 
They include the atmospheric optical thickness, 
surface albedo, and the three angles of Fig. (\ref{sketch3_fig}). 
The atmospheric single scattering albedo is taken to be one as corresponds to 
conservative scattering. 
In total, the battery comprises 15,876 test cases that explore 
both optically thin and thick atmospheres with viewing/illumination angles 
from zenith inclination to nearly horizontal pointing. 
The PBMC calculations were carried out with 
$n_{\rm{ph}}$=10$^4$, 10$^5$, 10$^6$ and 10$^7$ photon realisations. 
Figure (\ref{DISORTerror_fig}) shows the relative differences, defined as 
$\delta I$=$(I_{\rm{PBMC}}$$-$$I_{\rm{ref}})/I_{\rm{ref}}$$\times$100, 
between the computations with DISORT (reference model) and our PBMC algorithm. 
Median values for $|\delta I|$ are listed in Table (\ref{convergence_table}). 
The convergence rate is consistent with the expected $n^{-1/2}_{\rm{ph}}$ law 
for MC integration.

\begin{table}[h] 
\caption{\label{convergence_table} Median values for $|\delta I|$
in the PBMC test cases of \S\ref{rayleigh_disort} (scalar RTE) and \ref{pol_ray_rad_sec} (vRTE,
pre-conditioned sampling scheme) 
corresponding to conservative Rayleigh atmospheres. }
\begin{flushleft}
\begin{tabular}{ccc}
\hline
\hline
$n_{\rm{ph}}$ & Scalar [\%] & Vector [\%] \\
\hline
10$^4$ & 0.2835 & 0.2820  \\
10$^5$ & 0.0859 & 0.0894  \\
10$^6$ & 0.0302 & 0.0291  \\
10$^7$ & 0.0087 & 0.0093  \\
\hline
\hline 
 \end{tabular}
 \end{flushleft}
 \end{table}

\clearpage
\section{Comparison with Garcia \& Siewert (\cite{garciasiewert1986})}

Scattering in media inspired by the Venus atmosphere has been investigated by
Garcia \& Siewert (\cite{garciasiewert1986}) with a generalised spherical
harmonics method.
We refer to two of their study cases, namely: $L$=13 
($r_{\rm{eff}}$=0.2 $\mu$m,
 $v_{\rm{eff}}$=0.07,
 $\lambda$=0.951 $\mu$m, 
 refractive index of 1.44; akin to the mode-1 haze atop the Venus upper
 clouds) 
 and $L$=60 
($r_{\rm{eff}}$=1.05 $\mu$m,
 $v_{\rm{eff}}$=0.07,
 $\lambda$=0.782 $\mu$m, 
 refractive index of 1.43; akin to the mode-2 droplets that make up the Venus
 upper clouds).
Tables (\ref{garciasiewert1986_table1})--(\ref{garciasiewert1986_table16}) 
show the Garcia \& Siewert (\cite{garciasiewert1986}) solutions and 
our PBMC calculations for $n_{\rm{ph}}$ up to 10$^9$.

\begin{table*}[b]
\centering
\caption{\label{garciasiewert1986_table1} 
Stokes vector element $I$ at the top of the atmosphere for an $L$=13 atmosphere: 
$r_{\rm{eff}}$=0.2 $\mu$m, $v_{\rm{eff}}$=0.07, $\lambda$=0.951 $\mu$m, refractive index of 1.44, 
with optical thickness of one, single scattering albedo $\varpi$=0.99 and surface albedo of 0.1. 
Relative azimuth between the incident and emerging directions is 0 and cos(SPA)=0.2. 
The GS1986 results are extracted
from Table 1 of Garcia \& Siewert (\cite{garciasiewert1986}). For consistency with that reference,
we have preserved their criterion of signs for cos(OPA) (i.e. negative for outgoing radiation)
in Tables (\ref{garciasiewert1986_table1})--(\ref{garciasiewert1986_table16}). 
}
\begin{flushleft}
\begin{tabular}{ccccccc}
\hline
cos(OPA) & GS1986 & 10$^5$ & 10$^6$ & 10$^7$ & 10$^8$ & 10$^9$\\
\hline 
$-$1.0 & 5.4956(-2)  & 5.5020(-2) & 5.4893(-2) & 5.4942(-2) & 5.4964(-2) & 5.4960(-2)\\
$-$0.9 & 9.0491(-2)  & 9.0196(-2) & 9.0349(-2) & 9.0487(-2) & 9.0507(-2) & 9.0512(-2) \\
$-$0.8 & 1.2560(-1)  & 1.2587(-1) & 1.2584(-1) & 1.2572(-1) & 1.2561(-1) & 1.2562(-1) \\
$-$0.7 & 1.6781(-1)  & 1.6779(-1) & 1.6791(-1) & 1.6796(-1) & 1.6787(-1) & 1.6783(-1) \\
$-$0.6 & 2.1934(-1)  & 2.1954(-1) & 2.1939(-1) & 2.1941(-1) & 2.1933(-1) & 2.1936(-1) \\
$-$0.5 & 2.8294(-1)  & 2.8310(-1) & 2.8301(-1) & 2.8297(-1) & 2.8296(-1) & 2.8296(-1) \\
$-$0.4 & 3.6268(-1)  & 3.6377(-1) & 3.6262(-1) & 3.6263(-1) & 3.6269(-1) & 3.6271(-1) \\
$-$0.3 & 4.6523(-1)  & 4.6550(-1) & 4.6568(-1) & 4.6552(-1) & 4.6533(-1) & 4.6526(-1) \\
$-$0.2 & 6.0287(-1)  & 6.0369(-1) & 6.0313(-1) & 6.0298(-1) & 6.0291(-1) & 6.0289(-1) \\
$-$0.1 & 8.0223(-1)  & 8.0163(-1) & 8.0181(-1) & 8.0218(-1) & 8.0229(-1) & 8.0225(-1) \\
\hline
\hline
\end{tabular}
\end{flushleft}
\end{table*}

\begin{table*}[b]
\centering
\caption{\label{garciasiewert1986_table2} 
Same as Table (\ref{garciasiewert1986_table1}) for Stokes vector element $Q$. 
The GS1986 results are extracted
from Table 2 of Garcia \& Siewert (\cite{garciasiewert1986}). 
}
\begin{flushleft}
\begin{tabular}{ccccccc}
\hline
cos(OPA) & GS1986 & 10$^5$ & 10$^6$ & 10$^7$ & 10$^8$ & 10$^9$\\
\hline
$-$1.0 & $-$2.1609(-2)  & $-$2.1546(-2) & $-$2.1557(-2) & $-$2.1619(-2) & $-$2.1610(-2) & $-$2.1610(-2) \\
$-$0.9 & $-$3.2581(-2)  & $-$3.2345(-2) & $-$3.2454(-2) & $-$3.2553(-2) & $-$3.2584(-2) & $-$3.2588(-2) \\
$-$0.8 & $-$3.5048(-2)  & $-$3.4944(-2) & $-$3.5122(-2) & $-$3.5070(-2) & $-$3.5058(-2) & $-$3.5055(-2) \\
$-$0.7 & $-$3.4950(-2)  & $-$3.5194(-2) & $-$3.4992(-2) & $-$3.4998(-2) & $-$3.4962(-2) & $-$3.4953(-2) \\
$-$0.6 & $-$3.2768(-2)  & $-$3.2868(-2) & $-$3.2794(-2) & $-$3.2796(-2) & $-$3.2764(-2) & $-$3.2769(-2) \\
$-$0.5 & $-$2.8664(-2)  & $-$2.8876(-2) & $-$2.8734(-2) & $-$2.8663(-2) & $-$2.8666(-2) & $-$2.8666(-2) \\
$-$0.4 & $-$2.2754(-2)  & $-$2.3029(-2) & $-$2.2803(-2) & $-$2.2746(-2) & $-$2.2754(-2) & $-$2.2759(-2) \\
$-$0.3 & $-$1.5241(-2)  & $-$1.4574(-2) & $-$1.4991(-2) & $-$1.5184(-2) & $-$1.5228(-2) & $-$1.5238(-2) \\
$-$0.2 & $-$6.6429(-3)  & $-$6.4586(-3) & $-$6.6268(-3) & $-$6.6322(-3) & $-$6.6373(-3) & $-$6.6404(-3) \\
$-$0.1 & $+$1.4355(-3)  & $+$1.2341(-3) & $+$1.4922(-3) & $+$1.4053(-3) & $+$1.4481(-3) & $+$1.4408(-3) \\
\hline
\hline
\end{tabular}
\end{flushleft}
\end{table*}

\begin{table*}[h]
\caption{\label{garciasiewert1986_table3} 
Same as Table (\ref{garciasiewert1986_table1}) for relative azimuth between the 
incident and emerging directions equal to $\pi$/2. 
The GS1986 results are extracted
from Table 3 of Garcia \& Siewert (\cite{garciasiewert1986}).
}
\begin{flushleft}
\begin{tabular}{ccccccc}
\hline
cos(OPA) & GS1986 & 10$^5$ & 10$^6$ & 10$^7$ & 10$^8$ & 10$^9$\\
\hline 
$-$1.0 & 5.4956(-2)  & 5.5355(-2) & 5.5019(-2) & 5.4925(-2) & 5.4940(-2) & 5.4943(-2) \\
$-$0.9 & 6.2210(-2)  & 6.1875(-2) & 6.2079(-2) & 6.2196(-2) & 6.2215(-2) & 6.2208(-2) \\
$-$0.8 & 7.0553(-2)  & 7.0913(-2) & 7.0690(-2) & 7.0640(-2) & 7.0550(-2) & 7.0549(-2) \\
$-$0.7 & 8.0201(-2)  & 8.0182(-2) & 8.0114(-2) & 8.0210(-2) & 8.0213(-2) & 8.0199(-2) \\
$-$0.6 & 9.1434(-2)  & 9.1662(-2) & 9.1338(-2) & 9.1450(-2) & 9.1429(-2) & 9.1435(-2) \\
$-$0.5 & 1.0461(-1)  & 1.0419(-1) & 1.0439(-1) & 1.0462(-1) & 1.0463(-1) & 1.0462(-1) \\
$-$0.4 & 1.2018(-1)  & 1.2029(-1) & 1.2015(-1) & 1.2020(-1) & 1.2019(-1) & 1.2019(-1) \\
$-$0.3 & 1.3868(-1)  & 1.3899(-1) & 1.3914(-1) & 1.3882(-1) & 1.3874(-1) & 1.3869(-1) \\
$-$0.2 & 1.6070(-1)  & 1.6060(-1) & 1.6076(-1) & 1.6073(-1) & 1.6074(-1) & 1.6071(-1) \\
$-$0.1 & 1.8701(-1)  & 1.8711(-1) & 1.8712(-1) & 1.8701(-1) & 1.8704(-1) & 1.8702(-1) \\
\hline
\end{tabular}
\end{flushleft}
\end{table*}

\begin{table*}[h]
\caption{\label{garciasiewert1986_table4} 
Same as Table (\ref{garciasiewert1986_table3}) for Stokes vector element $Q$. 
The GS1986 results are extracted
from Table 4 of Garcia \& Siewert (\cite{garciasiewert1986}).
}
\begin{flushleft}
\begin{tabular}{ccccccc}
\hline
cos(OPA) & GS1986 & 10$^5$ & 10$^6$ & 10$^7$ & 10$^8$ & 10$^9$\\
\hline 
$-$1.0 & 2.1609(-2)  & 2.1860(-2) & 2.1643(-2) & 2.1590(-2) & 2.1604(-2) & 2.1608(-2) \\
$-$0.9 & 2.5704(-2)  & 2.5607(-2) & 2.5704(-2) & 2.5705(-2) & 2.5710(-2) & 2.5705(-2) \\
$-$0.8 & 3.0469(-2)  & 3.0671(-2) & 3.0429(-2) & 3.0517(-2) & 3.0475(-2) & 3.0472(-2) \\
$-$0.7 & 3.6046(-2)  & 3.6093(-2) & 3.6023(-2) & 3.6054(-2) & 3.6054(-2) & 3.6049(-2) \\
$-$0.6 & 4.2632(-2)  & 4.2779(-2) & 4.2559(-2) & 4.2631(-2) & 4.2628(-2) & 4.2633(-2) \\
$-$0.5 & 5.0505(-2)  & 5.0141(-2) & 5.0446(-2) & 5.0512(-2) & 5.0509(-2) & 5.0511(-2) \\
$-$0.4 & 6.0066(-2)  & 6.0063(-2) & 6.0026(-2) & 6.0067(-2) & 6.0065(-2) & 6.0068(-2) \\
$-$0.3 & 7.1913(-2)  & 7.2239(-2) & 7.2106(-2) & 7.1984(-2) & 7.1948(-2) & 7.1925(-2) \\
$-$0.2 & 8.6986(-2)  & 8.7084(-2) & 8.7023(-2) & 8.6990(-2) & 8.7004(-2) & 8.6992(-2) \\
$-$0.1 & 1.0690(-1)  & 1.0681(-1) & 1.0688(-1) & 1.0687(-1) & 1.0691(-1) & 1.0691(-1)\\
\hline
\end{tabular}
\end{flushleft}
\end{table*}

\begin{table*}[h]
\caption{\label{garciasiewert1986_table5} 
Same as Table (\ref{garciasiewert1986_table3}) for Stokes vector element $U$. 
The GS1986 results are extracted
from Table 5 of Garcia \& Siewert (\cite{garciasiewert1986}).
}
\begin{flushleft}
\begin{tabular}{ccccccc}
\hline
cos(OPA) & GS1986 & 10$^5$ & 10$^6$ & 10$^7$ & 10$^8$ & 10$^9$\\
\hline 
$-$1.0 & 0.0           & $+$2.9922(-5) & $+$4.1728(-7) & $-$1.6475(-5) & $-$4.7473(-6) & $-$2.6314(-6) \\
$-$0.9 & $-$5.9894(-3) & $-$5.9741(-3) &  $-$6.0376(-3) & $-$5.9877(-3) & $-$5.9862(-3) & $-$5.9921(-3)\\
$-$0.8 & $-$9.1368(-3) & $-$8.8194(-3) &  $-$9.1365(-3) &  $-$9.1444(-3) & $-$9.1407(-3) & $-$9.1363(-3)\\
$-$0.7 & $-$1.2109(-2) & $-$1.1913(-2) & $-$1.2042(-2) & $-$1.2125(-2) & $-$1.2101(-2) & $-$1.2112(-2) \\
$-$0.6 & $-$1.5187(-2) & $-$1.5355(-2) & $-$1.5094(-2) & $-$1.5189(-2) & $-$1.5180(-2) & $-$1.5186(-2) \\
$-$0.5 & $-$1.8526(-2) & $-$1.8432(-2) & $-$1.8539(-2) & $-$1.8551(-2) & $-$1.8527(-2) & $-$1.8530(-2) \\
$-$0.4 & $-$2.2261(-2) & $-$2.2511(-2) & $-$2.2263(-2) & $-$2.2252(-2) & $-$2.2266(-2) & $-$2.2264(-2) \\
$-$0.3 & $-$2.6534(-2) & $-$2.6487(-2) & $-$2.6515(-2) & $-$2.6531(-2) & $-$2.6532(-2) & $-$2.6538(-2) \\
$-$0.2 & $-$3.1534(-2) & $-$3.1663(-2) & $-$3.1532(-2) & $-$3.1545(-2) & $-$3.1543(-2) & $-$3.1537(-2) \\
$-$0.1 & $-$3.7631(-2) & $-$3.7731(-2) & $-$3.7529(-2) & $-$3.7606(-2) & $-$3.7620(-2) & $-$3.7628(-2) \\
\hline
\end{tabular}
\end{flushleft}
\end{table*}

\begin{table*}[h]
\caption{\label{garciasiewert1986_table6} 
Same as Table (\ref{garciasiewert1986_table3}) for Stokes vector element $V$. 
The GS1986 results are extracted
from Table 6 of Garcia \& Siewert (\cite{garciasiewert1986}).
}
\begin{flushleft}
\begin{tabular}{ccccccc}
\hline
cos(OPA) & GS1986 & 10$^5$ & 10$^6$ & 10$^7$ & 10$^8$ & 10$^9$\\
\hline 
$-$1.0 & 0.0           & $-$4.0359(-6)  & $-$2.8272(-6) & $-$1.2913(-6)  & $-$3.3342(-7) & $-$7.6852(-8) \\
$-$0.9 & $-$5.6876(-5) & $-$5.8781(-5) & $-$5.8552(-5) & $-$5.6437(-5) & $-$5.7022(-5)  & $-$5.6950(-5)\\
$-$0.8 & $-$6.8062(-5) & $-$5.7527(-5) & $-$6.6774(-5) & $-$6.8001(-5) & $-$6.8242(-5) & $-$6.8138(-5)\\
$-$0.7 & $-$6.7491(-5) & $-$7.8796(-5) & $-$6.9653(-5) & $-$6.8228(-5) & $-$6.7184(-5) & $-$6.7626(-5)\\
$-$0.6 & $-$5.8655(-5) & $-$6.6799(-5) & $-$6.0468(-5) & $-$5.8863(-5) & $-$5.8413(-5) & $-$5.8706(-5) \\
$-$0.5 & $-$4.2700(-5) & $-$3.3163(-5) & $-$4.4207(-5) & $-$4.3677(-5) & $-$4.2733(-5) & $-$4.2723(-5) \\
$-$0.4 & $-$1.9781(-5) & $-$2.0763(-5) & $-$2.1645(-5) & $-$2.0016(-5) & $-$1.9869(-5) & $-$1.9829(-5) \\
$-$0.3 & $+$1.0762(-5) & $+$1.5177(-5) & $+$1.3012(-5) & $+$1.1121(-5) & $+$1.0477(-5) & $+$1.0727(-5) \\
$-$0.2 & $+$5.0591(-5) & $+$5.4362(-5) & $+$4.9086(-5) & $+$5.0666(-5) & $+$5.0576(-5) & $+$5.0594(-5) \\
$-$0.1 & $+$1.0277(-4) & $+$9.5783(-5) & $+$1.0541(-4) & $+$1.0422(-4) & $+$1.0323(-4) & $+$1.0281(-4)\\
\hline
\end{tabular}
\end{flushleft}
\end{table*}

\begin{table*}[h]
\caption{\label{garciasiewert1986_table7} 
Same as Table (\ref{garciasiewert1986_table3}) for relative azimuth between the 
incident and emerging directions equal to $\pi$. 
The GS1986 results are extracted
from Table 7 of Garcia \& Siewert (\cite{garciasiewert1986}).
}
\begin{flushleft}
\begin{tabular}{ccccccc}
\hline
cos(OPA) & GS1986 & 10$^5$ & 10$^6$ & 10$^7$ & 10$^8$ & 10$^9$\\
\hline 
$-$1.0 & 5.4956(-2) & 5.5381(-2) & 5.4887(-2) & 5.4938(-2) & 5.4940(-2) & 5.4954(-2) \\
$-$0.9 & 5.3085(-2) & 5.2828(-2) & 5.3125(-2) & 5.3082(-2) & 5.3088(-2) & 5.3091(-2) \\ 
$-$0.8 & 5.8688(-2) & 5.8651(-2) & 5.8711(-2) & 5.8702(-2) & 5.8683(-2) & 5.8687(-2) \\ 
$-$0.7 & 6.5653(-2) & 6.5626(-2) & 6.5606(-2) & 6.5650(-2) & 6.5650(-2) & 6.5652(-2) \\ 
$-$0.6 & 7.3678(-2) & 7.3572(-2) & 7.3684(-2) & 7.3720(-2) & 7.3686(-2) & 7.3680(-2) \\ 
$-$0.5 & 8.2754(-2) & 8.2726(-2) & 8.2727(-2) & 8.2719(-2) & 8.2742(-2) & 8.2755(-2) \\ 
$-$0.4 & 9.2933(-2) & 9.2400(-2) & 9.2931(-2) & 9.2896(-2) & 9.2931(-2) & 9.2936(-2) \\ 
$-$0.3 & 1.0421(-1) & 1.0402(-1) & 1.0454(-1) & 1.0425(-1) & 1.0424(-1) & 1.0422(-1) \\ 
$-$0.2 & 1.1641(-1) & 1.1681(-1) & 1.1645(-1) & 1.1640(-1) & 1.1643(-1) & 1.1641(-1) \\ 
$-$0.1 & 1.2913(-1) & 1.2987(-1) & 1.2903(-1) & 1.2913(-1) & 1.2915(-1) & 1.2913(-1) \\
\hline
\end{tabular}
\end{flushleft}
\end{table*}

\begin{table*}[h]
\caption{\label{garciasiewert1986_table8} 
Same as Table (\ref{garciasiewert1986_table7}) for Stokes vector element $Q$. 
The GS1986 results are extracted
from Table 8 of Garcia \& Siewert (\cite{garciasiewert1986}).
}
\begin{flushleft}
\begin{tabular}{ccccccc}
\hline
cos(OPA) & GS1986 & 10$^5$ & 10$^6$ & 10$^7$ & 10$^8$ & 10$^9$\\
\hline 
$-$1.0 & $-$2.1609(-2) & $-$2.1667(-2) & $-$2.1508(-2) & $-$2.1620(-2) & $-$2.1604(-2)  & $-$2.1609(-2) \\
$-$0.9 & $-$8.9018(-3) & $-$8.8077(-3) & $-$8.8265(-3) & $-$8.8926(-3) & $-$8.9011(-3)  & $-$8.9020(-3) \\
$-$0.8 & $-$3.3079(-3) & $-$3.0655(-3) & $-$3.3472(-3) & $-$3.3118(-3) & $-$3.3080(-3)  & $-$3.3078(-3) \\
$-$0.7 & $+$1.2155(-3) & $+$1.2320(-3) & $+$1.2018(-3) & $+$1.2004(-3) & $+$1.2182(-3)  & $+$1.2157(-3)\\
$-$0.6 & $+$5.2168(-3) & $+$5.1933(-3) & $+$5.2157(-3) & $+$5.2033(-3) & $+$5.2184(-3)  & $+$5.2183(-3)\\
$-$0.5 & $+$8.8753(-3) & $+$8.7096(-3) & $+$8.8633(-3) & $+$8.8735(-3) & $+$8.8756(-3)  & $+$8.8752(-3)\\
$-$0.4 & $+$1.2230(-2) & $+$1.1998(-2) & $+$1.2262(-2) & $+$1.2239(-2) & $+$1.2237(-2)  & $+$1.2232(-2)\\
$-$0.3 & $+$1.5202(-2) & $+$1.5458(-2) & $+$1.5417(-2) & $+$1.5247(-2) & $+$1.5219(-2)  & $+$1.5206(-2)\\
$-$0.2 & $+$1.7500(-2) & $+$1.7913(-2) & $+$1.7470(-2) & $+$1.7479(-2) & $+$1.7516(-2)  & $+$1.7503(-2)\\
$-$0.1 & $+$1.8225(-2) & $+$1.8193(-2) & $+$1.8276(-2) & $+$1.8183(-2) & $+$1.8234(-2)  & $+$1.8226(-2)\\
\hline
\end{tabular}
\end{flushleft}
\end{table*}

\begin{table*}[h]
\caption{\label{garciasiewert1986_table9} 
Stokes vector element $I$ at the top of the atmosphere for an $L$=60 atmosphere: 
$r_{\rm{eff}}$=1.05 $\mu$m, $v_{\rm{eff}}$=0.07, $\lambda$=0.782 $\mu$m, refractive index of 1.43, 
with 
optical thickness of one, single scattering albedo $\varpi$=0.99 and surface albedo of 0.1. 
Relative azimuth between the incident and emerging directions is 0 and cos(SPA)=0.2. 
The GS1986 results are extracted from Table 9 of Garcia \& Siewert (\cite{garciasiewert1986}). 
}
\begin{flushleft}
\begin{tabular}{lllllll}
\hline
cos(OPA) & GS1986 & 10$^5$ & 10$^6$ & 10$^7$ & 10$^8$ & 10$^9$\\
\hline 
$-$1.0 & 3.8783(-2) & 4.0259(-2) & 3.8870(-2) & 3.8890(-2) & 3.8815(-2) & 3.8787(-2) \\
$-$0.9 & 6.3881(-2) & 6.5155(-2) & 6.4312(-2) & 6.3962(-2) & 6.3903(-2) & 6.3911(-2) \\ 
$-$0.8 & 9.3567(-2) & 9.3881(-2) & 9.3393(-2) & 9.3590(-2) & 9.3609(-2) & 9.3614(-2) \\ 
$-$0.7 & 1.3570(-1) & 1.3459(-1) & 1.3554(-1) & 1.3575(-1) & 1.3575(-1) & 1.3577(-1) \\ 
$-$0.6 & 1.9652(-1) & 1.9709(-1) & 1.9670(-1) & 1.9659(-1) & 1.9655(-1) & 1.9660(-1) \\ 
$-$0.5 & 2.8490(-1) & 2.8667(-1) & 2.8553(-1) & 2.8539(-1) & 2.8507(-1) & 2.8499(-1) \\ 
$-$0.4 & 4.1401(-1) & 4.1261(-1) & 4.1488(-1) & 4.1407(-1) & 4.1421(-1) & 4.1413(-1) \\ 
$-$0.3 & 6.0620(-1) & 6.1091(-1) & 6.0618(-1) & 6.0609(-1) & 6.0630(-1) & 6.0639(-1) \\ 
$-$0.2 & 9.3026(-1) & 9.2723(-1) & 9.2921(-1) & 9.2998(-1) & 9.3044(-1) & 9.3055(-1) \\ 
$-$0.1 & 1.7498 & 1.7453 & 1.7480 & 1.7504 & 1.7504 & 1.7503 \\ 
\hline
\end{tabular}
\end{flushleft}
\end{table*}

\begin{table*}[h]
\caption{\label{garciasiewert1986_table10} 
Same as Table (\ref{garciasiewert1986_table9}) for Stokes vector element $Q$. 
The GS1986 results are extracted from Table 10 of Garcia \& Siewert (\cite{garciasiewert1986}). }
\begin{flushleft}
\begin{tabular}{lllllll}
\hline
cos(OPA) & GS1986 & 10$^5$ & 10$^6$ & 10$^7$ & 10$^8$ & 10$^9$\\
\hline 
$-$1.0 & $+$3.2087(-3) & $+$3.4480(-3) & $+$3.2325(-3) & $+$3.2196(-3) & $+$3.2120(-3) & $+$3.2096(-3) \\
$-$0.9 & $+$5.6437(-3) & $+$5.7535(-3) & $+$5.6839(-3) & $+$5.6392(-3) & $+$5.6471(-3) & $+$5.6488(-3) \\ 
$-$0.8 & $+$7.8901(-3) & $+$7.9410(-3) & $+$7.8199(-3) & $+$7.8853(-3) & $+$7.8859(-3) & $+$7.8946(-3) \\ 
$-$0.7 & $+$9.9943(-3) & $+$9.8596(-3) & $+$9.9632(-3) & $+$9.9880(-3) & $+$1.0001(-2) & $+$9.9992(-3) \\ 
$-$0.6 & $+$1.1613(-2) & $+$1.1428(-2) & $+$1.1618(-2) & $+$1.1636(-2) & $+$1.1611(-2) & $+$1.1616(-2) \\ 
$-$0.5 & $+$1.2199(-2) & $+$1.2363(-2) & $+$1.2207(-2) & $+$1.2207(-2) & $+$1.2203(-2) & $+$1.2199(-2) \\ 
$-$0.4 & $+$1.1091(-2) & $+$1.1064(-2) & $+$1.1108(-2) & $+$1.1073(-2) & $+$1.1094(-2) & $+$1.1093(-2) \\ 
$-$0.3 & $+$8.3270(-3) & $+$8.3240(-3) & $+$8.3002(-3) & $+$8.3205(-3) & $+$8.3292(-3) & $+$8.3296(-3)\\ 
$-$0.2 & $+$7.3504(-3) & $+$7.5133(-3) & $+$7.4235(-3) & $+$7.3600(-3) & $+$7.3618(-3) & $+$7.3547(-3) \\ 
$-$0.1 & $+$1.9182(-2) & $+$1.9353(-2) & $+$1.9177(-2) & $+$1.9211(-2) & $+$1.9189(-2) & $+$1.9188(-2) \\ 
\hline
\end{tabular}
\end{flushleft}
\end{table*}

\begin{table*}[h]
\caption{\label{garciasiewert1986_table11} 
Same as Table (\ref{garciasiewert1986_table9}) for relative azimuth between the incident and
emerging directions equal to $\pi$/2.
The GS1986 results are extracted from Table 11 of Garcia \& Siewert (\cite{garciasiewert1986}). }
\begin{flushleft}
\begin{tabular}{lllllll}
\hline
cos(OPA) & GS1986 & 10$^5$ & 10$^6$ & 10$^7$ & 10$^8$ & 10$^9$\\
\hline 
$-$1.0 & 3.8783(-2) & 3.8832(-2) & 3.8936(-2) & 3.8833(-2) & 3.8784(-2) & 3.8769(-2) \\
$-$0.9 & 4.3702(-2) & 4.4171(-2) & 4.3742(-2) & 4.3753(-2) & 4.3693(-2) & 4.3691(-2) \\ 
$-$0.8 & 4.9701(-2) & 5.0634(-2) & 4.9657(-2) & 4.9693(-2) & 4.9695(-2) & 4.9677(-2)  \\ 
$-$0.7 & 5.7037(-2) & 5.7559(-2) & 5.7175(-2) & 5.6978(-2) & 5.7033(-2) & 5.7017(-2) \\ 
$-$0.6 & 6.6034(-2) & 6.5713(-2) & 6.5610(-2) & 6.6049(-2) & 6.6014(-2) & 6.6043(-2) \\ 
$-$0.5 & 7.7098(-2) & 7.5906(-2) & 7.6959(-2) & 7.7383(-2) & 7.7145(-2) & 7.7099(-2) \\ 
$-$0.4 & 9.0697(-2) & 9.0893(-2) & 9.0944(-2) & 9.0729(-2) & 9.0789(-2) & 9.0726(-2) \\ 
$-$0.3 & 1.0709(-1) & 1.0592(-1) & 1.0761(-1) & 1.0725(-1) & 1.0713(-1) & 1.0714(-1) \\ 
$-$0.2 & 1.2517(-1) & 1.2588(-1) & 1.2570(-1) & 1.2517(-1) & 1.2524(-1) & 1.2522(-1) \\ 
$-$0.1 & 1.3935(-1) & 1.4012(-1) & 1.3925(-1) & 1.3937(-1) & 1.3941(-1) & 1.3942(-1) \\ 
\hline
\end{tabular}
\end{flushleft}
\end{table*}

\begin{table*}[h]
\caption{\label{garciasiewert1986_table12} 
Same as Table (\ref{garciasiewert1986_table11}) for Stokes vector element $Q$.
The GS1986 results are extracted from Table 12 of Garcia \& Siewert (\cite{garciasiewert1986}). }
\begin{flushleft}
\begin{tabular}{lllllll}
\hline
cos(OPA) & GS1986 & 10$^5$ & 10$^6$ & 10$^7$ & 10$^8$ & 10$^9$\\
\hline 
$-$1.0 & $-$3.2087(-3) & $-$3.2246(-3) & $-$3.1961(-3) & $-$3.2089(-3) & $-$3.2079(-3) & $-$3.2085(-3) \\
$-$0.9 & $-$3.7776(-3) & $-$3.7995(-3) & $-$3.8133(-3) & $-$3.7942(-3) & $-$3.7817(-3) & $-$3.7775(-3) \\ 
$-$0.8 & $-$4.5173(-3) & $-$4.6534(-3) & $-$4.5095(-3) & $-$4.5168(-3) & $-$4.5187(-3) & $-$4.5158(-3) \\ 
$-$0.7 & $-$5.4695(-3) & $-$5.4574(-3) & $-$5.4403(-3) & $-$5.4509(-3) & $-$5.4674(-3) & $-$5.4675(-3) \\ 
$-$0.6 & $-$6.6841(-3) & $-$6.6171(-3) & $-$6.6422(-3) & $-$6.6838(-3) & $-$6.6841(-3) & $-$6.6855(-3) \\ 
$-$0.5 & $-$8.2210(-3) & $-$8.0785(-3) & $-$8.2276(-3) & $-$8.2521(-3) & $-$8.2255(-3) & $-$8.2208(-3) \\ 
$-$0.4 & $-$1.0153(-2) & $-$1.0132(-2) & $-$1.0199(-2) & $-$1.0163(-2) & $-$1.0166(-2) & $-$1.0157(-2) \\ 
$-$0.3 & $-$1.2551(-2) & $-$1.2301(-2) & $-$1.2697(-2) & $-$1.2583(-2) & $-$1.2555(-2) & $-$1.2556(-2) \\ 
$-$0.2 & $-$1.5362(-2) & $-$1.5522(-2) & $-$1.5402(-2) & $-$1.5368(-2) & $-$1.5375(-2) & $-$1.5371(-2) \\ 
$-$0.1 & $-$1.7919(-2) & $-$1.8065(-2) & $-$1.7879(-2) & $-$1.7918(-2) & $-$1.7922(-2) & $-$1.7929(-2) \\ 
\hline
\end{tabular}
\end{flushleft}
\end{table*}

\begin{table*}[h]
\caption{\label{garciasiewert1986_table13} 
Same as Table (\ref{garciasiewert1986_table11}) for Stokes vector element $U$.
The GS1986 results are extracted from Table 13 of Garcia \& Siewert (\cite{garciasiewert1986}). }
\begin{flushleft}
\begin{tabular}{lllllll}
\hline
cos(OPA) & GS1986 & 10$^5$ & 10$^6$ & 10$^7$ & 10$^8$ & 10$^9$\\
\hline 
$-$1.0 & 0.0 & 4.1956(-5) & $-$1.2981(-5) & 6.1562(-6) & 1.1299(-6) & $-$4.9001(-7) \\
$-$0.9 & 8.2963(-4) & 8.3503(-4) & 8.0867(-4) & 8.3203(-4) & 8.2784(-4) & 8.2941(-4) \\ 
$-$0.8 & 1.3068(-3) & 1.3137(-3) & 1.3230(-3) & 1.3182(-3) & 1.3095(-3) & 1.3064(-3) \\ 
$-$0.7 & 1.7878(-3) & 1.8250(-3) & 1.7629(-3) & 1.7827(-3) & 1.7880(-3) & 1.7876(-3) \\ 
$-$0.6 & 2.3127(-3) & 2.3901(-3) & 2.2889(-3) & 2.3131(-3) & 2.3111(-3) & 2.3136(-3) \\ 
$-$0.5 & 2.9041(-3) & 2.8899(-3) & 2.9072(-3) & 2.9146(-3) & 2.9066(-3) & 2.9046(-3) \\ 
$-$0.4 & 3.5776(-3) & 3.6875(-3) & 3.5900(-3) & 3.5616(-3) & 3.5790(-3) & 3.5787(-3) \\ 
$-$0.3 & 4.3260(-3) & 4.2575(-3) & 4.3401(-3) & 4.3411(-3) & 4.3322(-3) & 4.3279(-3) \\ 
$-$0.2 & 5.0624(-3) & 5.0706(-3) & 4.9985(-3) & 5.0585(-3) & 5.0599(-3) & 5.0640(-3) \\ 
$-$0.1 & 5.5033(-3) & 5.6343(-3) & 5.4692(-3) & 5.5019(-3) & 5.5060(-3) & 5.5049(-3) \\ 
\hline
\end{tabular}
\end{flushleft}
\end{table*}

\begin{table*}[h]
\caption{\label{garciasiewert1986_table14} 
Same as Table (\ref{garciasiewert1986_table11}) for Stokes vector element $V$.
The GS1986 results are extracted from Table 14 of Garcia \& Siewert (\cite{garciasiewert1986}). }
\begin{flushleft}
\begin{tabular}{lllllll}
\hline
cos(OPA) & GS1986 & 10$^5$ & 10$^6$ & 10$^7$ & 10$^8$ & 10$^9$\\
\hline 
$-$1.0 & $+$0.0 & $-$4.3645(-6) & $-$1.8591(-7) & $-$6.3359(-7) & $+$4.3425(-7) & $+$9.3999(-9) \\
$-$0.9 & $-$2.3298(-5) & $-$3.8504(-5) & $-$2.3586(-5) & $-$2.4567(-5) & $-$2.3974(-5) & $-$2.3709(-5) \\ 
$-$0.8 & $-$3.3159(-5) & $-$1.7557(-5) & $-$2.8954(-5) & $-$2.9475(-5) & $-$3.2471(-5) & $-$3.3215(-5) \\  
$-$0.7 & $-$4.0605(-5) & $-$4.3482(-5) & $-$3.3880(-5) & $-$4.1344(-5) & $-$4.0458(-5) & $-$4.0574(-5) \\  
$-$0.6 & $-$4.6329(-5) & $-$1.7786(-5) & $-$3.7889(-5) & $-$4.5658(-5) & $-$4.5602(-5) & $-$4.6485(-5) \\   
$-$0.5 & $-$5.0166(-5) & $-$9.0685(-5) & $-$5.0612(-5) & $-$4.9531(-5) & $-$4.8829(-5) & $-$4.9981(-5) \\   
$-$0.4 & $-$5.2027(-5) & $-$3.5571(-5) & $-$5.8161(-5) & $-$5.6519(-5) & $-$4.9828(-5) & $-$5.1378(-5) \\   
$-$0.3 & $-$5.3865(-5) & $-$6.7988(-5) & $-$3.6750(-5) & $-$4.8297(-5) & $-$5.3102(-5) & $-$5.4228(-5) \\   
$-$0.2 & $-$6.1686(-5) & $-$4.4539(-5) & $-$6.4015(-5) & $-$6.1808(-5) & $-$6.1938(-5) & $-$6.1856(-5) \\   
$-$0.1 & $-$8.0698(-5) & $-$1.1810(-4) & $-$9.4732(-5) & $-$8.1532(-5) & $-$8.0507(-5) & $-$8.0468(-5) \\   
\hline
\end{tabular}
\end{flushleft}
\end{table*}

\begin{table*}[h]
\caption{\label{garciasiewert1986_table15} 
Same as Table (\ref{garciasiewert1986_table9}) for relative azimuth between the incident and
emerging directions equal to $\pi$.
The GS1986 results are extracted from Table 15 of Garcia \& Siewert (\cite{garciasiewert1986}). }
\begin{flushleft}
\begin{tabular}{lllllll}
\hline
cos(OPA) & GS1986 & 10$^5$ & 10$^6$ & 10$^7$ & 10$^8$ & 10$^9$\\
\hline 
$-$1.0 & 3.8783(-2) & 3.8659(-2) & 3.8838(-2) & 3.8753(-2) & 3.8783(-2) & 3.8784(-2) \\
$-$0.9 & 4.1409(-2) & 4.2703(-2) & 4.1395(-2) & 4.1426(-2) & 4.1439(-2) & 4.1412(-2) \\
$-$0.8 & 5.1943(-2) & 5.2001(-2) & 5.1773(-2) & 5.1933(-2) & 5.1942(-2) & 5.1937(-2) \\
$-$0.7 & 6.8133(-2) & 6.9284(-2) & 6.8124(-2) & 6.8141(-2) & 6.8127(-2) & 6.8136(-2) \\
$-$0.6 & 9.5937(-2) & 9.6150(-2) & 9.6270(-2) & 9.5912(-2) & 9.5951(-2) & 9.5959(-2) \\
$-$0.5 & 1.3265(-1) & 1.3250(-1) & 1.3319(-1) & 1.3305(-1) & 1.3273(-1) & 1.3267(-1) \\
$-$0.4 & 1.3780(-1) & 1.3838(-1) & 1.3761(-1) & 1.3781(-1) & 1.3788(-1) & 1.3783(-1) \\
$-$0.3 & 1.4338(-1) & 1.4424(-1) & 1.4402(-1) & 1.4344(-1) & 1.4343(-1) & 1.4342(-1) \\
$-$0.2 & 1.9176(-1) & 1.9157(-1) & 1.9172(-1) & 1.9181(-1) & 1.9182(-1) & 1.9182(-1) \\
$-$0.1 & 1.9322(-1) & 1.9453(-1) & 1.9322(-1) & 1.9314(-1) & 1.9325(-1) & 1.9325(-1) \\
\hline
\end{tabular}
\end{flushleft}
\end{table*}

\begin{table*}[h]
\caption{\label{garciasiewert1986_table16} 
Same as Table (\ref{garciasiewert1986_table15}) for Stokes vector element $Q$.
The GS1986 results are extracted from Table 16 of Garcia \& Siewert (\cite{garciasiewert1986}). }
\begin{flushleft}
\begin{tabular}{lllllll}
\hline
cos(OPA) & GS1986 & 10$^5$ & 10$^6$ & 10$^7$ & 10$^8$ & 10$^9$\\
\hline 
$-$1.0 & 3.2087(-3) & 3.2612(-3) & 3.2480(-3) & 3.1974(-3) & 3.2063(-3) & 3.2083(-3) \\
$-$0.9 & 3.4697(-3) & 3.6107(-3) & 3.4775(-3) & 3.4732(-3) & 3.4767(-3) & 3.4697(-3) \\
$-$0.8 & 3.9993(-3) & 4.1837(-3) & 3.9768(-3) & 4.0047(-3) & 3.9993(-3) & 3.9993(-3) \\
$-$0.7 & 3.2661(-3) & 3.3228(-3) & 3.2461(-3) & 3.2540(-3) & 3.2657(-3) & 3.2671(-3) \\
$-$0.6 & 1.0031(-3) & 7.4623(-4) & 1.0070(-3) & 1.0129(-3) & 9.9955(-4) & 1.0025(-3) \\
$-$0.5 & 3.4663(-3) & 3.5630(-3) & 3.4973(-3) & 3.4719(-3) & 3.4727(-3) & 3.4655(-3) \\
$-$0.4 & 1.1649(-2) & 1.1922(-2) & 1.1758(-2) & 1.1639(-2) & 1.1641(-2) & 1.1652(-2) \\
$-$0.3 & 7.3601(-3) & 7.6231(-3) & 7.5405(-3) & 7.3596(-3) & 7.3704(-3) & 7.3621(-3) \\
$-$0.2 & $-$3.5796(-3) & $-$3.7610(-3) & $-$3.6178(-3) & $-$3.5574(-3) & $-$3.5784(-3) & $-$3.5792(-3) \\
$-$0.1 & 1.2076(-2) & 1.2413(-2) & 1.2175(-2) & 1.2137(-2) & 1.2085(-2) & 1.2081(-2) \\
\hline
\end{tabular}
\end{flushleft}
\end{table*}

\clearpage

\section{Rayleigh phase curves}

\begin{figure}[b]
\centering
\includegraphics[width=9cm]{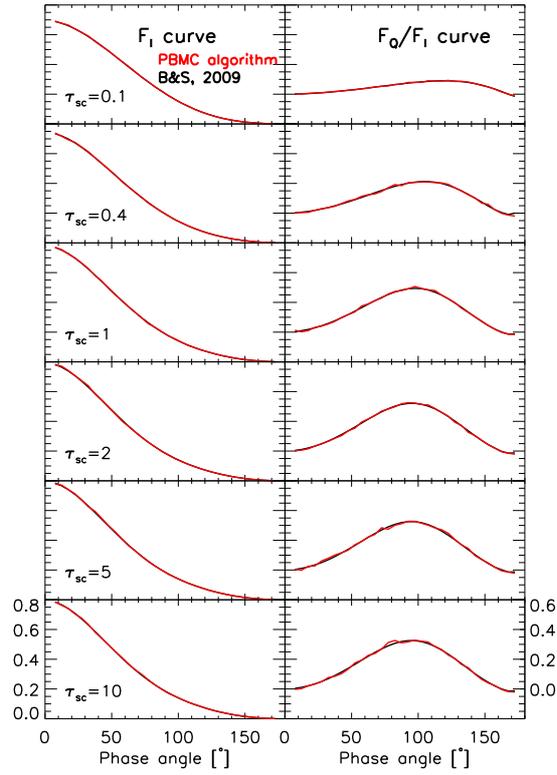}
\caption{\label{rayleighLC_fig} Disk-integrated phase curves for $F_I$ and $F_Q/F_I$ for a
Rayleigh-scattering atmosphere with $\varpi$=1 and $r_g$=1 and 
optical thicknesses from 0.1 to 10. Our PBMC calculations for $n_{\rm{ph}}$=10$^4$
(red curves) are compared
to the calculations by Buenzli \& Schmid (\cite{buenzlischmid2009}) using a forward
Monte Carlo algorithm (black curves) and at least 2$\times$10$^6$ photons per 
exit direction bin over the 30--120$^{\circ}$ $\alpha$-range. 
Both sets of curves are nearly undistinguishable at the scale of the
graph, even for the relatively small number of photon realisations used in 
our PBMC calculations.      
}
\end{figure}


\begin{thebibliography}{}
	
\bibitem[2007]{bailey2007} 
Bailey, J. 2007,
Astrobiol., 7, 320 

\bibitem[2000]{bartelhielscher2000} 
Bartel, S. \&
Hielscher, A.H. 2000,
Appl. Optics, 39, 1580  

\bibitem[2008]{berdyuginaetal2008} 
Berdyugina, S.V.,
Berdyugin, A.V.,
Fluri, D.M., \&
Piirola, V. 2008,
Astrophys. J., 673, L83

\bibitem[2011]{berdyuginaetal2011} 
Berdyugina, S.V.,
Berdyugin, A.V.,
Fluri, D.M., \&
Piirola, V. 2011,
Astrophys. J. Lett., 728, L6

\bibitem[2008]{beuzitetal2008} 
Beuzit, J.-L., 
Feldt, M.,
Dohlen, K.,
Mouillet, D.,
Puget, P. et al. 2008, 
Proc. SPIE, 7014, id. 701418

\bibitem[1996]{bianchietal1996} 
Bianchi, S.,
Ferrara, A. \&
Giovanardi, C. 1996,
Astrophys. J., 465, 127

\bibitem[2012]{boccalettietal2012} 
Boccaletti, A.,
Schneider, J.,
Traub, W.,
Lagage, P.-O., 
Stam, D., et al. 2012, 
Exp. Astr., 34, 355

\bibitem[2009]{buenzlischmid2009} 
Buenzli, E. \&
Schmid, H. M. 2009,
A\&A, 504, 259

\bibitem[1978]{carteretal1978} 
Carter, L.L.,
Horak, H.G. \&
Sandford II, M.T. 1978,
J. Comput. Phys. 26, 119

\bibitem[1959]{cashwelleverett1959} 
Cashwell, E.D. \&
Everett, C.J. 1959,
A practical manual on the Monte Carlo method for random walk problems 
   (Pergamon Press, New York)

\bibitem[1969]{coffeen1969} 
Coffeen, D.A. 1969,
Astron. J., 74, 446

\bibitem[1972]{collinsetal1972} 
Collins, D.G.,
Bl\"attner, W.G., 
Wells, M.B. \& 
Horak, H.G. 1972,
Appl. Optics, 11, 2684

\bibitem[2010]{cornetetal2010} 
Cornet, C., 
C-Labonnote, L. \& 
Szczap, F. 2010, 
JQSRT, 111, 174

\bibitem[1960]{coulsonetal1960} 
Coulson, K.L., 
Dave, J.V. \&  
Sekera, Z. 1960,
Tables related to radiation emerging from a planetary atmosphere with Rayleigh scattering 
      (University of California Press, Berkeley \& Los Angeles)

 \bibitem[1987]{dehaanetal1987} 
 de Haan, J.F.,
 Bosma, P.B. \&
 Hovenier, J.W. 1987,
 A\&A, 183, 371

\bibitem[1957]{dollfus1957} 
Dollfus, A. 1957,
Supplements aux Annales d'Astrophysique, 4, 3

 \bibitem[2010]{emdeetal2010} 
 Emde, C.,
 Buras, R.,
 Mayer, B. \&
 Blumthaler, M. 2010,
 Atmos. Chem. Phys., 10, 383 
 
\bibitem[1994]{fischeretal1994} 
Fischer, O.,
Henning, Th. \& 
Yorke, H.W. 1994,
A\&A, 284, 187

\bibitem[2001]{fordetal2001} 
Ford, E.B.,
Seager, S. \& 
Turner, E.L. 2001,
Nature, 412, 885

\bibitem[2010]{fluriberdyugina2010} 
Fluri, D.M. \& 
Berdyugina, S.V. 2010,
A\&A, 512, A59

\bibitem[2011]{garciamunozpalle2011}
Garc\'ia Mu\~noz, A. \&
Pall\'e, E. 2011, 
JQSRT, 112, 1609 

\bibitem[2011]{garciamunozetal2011}
Garc\'ia Mu\~noz, A.,
Pall\'e, E., 
Zapatero Osorio, M.R. \&
Mart\'in, E.L. 2011,
Geophys. Res. Lett., 38, L14805

\bibitem[2012]{garciamunozetal2012}
Garc\'ia Mu\~noz, A.,
Zapatero Osorio, M.R.,
Barrena, R.,
Monta\~n\'es-Rodr\'iguez, P.,
Mart\'in, E.L. \&
Pall\'e, E. 2012,
Astrophys. J., 755, 103

\bibitem[2014]{garciamunozetal2014}
Garc\'ia Mu\~noz, A.,
P\'erez-Hoyos, S. \&
S\'anchez-Lavega, A. 2014,
A\&A, \textit{in press}, 
DOI: 10.1051/0004-6361/201423531 

\bibitem[2012]{garciamunozmills2012}
Garc\'ia Mu\~noz, A. \&
Mills, F.P. 2012,
A\&A, 547, A22

\bibitem[1986]{garciasiewert1986}
Garcia, R.D.M. \&
Siewert, C.E. 1986, 
JQSRT, 36, 401 

\bibitem[2010]{gayetal2010}
Gay, B.,
Vaillon, R. \&
Meng\"u\c{c}, M.P. 2010,
JQSRT, 111, 287

 \bibitem[1974]{hansenhovenier1974} 
 Hansen, J. E. \& Hovenier, J. W. 1974,
 J. Atmos. Science, 31, 1137

 \bibitem[1974]{hansentravis1974} 
 Hansen, J.E. \& Travis, L.D. 1974,
 Space Science Reviews, 16, 527 

 \bibitem[2000]{hopcraftetal2000} 
 Hopcraft, K.I.,
 Chang, P.C.Y.,
 Walker, J.G. \& 
 Jakeman, E. 2000,
 in Light Scattering from Microstructures,
 ed.\  F. Moreno \& F. Gonz\'alez (Eds.): Lectures 1998, LNP 534 (Springer) 135
  
\bibitem[1950]{horak1950} 
Horak, H.G. 1950,
Astrophys. J., 112, 445

 \bibitem[2007]{joosschmid2007} 
 Joos, F. \&
 Schmid, H.M. 1980,
 A\&A, 463, 1201

 \bibitem[2001]{kaplanetal2001} 
 Kaplan, B.,
 Ledanois, G. \&
 Dr\'evillon, B. 2001,
 Appl. Optics, 40, 2769

 \bibitem[2011]{karalidietal2011} 
 Karalidi, T.,
 Stam, D.M. \& 
 Hovenier, J.W. 2011,
 A\&A, 530, A69 

 \bibitem[2012]{karalidistam2012} 
 Karalidi, T. \&
 Stam, D.M. 2012,
 A\&A, 546, A56


 \bibitem[2012]{karalidietal2012} 
 Karalidi, T.,
 Stam, D.M. \& 
 Hovenier, J.W. 2012,
 A\&A, 548, A90 

 \bibitem[2013]{karalidietal2013} 
 Karalidi, T.,
 Stam, D.M. \& 
 Guirado, D. 2013,
 A\&A, 555, A127 
 
\bibitem[2008]{kasperetal2008} 
Kasper, M.E.,
Beuzit, J.-L.,
Verinaud, C.,
Yaitskova, N.,
Baudoz, P. et al. 2008,
Proc. SPIE, 7015, id. 70151S

\bibitem[1966]{kastner1966} 
Kastner, S.O. 1966, 
JQSRT, 6, 317
 
\bibitem[1971]{kattawaradams1971} 
Kattawar, G.W. \& Adams, C.N. 1971,
Astrophys. J., 167, 183

 \bibitem[2004]{loughmanetal2004}
 Loughman, R.P., 
 Griffioen, E., 
 Oikarinen, L., 
 Postylyakov,  O.V., 
 Rozanov, A. et al. 2004,
 J. Geophys. Res., 109, D06303,
 doi:10.1029/2003JD003854

 \bibitem[1985]{luxkoblinger1991} 
 Lux, I. \&
 Koblinger, L. 1991, 
 Monte Carlo particle transport methods: Neutron and photon calculations. 
 CRC Press Inc., Boca Raton, Florida.

\bibitem[2006]{macintoshetal2006} 
Macintosh, B.,
Graham, J.,
Palmer, D.,
Doyon, R.,
Gavel, D. et al. 2006,
Proc. of the SPIE, 6272, 62720L

\bibitem[2012]{madhusudhanburrows2012} 
Madhusudhan, N. \&
Burrows, A. 2012,
Astrophys. J., 747, 25

\bibitem[1980]{marchuketal1980} 
Marchuk, G.I.,
Mikhailov, G.A.,  
Nazaraliev, M.A.,  
Darbinjan, R.A.,  
Kargin, B.A. \&
Elepov, B.S. 1980, 
The Monte Carlo methods in atmospheric optics
(Springer-Verlag, Berlin Heidelberg).

\bibitem[1977]{michalskystokes1977}
Michalsky, J.J. \& Stokes, R.A. 1977,
Astrophys. J. Lett., 213, L135
	
\bibitem[2002]{mishchenkoetal2002}
Mishchenko, M.I., 
Travis, L.D. \& 
Lacis, A.A. 2002,
Scattering, absorption and emission of light by small particles
(Cambridge University Press, Cambridge)

\bibitem[2003]{modest2003} 
Modest, M.F. 2003,
J. Heat Transfer, 125, 57

\bibitem[1973]{morozhenkoyanovitskii1973} 
Morozhenko, A.V. \& 
Yanovitskii, E.G. 1973,
Icarus, 18, 583
	
\bibitem[2009]{natrajetal2009} 
Natraj, V.,  
Li, K.-F. \&
Yung, Y. L. 2009,
Astrophys. J., 691, 1909

\bibitem[2012]{natrajhovenier2012} 
Natraj, V. \&
Hovenier, J.W. 2012, 
Astrophys. J., 748, 28

\bibitem[1992]{obrien1992} 
O'Brien, D.M. 1992, JQSRT, 48, 41

\bibitem[1998]{obrien1998}
O'Brien, D. M. Monte Carlo integration of the radiative transfer equation in a
scattering medium with stochastic reflecting boundary. JQSRT 1998; 60:573--583. 

 \bibitem[2001]{oikarinen2001}
 Oikarinen, L. 2001,
 J. Geophys. Res., 106, 1533

\bibitem[2004]{postylyakov2004}
Postylyakov, O.V. 2004,
JQSRT, 88, 297

 \bibitem[1985]{santeretal1985} 
 Santer, R.,
 Deschamps, M.,
 Ksanformaliti, L.V. \&
 Dollfus, A. 1985,
 A\&A, 150, 217

 \bibitem[1992]{schmid1992}  
 Schmid,  H.M. 1992,
 A\&A, 254, 224

 \bibitem[2006]{schmidetal2006}  
 Schmid, H.M.,  
 Joos, F. \&
 Tschan, D. 2006,
 A\&A, 452, 657

 \bibitem[2011]{schmidetal2011} 
 Schmid, H.M.,  
 Joos, F.,  
 Buenzli, E. \&
 Gisler, D. 2011,
 Icarus, 212, 701

 \bibitem[2000]{seageretal2000} 
 Seager, S.,  
 Whitney, B.A.  \&
 Sasselov, D.D. 2000,
 Astrophys. J., 540, 504

 \bibitem[2004]{stametal2004}  
 Stam, D.M.,
 Hovenier, J.W. \&
 Waters, L.B.F.M. 2004,
 A\&A, 428, 663

 \bibitem[2008]{stam2008}  
 Stam, D.M. 2008,
 A\&A, 482, 989

 \bibitem[2006]{stametal2006}  
 Stam, D.M., de Rooij, W.A., Cornet, G. \& Hovenier, J. W. 2006,
 A\&A, 452, 669

 \bibitem[1988]{stamnesetal1988}  
 Stamnes, K.,
 Tsay, S.-C.,
 Jayaweera, K. \&
 Wiscombe, W. 1988,
 Appl. Optics, 27, 2502

 \bibitem[1973]{veverka1973} 
 Veverka, J. 1973,
 Icarus, 18, 657
	
 \bibitem[1983]{westetal1983} 
 West, R.A.,
 Hart, H.,
 Hord, C.W.,
 Simmons, K.E.,
 Esposito, L.W., et al. 1983, 
 J. Geophys. Res., 88, 8679

 \bibitem[1991]{westsmith1991} 
 West, R.A. \&
 Smith, P.H. 1991,
 Icarus, 90, 330

 \bibitem[2011]{whitney2011} 
 Whitney, B.A. 2011,
 Bull. Astron. Society of India, 39, 101

\bibitem[2009]{wiktorowicz2009} 
Wiktorowicz, S.J. 2009,
Astrophys. J., 696, 1116

\bibitem[2008]{williamsgaidos2008} 
Williams, D.M. \&
Gaidos, E. 2008,
Icarus, 195, 927

\bibitem[2010]{zuggeretal2010} 
Zugger, M.E.,
Kasting, J.F.,
Williams, D.M.,
Kane, T.J. \&  
Philbrick, C.R. 2010,
Astrophys. J., 723, 1168

\bibitem[2011]{zuggeretal2011} 
Zugger, M.E.,
Kasting, J.F.,
Williams, D.M.,
Kane, T.J. \&  
Philbrick, C.R. 2011,
Astrophys. J., 739, id. 12

\end{thebibliography}
\end{document}